\documentclass[%
 reprint,
superscriptaddress,
nofootinbib,
 amsmath,amssymb,
 aps,
 prr,
longbibliograohy,
]{revtex4-2}
\usepackage{bbold}
\usepackage[breaklinks=true,colorlinks=true
,urlcolor=blue
,anchorcolor=blue
,citecolor=blue
,filecolor=blue
,linkcolor=blue
,menucolor=blue
,pagecolor=blue
,linktocpage=true
,pdfproducer=medialab
,pdfa=true
]{hyperref}
\usepackage{graphicx,hyperref,color}
\usepackage{color}
\usepackage[dvipsnames,svgnames]{xcolor}
\usepackage{graphicx}
\usepackage{dcolumn}
\usepackage{bm}
\usepackage{bbm}
\usepackage{mathtools}
\usepackage{amsmath,braket}
\usepackage{amssymb,physics}
\usepackage{float}

\usepackage{tikz}
\usepackage{circuitikz}
\usetikzlibrary{shapes.geometric} 
\usetikzlibrary{3d,calc}
\usepackage{pgfplots}
\pgfplotsset{compat=1.17}

\usetikzlibrary{decorations.pathreplacing} 
\usetikzlibrary{decorations.markings}
\tikzset{snake it/.style={decorate, decoration=snake}}

\def\w{{\omega}}

\def\ep{{\epsilon}}

\def\rh{\rho}

\def\frac#1#2{{#1\over #2}}

\def\s{\sqrt}

\def\L{{\Lambda}}

\def\be{\begin{equation}}
\def\ee{\end{equation}}
\def\ba{\begin{eqnarray}}
\def\ea{\end{eqnarray}}
\def\de{\partial}

\def\f {\frac}
\def\ti{\tilde}
\def\ap{\alpha}

\def\no{\nonumber \\}

\def\la{\langle}
\def\lb{\rangle}
\def\ep{\epsilon}

\def\vp{\varphi}

\newcommand{\R}{\mathbb{R}}

\newcommand{\bTr}[1]{\mathrm{Tr}\left[#1\right]}

\newcommand{\mint}{\mathrm{int}}

\newcommand{\TFD}{\mathrm{TFD}}

\newcommand{\rmi}{\mathrm{i}}

\newcommand{\mO}{\mathcal{O}}

\usepackage{braket}

\allowdisplaybreaks[4]

\begin{document}
\begin{flushright}
YITP-25-186
\\
\end{flushright}
\title{{\large Non-hermitian Density Matrices from Time-like Entanglement and Wormholes}}

\author{Jonathan Harper}
\affiliation{Center for Gravitational Physics and Quantum Information, Yukawa Institute for Theoretical Physics, Kyoto University, Kitashirakawa Oiwakecho, Sakyo-ku, Kyoto 606-8502, Japan}

\author{Taishi Kawamoto}
\affiliation{Center for Gravitational Physics and Quantum Information, Yukawa Institute for Theoretical Physics, Kyoto University, Kitashirakawa Oiwakecho, Sakyo-ku, Kyoto 606-8502, Japan}

\author{Ryota Maeda}
\affiliation{Center for Gravitational Physics and Quantum Information, Yukawa Institute for Theoretical Physics, Kyoto University, Kitashirakawa Oiwakecho, Sakyo-ku, Kyoto 606-8502, Japan}

\author{Nanami Nakamura}
\affiliation{Center for Gravitational Physics and Quantum Information, Yukawa Institute for Theoretical Physics, Kyoto University, Kitashirakawa Oiwakecho, Sakyo-ku, Kyoto 606-8502, Japan}

\author{Tadashi Takayanagi}
\affiliation{Center for Gravitational Physics and Quantum Information, Yukawa Institute for Theoretical Physics, Kyoto University, Kitashirakawa Oiwakecho, Sakyo-ku, Kyoto 606-8502, Japan}
\affiliation{Inamori Research Institute for Science, 620 Suiginya-cho, Shimogyo-ku, Kyoto 600-8411, Japan}


\begin{abstract}

We extensively explore the connections between time-like entanglement and non-hermitian density matrices in quantum many-body systems. We classify setups where we encounter non-hermitian density matrices into two types: one is due to causal influences under unitary evolutions, and the other is due to non-unitary evolutions in non-hermitian systems. We provide various examples of these setups including interacting harmonic oscillators, two dimensional conformal field theories and holographic dualities. In them, we compute the time-like entanglement entropy and imagitivity, which measures how much density matrices are non-hermitian. In both two classes, typical holographic examples are given by traversable AdS wormholes. We explain how causal influences in a wormhole dual to a pair of non-hermitian quantum systems is possible even without interactions 
between them. We argue that to realize a traversable wormhole we need not only ordinary quantum entanglement but also time-like entanglement.

\end{abstract}

\maketitle


\tableofcontents


\section{Introduction}

Quantum entanglement reveals the essential microscopic structures of a given quantum many-body system. Quantum entanglement describes quantum correlations between two causally disconnected subsystems $A$ and $B$. For a pure quantum state, quantum entanglement is quantified by entanglement entropy, which is defined by the von-Neumann entropy of the reduced density matrix $\rho_A$. Since quantum entanglement is directly related to the vacuum fluctuations, its amount provides a useful measure of degrees of freedom in a quantum many-body system. For example, entanglement entropy plays the role of order parameters in quantum phase transitions
\cite{Vidal:2002rm} and in topological quantum phases \cite{Kitaev:2005dm,Levin:2006zz}. It provides the characterization of degrees of freedom in quantum field theories at fixed points and RG flows \cite{Holzhey:1994we,Calabrese:2004eu,Casini:2004bw,Casini:2012ei}. In gravitational holography \cite{tHooft:1993dmi,Susskind:1994vu}, entanglement entropy in conformal field theories (CFTs) 
can be computed as the area of space-like extremal surfaces in an anti de Sitter space (AdS) \cite{Ryu:2006bv,Ryu:2006ef,Hubeny:2007xt} via the AdS/CFT \cite{Maldacena:1997re}. This implies a novel idea that quantum entanglement in a quantum many-body system gets geometrized by the holography and a spacetime in gravity emerges from quantum entanglement \cite{Swingle:2009bg,VanRaamsdonk:2010pw}. In this context, we can understand that the size in the space-like directions of a gravitational spacetime is equivalent to the amount of quantum entanglement. 

This consideration raises a natural question: can we treat the quantum correlations in the time-like direction as a generalization of quantum entanglement?  This is motivated partly because time evolutions of quantum many-body systems are crucially important to understand the dynamics under non-equilibrium quantum processes and partly because we would also like to understand the emergence of time-like coordinate in holography from a quantum information theoretic idea. In the past literature, the correlation along time direction is highly related to the quantum chaos and ergodicity \cite{Vikram:2022xiq,Fava:2023pac,Kawamoto:2024vzd,Camargo:2025zxr}. However, the probes discussed are correlation functions and highly depends on the choice operators. In order to incorporate the time-like correlations in universal manner, we need to generalize the definition of density matrices and entanglement entropy. 

Recently, the idea of time-like entanglement entropy has been introduced in \cite{Doi:2022iyj,Doi:2023zaf} by allowing the subsystem $A$ to be time-like such that it has causal influences from itself. Refer also to \cite{Narayan:2022afv,Foligno:2023dih,Narayan:2023ebn,Carignano:2023xbz,Chu:2023zah,Grieninger:2023knz,Heller:2024whi,Guo:2024lrr,Carignano:2024jxb,Kawamoto:2025oko,Milekhin:2025ycm,Heller:2025kvp,Chu:2025sjv,Bou-Comas:2024pxf,Xu:2024yvf,Afrasiar:2024ldn,Afrasiar:2024lsi,Nunez:2025gxq,Nunez:2025puk,Nunez:2025ppd,Guo:2025dtq,Bohra:2025mhb,Anastasiou:2025rvz,Li:2025tud,Dulac:2025owj,Das:2025fcd,Giataganas:2025div}
for a partial list of developments of time-like entanglement entropy which are relevant to the present paper. The generalized density matrices $\rho_A$ for a subsystem $A$ now becomes non-hermitian in general and the entropy has a non-vanishing imaginary part. Also it is interesting to note that the non-hermitian density matrices naturally arise in non-hermitian quantum systems, which provide effective descriptions of open quantum systems \cite{Hatano_1996,Bender:2007nj,Castro-Alvaredo:2017udm,Bender:2023cem,Bergholtz:2019deh,Ashida:2020dkc,Shimizu:2025kse}.  Same structure also appears in non-unitary CFTs \cite{Castro-Alvaredo:2017udm,Fukusumi:2025xrj,Fukusumi:2025fir}.
In holography, if we consider a holographic dual to a de Sitter space (dS), so called dS/CFT \cite{Strominger:2001pn}, we encounter non-hermitian CFTs such as three dimensional CFTs with ghost fields \cite{Anninos:2011ui} and the two dimensional CFTs with imaginary valued central charges   \cite{Hikida:2021ese,Hikida:2022ltr}.

A generalization of entanglement entropy to the case where $\rho_A$ is not hermitian was introduced in \cite{Nakata:2021ubr}, so called pseudo entropy. The time-like entanglement entropy is considered as a special example of pseudo entropy.
In holography, it is remarkable that we can still calculate pseudo entropy by computing the area of extremal surfaces \cite{Nakata:2021ubr}, which now include surfaces extending into the time-like or complex directions \cite{Heller:2024whi,Heller:2025kvp,Nunez:2025gxq}. 
In the context of quantum many-body systems, the temporal density matrices were already introduced in earlier time \cite{Banuls:2009jmn,Muller-Hermes:2012irk,Hastings:2014qqa}, aiming at efficient tensor network calculations.  In quantum information theory, temporal density matrices have also been studied in \cite{Fullwood:2022rjd,Parzygnat:2022ldx, Parzygnat:2022pax, Wu:2024ucp, Milekhin:2025ycm}. Other approaches to the temporal correlations has been discussed in \cite{Cotler:2017anu,Fitzsimons:2013gga,Glorioso:2024xan}, where only hermitian density matrices have been employed.

As a further progress, it was noted in \cite{Kawamoto:2025oko} through the calculations of time-like entanglement entropy in several explicit examples that the non-hermitian properties of density matrices are responsible for the causal influences in quantum many-body systems. A clear argument which proves the general connection between the non-hermitian properties of density matrix and the causal influences was provided soon later in the remarkable paper \cite{Milekhin:2025ycm}, where a quantity called imagitivity, which measures how much a given density matrix is non-hermitian was also introduced. 

In the AdS/CFT, when two CFTs are entangled, the total system is equivalent to an eternal AdS black hole \cite{Maldacena:2001kr} where two asymptotically AdS boundaries, where two CFTs each live, are connected through a wormhole so called the Einstein-Rosen bridge. Since there is a horizon between them, the two boundaries are not causally connected. However, if we add interactions between two CFTs, the corresponding wormhole becomes traversable as first discovered in \cite{Gao:2016bin}. As shown in \cite{Kawamoto:2025oko}, we can confirm that the generalized density matrix in the CFTs becomes non-hermitian, being consistent with the presence of causal influences, which we call the class 1 setup. In addition there is another way to construct the traversable wormhole as found in \cite{Kawamoto:2025oko}. This is to consider a non-hermitian deformation (imaginary Janus deformation) of the Hamiltonian of the two CFTs without introducing any interactions between them, called the class 2 setup. In this second example, one may wonder why we can send signals from one side to the other, which we will resolve later. Another exotic feature is that the horizon entropy, which gives the entanglement pseudo entropy between the two CFTs, becomes larger than the undeformed one after the deformation. This also looks at first puzzling as the undeformed state is the thermofield double state and is expected to be maximally entangled for a fixed temperature. 

Motivated by these developments of time-like entanglement, the main purpose of this paper is to provide a general framework to treat time-like entanglement, stimulated by the holography for traversable wormholes and to work out many interesting examples. 
We will classify the non-hermitian density matrices into the above mentioned two classes and give general arguments in each of them. In the class 1, we can explicitly confirm that the causal influences both in the CFTs and in the gravity are equivalent to the non-hermitian properties of the generalized density matrices. We will compute the imagitivity in a quantum mechanics of interacting harmonic oscillators and in a two dimensional free CFT under local excitations. We will show that its agrees with our expectation based on quasi-particle excitations. We will also compute the time-like entanglement entropy and imagitivity in two dimensional CFTs for a subsystem which consist of causally connected double intervals. We will perform this analysis explicitly for both the free fermion CFT and the holographic CFT, which reveals how the behaviors depend on the interactions in CFTs. In the class 2, we will explain how to compute the generalized density matrices in non-hermitian quantum systems where a modified form of conjugation, which we define, plays a crucial role. We will show how the influences without interactions can be possible in non-hermitian systems and use this to resolve the previously mentioned exotic behavior found for traversable AdS wormhole. We will also give a free scalar field theory counter part of the class 2 setup and explicitly compute the Renyi pseudo entropy between two CFTs and the imagitivity. The former turns out to increase under the non-hermitian deformation and agrees with the gravity dual calculation qualitatively, resolving the above mentioned puzzle. Our holographic calculation will also predict one more novel behavior that the pseudo entropy gets linearly decreasing under the time evolution in the presence of non-hermitian deformation. 

This paper is organized as follows. In section \ref{sec:gene}, we will present the general framework of non-hermitian density matrices, classifying it to two classes. For class 1, we will explain how time-like entanglement arises and show how we find causal influences from non-hermitian density matrices. For class 2, we will show how we can construct generalized density matrices in non-hermitian systems by introducing the modified conjugation and explain that we can find influences even without interactions.

In section \ref{sec:LO}, we will compute the second Renyi pseudo entropy and imagitivity in a two dimensional free CFT under local excitations and show that the results agrees with a quasi-particle picture. 

In section \ref{sec:classone}, we will study explicit examples of the class 1 setup. First we calculate the second Renyi pseudo entropy and imagitivity in the coupled harmonic oscillators when the subsystem has internal causal influences. Next we will analyze these quantities in the two dimensional free fermion and holographic CFTs for a double interval subsystem which are causally connected. We will give physical interpretations for both examples. We also find how interactions in CFTs affect these results.

In section \ref{sec:classtwo}, we will study a non-hermitian deformation of thermofield double state, called imaginary Janus deformation. After we give a general prescription, we will analyze an example of the free scalar CFT with this deformation. We will compute the second Renyi entropy and imagitivity in this CFT and give a heuristic interpretation. Finally, we will examine the gravity dual of this imaginary Janus deformation, which is given by a traversable AdS wormhole solution. We will analyze the holographic pseudo entropy and show the results qualitatively agree with the ones from the dual CFT. 

In section \ref{sec:concl}, we will summarize our conclusions and discuss future problems. 

In appendix \ref{app:2cftdets}, we will present details of the calculations of time-like entanglement entropy in two dimensional CFTs for the causally connected double intervals, used in section \ref{sec:TEE}.
In appendix \ref{ap:CHO}, we will give the detailed analysis of the coupled harmonic oscillators presented in section \ref{sec:harmoc}. In appendix \ref{ap:janus}, we will explain the calculations of the Janus deformed CFT which was fully employed in section \ref{sec:imjan}.
In appendix \ref{janusent},  we will provide explicit calculations of holographic entanglement/pseudo entropy in Janus wormhole discussed in Sec.\ref{janholo}.

\section{non-hermitian density matrices in quantum systems}\label{sec:gene}

In quantum mechanics, a quantum state is described by a density matrix $\rho$, which is positive semi-definite and hermitian $\rho^\dagger=\rho$, acting on the Hilbert space of the system ${\cal H}$. One of the most important properties of quantum states in many-body systems is quantum entanglement, which is quantum correlation between two subsystems. We decompose the total Hilbert space into those of subsystems $A$ and $\bar{A}$ such that ${\cal H}={\cal H}_A\otimes {\cal H}_{\bar{A}}$. We assume that the total state is given by a pure state described by the wave function $|\Psi\lb$. We introduce the reduced density matrix for $A$ by $\rho_A=\mbox{Tr}_{\bar{A}}[|\Psi\lb\la\Psi|]$. 
The amount of quantum entanglement for a given pure state is quantified by entanglement entropy and it is defined by
\ba
S_A=-\mbox{Tr}[\rho_A\log\rho_A].  \label{ENT}
\ea

Recently, non-hermitian extensions of density matrices have been introduced in order to study properties of various quantum systems in diverse fields such as quantum information \cite{Fullwood:2022rjd,Parzygnat:2022ldx, Parzygnat:2022pax, Wu:2024ucp}, quantum many-body systems \cite{Banuls:2009jmn,Muller-Hermes:2012irk,Hastings:2014qqa}, quantum field theories \cite{Nakata:2021ubr,Mollabashi:2020yie,Mollabashi:2021xsd,Doi:2022iyj,Doi:2023zaf}. In this paper we would like to investigate the connections between the non-hermitian density matrices and temporal quantum correlations, called
time-like entanglement. From this viewpoint we may classify them into the following two classes: {\it Class 1: causal influences under unitary evolutions}, 
and {\it Class 2: non-unitary evolutions}.

\subsubsection{Generalized density matrices, pseudo entropy and imagitivity}
 Before we get into the two classes of non-hermitian density matrices,  we would like to explain a general framework which can be applied to both. When we consider pure states, the generalized density matrix (or so-called transition matrix) \cite{Nakata:2021ubr} is expressed as 
\ba
\rho=\frac{|\psi\lb\la \vp|}{\la \vp|\psi\lb}, \label{PEden}
\ea
which is not hermitian when $|\psi\lb\neq |\vp\lb$ and is normalized such that Tr$\rho=1$.
We can interpret this as a post-selection, starting with the initial state $|\psi\lb$ and then performing the projection to the final state $|\vp\lb$. 

For a hermitian operator $\mO$ which corresponds to a physical quantity, we introduce the expectation value of the post-selection, called the weak value \cite{Dressel:2014kub}:
\ba
\la \mO\lb_\rho=\mbox{Tr}[\rho \mO]=\frac{\la \vp|\mO|\psi\lb}{\la \vp|\psi\lb}.
\ea
By introducing the density matrices or projectors for the initial and final state $\Pi_\psi=|\psi\lb\la\psi|$ and $\Pi_\vp=|\vp\lb\la\vp|$, we can rewrite this as 
\ba
\la \mO\lb_\rho=\frac{\mbox{Tr}[\Pi_\vp \mO \Pi_\psi]}
{\mbox{Tr}[\Pi_\vp \Pi_\psi]},
\ea
which makes its meaning clearer. 

\begin{figure}[ttt]
   \centering
   \includegraphics[width=4cm]{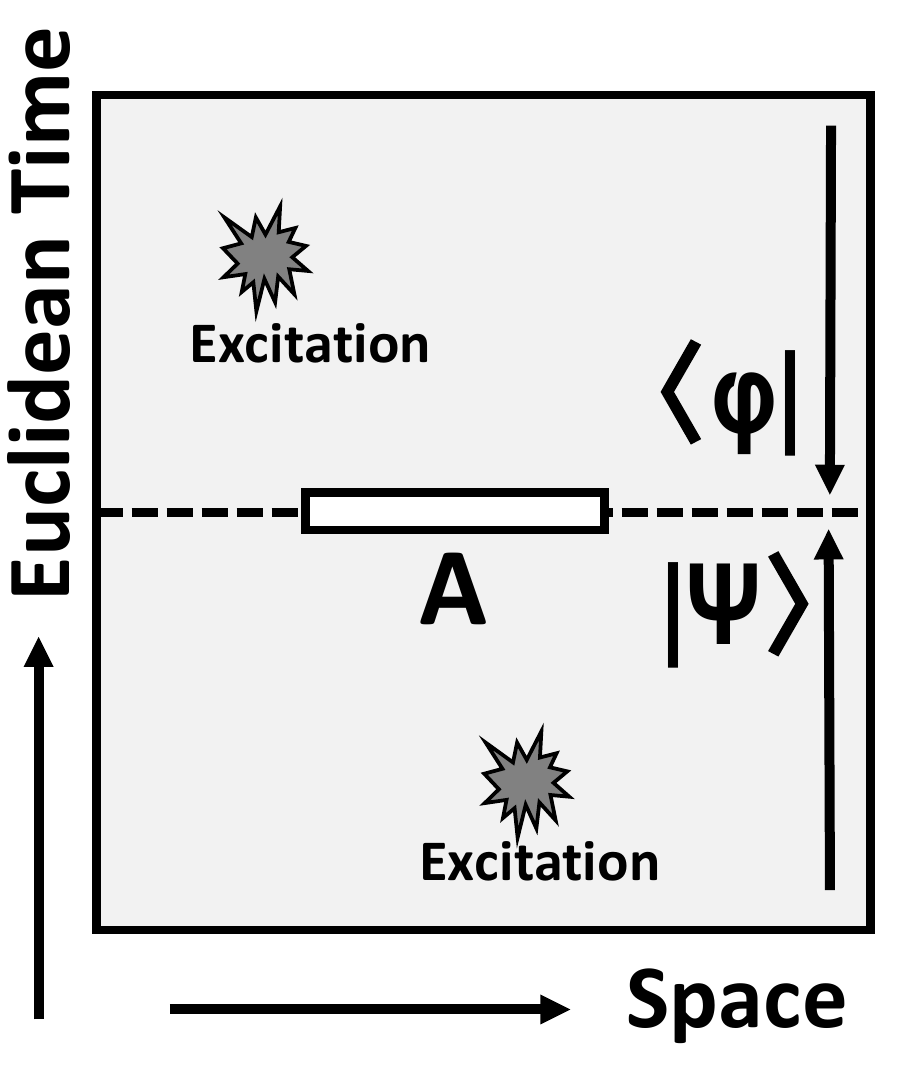}
   \caption{non-hermitian density matrix from local operator insertions.}
   \label{fig:LO}
\end{figure}

The reduced generalized density matrix is also defined by tracing out the compliment $\bar{A}$ as $\rho_A=\mbox{Tr}_{\bar{A}}\rho$ and the pseudo entropy is given by (\ref{ENT}) as first introduced in \cite{Nakata:2021ubr}.  A simple example of a state (\ref{PEden}) in quantum field theory is constructed by considering local operator excitations at different locations in bra and ket state, as depicted in Fig.\ref{fig:LO}. We will present the detailed analysis in section \ref{sec:LO}.

Generally, in non-hermitian examples,  we can introduce generalized density matrices  (also called transition matrices), which will be expressed by the same symbol $\rho$ in this paper. We can also introduce the generalized reduced density matrices $\rho_A=\mbox{Tr}_{\bar{A}}\rho$. Notice that in this context, $\rho$ and $\rho_A$ are not hermitian in general. As we mentioned, it is useful to introduce an extension of the von-Neumann entropy defined by the same formula (\ref{ENT}), called the pseudo entropy \cite{Nakata:2021ubr}. The pseudo entropy generally takes complex values for non-hermitian density matrices. In other words, the presence of imaginary part of pseudo entropy tells us that the density matrix is not hermitian, though the converse is not always true.

The imagitivity introduced in \cite{Milekhin:2025ycm} is also an interesting quantity which directly measures how much a given density matrix is not hermitian. This is defined by $||\rho_A-\rho^\dagger_A||_p$ using the Schatten $p$-norm $||*||_p$. The non-vanishing imagitivity is clearly equivalent to the fact that $\rho_A$ is not hermitian. In this paper, we define imagitivity as the one with $p=2$
\begin{equation}
\begin{split}
    \mathrm{Imagitivity}[\rho_A] &= (||\rho_A-\rho^\dagger_A||_2 )^2\\
    &= 2 \bTr{\rho_A\rho_A^\dag}-2\mathrm{Re}\qty[\bTr{\rho_A^2}].\label{imgt}
\end{split}
\end{equation}

\subsection{Class 1: Influences under unitary evolutions}\label{sec:classaa}

Here we explain how a non-hermitian density matrix leads to the causal influences under unitary time evolutions. Here the time evolution is described by the Hamiltonian $H$, which is 
hermitian. In our argument below, we will try to be general as much as we can. We will present explicit examples in section \ref{sec:classone}.

\subsubsection{Time-like entanglement}\label{sec:tlef}

\begin{figure}[ttt]
   \centering
   \includegraphics[width=7cm]{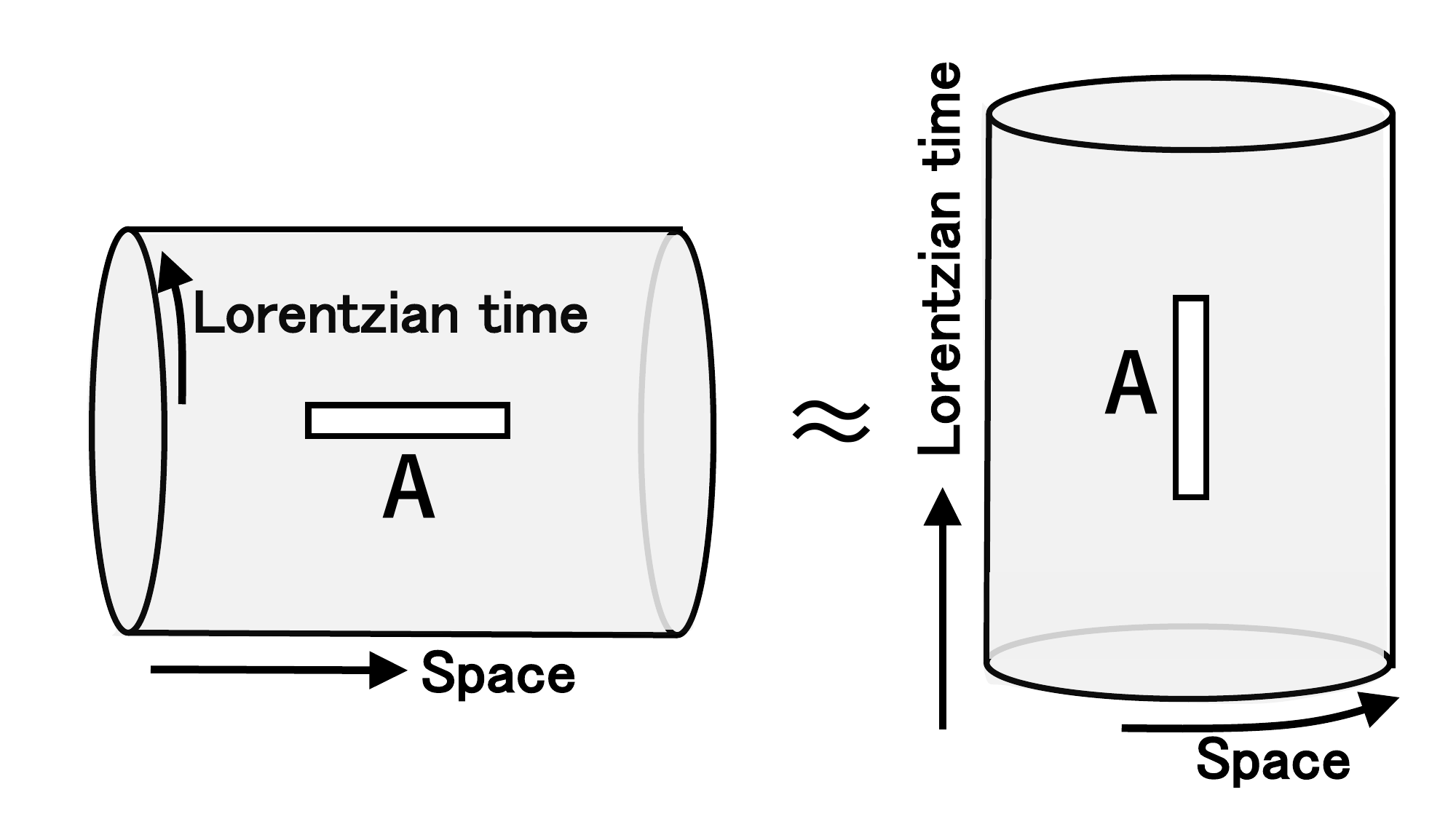}
   \caption{non-hermitian density matrix from the periodic Lorentzian time (left) and and a time-like interval (right).}
   \label{fig:TEE}
\end{figure}

One of the simplest setups of a non-hermitian density matrix will be the one where the state at $t=t_2$ is identified with that at $t=t_1$ (refer to the left panel of Fig.\ref{fig:TEE}). The generalized density matrix at $t=t_1$ simply given by the unitary time evolution by the Hamiltonian $H$:
\ba
\rho=e^{-\rmi (t_2-t_1)H},
\ea
which is clearly non-hermitian and can be regarded as a mixed state version of 
(\ref{PEden}). We can also interpret this as a canonical distribution at imaginary valued temperature, which is a special case of the thermal pseudo entropy considered in \cite{Caputa:2024gve}.

The same setup occurs when we consider a time-like slice in a Lorentzian spacetime. For this, we compactify one of space coordinates, denoted by $y$, on a circle with the circumference $L$ (refer to the right panel of Fig.\ref{fig:TEE}). Then we cut the spacetime along $y=0$ and define the generalized density matrix on this cut, which is time-like. Then the evolution along $y$ direction turns out to be anti-hermitian which can be written as $e^{-\rmi LH}$ \cite{Doi:2022iyj,Doi:2023zaf}, where $H$ is hermitian. If we focus on an interval $A$ on this time-like slice, then its generalized reduced density matrix looks like
\ba
\rho_A=\mbox{Tr}_{\bar{A}} [e^{-\rmi LH}].
\ea
Its pseudo entropy coincides with time-like entanglement entropy \cite{Doi:2022iyj,Doi:2023zaf}.

Another important example will be time evolution in quantum systems. Consider a quantum many-body system with a Hamiltonian $H$ and write its ground state as $|0\lb$. We consider its (trivial) time evolution from $t=t_1$ to $t=t_2$. 
 The generalized density matrix reads
\ba
\rho=|0\lb_1 [e^{-\rmi (t_2-t_1)H}]_{2\bar{1}}\la 0|_{\bar{2}},
\ea
where we write the ket and bra state at $t=t_1$ (or $t=t_2$) as $1$ and $\bar{1}$ (or $2$ and $\bar{2}$). In terms of energy eigen basis, we can write $\rho$ as
\ba
\rho=\sum_{n}e^{-\rmi(t_2-t_1)E_n}|0\lb_1 |n\lb_2  \la n|_{\bar{1}}\la 0|_{\bar{2}}.
\ea
This is clearly non-hermitian due to the causal Unitary evolution. We can regard this an superposition of (\ref{PEden}).

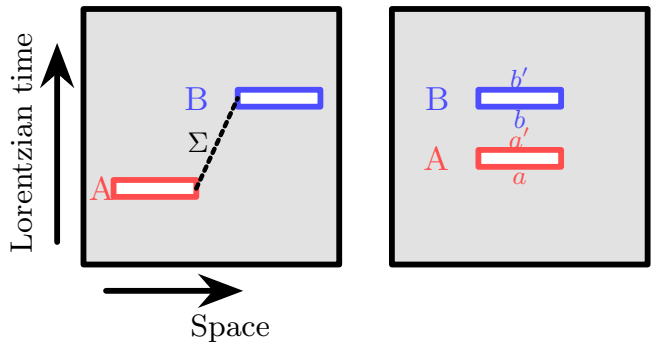
\begin{figure}
\centering
\begin{tikzpicture}[
  scale=0.5,
  transform shape,
  line cap=round,
  line join=round
]

\def\L{6.8}
\def\gap{1.4}
\def\w{2.2}
\def\h{0.45}
\def\off{0.32} 

\begin{scope}[shift={(0,0)}]

  \fill[lightgray!45] (0,0) rectangle (\L,\L);

  \fill[white] (0.8,1.8) rectangle ++(\w,\h);
  \draw[line width=2.2pt,red!70]
    (0.8,1.8) rectangle ++(\w,\h);
  \node[scale=2.8,red!70,anchor=east]
    at (0.8+0.35,1.8+0.5*\h) {A};

  \fill[white] (4.1,4.2) rectangle ++(\w,\h);
  \draw[line width=2.2pt,blue!70]
    (4.1,4.2) rectangle ++(\w,\h);
  \node[scale=2.8,blue!70,anchor=east]
    at (4.1-0.45,4.2+0.5*\h) {B};

  \draw[line width=2.2pt] (0,0) rectangle (\L,\L);

  \draw[line width=1.7pt, dotted]
    (0.8+\w,1.8+0.5*\h) -- (4.1,4.2+0.5*\h);
  \node[scale=2.4] at (3.05,3.25) {$\Sigma$};

\end{scope}

\begin{scope}[shift={(\L+\gap,0)}]

  \fill[lightgray!45] (0,0) rectangle (\L,\L);

  \fill[white] (2.3,4.2) rectangle ++(\w,\h);
  \draw[line width=2.2pt,blue!70]
    (2.3,4.2) rectangle ++(\w,\h);
  \node[scale=2.8,blue!70,anchor=east]
    at (2.3-0.45,4.2+0.5*\h) {B};

  \node[blue!70,scale=2.2]
    at (2.3+0.5*\w,4.2+\h+\off) {$b'$};
  \node[blue!70,scale=2.2]
    at (2.3+0.5*\w,4.2-\off) {$b$};

  \fill[white] (2.3,2.6) rectangle ++(\w,\h);
  \draw[line width=2.2pt,red!70]
    (2.3,2.6) rectangle ++(\w,\h);
  \node[scale=2.8,red!70,anchor=east]
    at (2.3-0.45,2.6+0.5*\h) {A};

  \node[red!70,scale=2.2]
    at (2.3+0.5*\w,2.6+\h+\off) {$a'$};
  \node[red!70,scale=2.2]
    at (2.3+0.5*\w,2.6-\off) {$a$};

  \draw[line width=2.2pt] (0,0) rectangle (\L,\L);

\end{scope}

\draw[-{Stealth[length=6mm]}, line width=2.2pt]
  (-0.7,0.6) -- (-0.7,5.9);
\node[rotate=90, scale=2.6]
  at (-1.65,3.25) {Lorentzian time};

\draw[-{Stealth[length=6mm]}, line width=2.2pt]
  (0.6,-0.7) -- (4.25,-0.7);
\node[scale=2.6]
  at (3.9,-1.7) {Space};

\end{tikzpicture}
   \caption{non-hermitian density matrices from time-like entanglement for double intervals. In the left panel, if the interval $\Sigma$ becomes time-like, the density matrix $\rho_{AB}$ becomes non-hermitian as $A$ and $B$ are not on a common time slice and thus are causally connected. In the right panel, clearly $A$ and $B$ are causally connected and thus $\rho_{AB}$ is non-hermitian.  }
   \label{fig:Double}
\end{figure}

Now we take a subregion $A$ at $t=t_1$ and $B$ at $t=t_2$.
We can define the generalized reduced density matrix $\rho_{AB}$ by tracing out $\bar{A}$ at $t=t_1$ and $\bar{B}$ at $t=t_2$, which looks like
\ba
\rho_{AB}={}_{\bar{2}}\la 0|b'\lb_B \la b|e^{-\rmi (t_2-t_1)H}|a' \lb_A \la a |0\lb_1,
\label{rasz}
\ea
where $(a,a')$ and $(b,b')$ are the labels of bra/ket basis in ${\cal H}_A$ 
${\cal H}_B$, respectively. We find $\rho_{AB}$ is hermitian when $A$ and $B$ are space-like separated, while in the other cases, it should be non-hermitian in general. This non-hermitian density matrix for a double interval was considered in \cite{Kawamoto:2025oko} when $B$ is situated in the left of $A$ (illustrated in the left panel of Fig. \ref{fig:Double}) and in  
\cite{Milekhin:2025ycm} when $B$ is the time translation of $B$ (the right panel of Fig. \ref{fig:Double}). 

It is also useful to note that even if we consider Euclidean time evolution by setting $t_1=-\rmi\tau_1$ and $t_2=-\rmi\tau_2$, $\rho$ is still non-hermitian. Moreover, in this case $\rho_{AB}$ is always non-hermitian even if $A$ and $B$ are far apart.

More generally, if we consider a time evolution of a generic mixed state $\rho^{(0)}_{12}$ at time $t=0$, then the generalized density matrix reads 
\ba
&&\rho_{AB}\no
&&=\mbox{Tr}\left[\left(e^{\rmi t_2H}|b'\lb_B \la b|e^{-\rmi t_2H}\right)\left(e^{\rmi t_1H}
|a' \lb_A \la a| e^{-\rmi t_1H}\right)\rho^{(0)}\right],\no \label{rabg}
\ea
which generalizes (\ref{rasz}). This is not hermitian because 
$e^{\rmi t_2H}|b'\lb_B \la b|e^{-\rmi t_2H}$ does not commute with $e^{\rmi t_1H}
|a' \lb_A \la a| e^{-\rmi t_1H}$ in general. This already connects the causal influences and the non-hermiticity of the density matrix $\rho_{AB}$. We will explain this more explicitly soon later.

\subsubsection{Causal influences due to interactions}
When we consider the setups explained in the above and introduce two subsystems $A$ and $B$, we can define corresponding Hilbert spaces ${\cal H}_A$ and ${\cal H}_B$.
However, since we have in mind the cases when $A$ and $B$ are causally connected or not completely space-like to each other, we cannot take a common time slice of the whole system such that the total Hilbert gets factorized as 
${\cal H}={\cal H}_A\otimes {\cal H}_B\otimes {\cal H}_{C}$ for a certain subsystem $C$ (see Fig. \ref{fig:Double}). Though we can define the generalized density matrix (or transition matrix) $\rho_{AB}$ in an obvious way by tracing out $\bar{A}$ and $\bar{B}$, we expect that $\rho_{AB}$ is no longer hermitian, as suggested from the analysis of pseudo entropy \cite{Kawamoto:2025oko}. Indeed we can manifestly relate the commutators of operators on $A$ and $B$ to the non-hermiticity of $\rho_{AB}$ as shown in \cite{Milekhin:2025ycm} (see also \cite{Parzygnat:2022pax,Parzygnat:2022ldx}):
\ba
\la [\mO_A(t_1),\mO_B(t_2)]\lb=\mbox{Tr}\left[(\rho_{AB}-\rho^\dagger_{AB})\mO_A\mO_B\right].
\label{Mil}
\ea

We can also explicitly see how an unitary operation $U_A$ localized on $A$ influences the measurement on $B$ as follows. We take $U_A$ to be very close to the identity such that $U_A=e^{\rmi\eta \mO_A}$, where $\eta$ is an infinitesimally small parameter and $\mO_A$ is an hermitian operator. We would like to compute how the expectation value of a hermitian operator $\mO_B$ localized on $B$ is changed by $U_A$. By acting $U_A$ on $A$ in (\ref{rabg}) we find 
\ba
\rho_B=\mbox{Tr}_{\bar{B}}\left[e^{-\rmi (t_2-t_1)H}U_A e^{-\rmi t_1 H}\rho^{(0)}e^{\rmi t_1H}U^\dagger_A e^{i(t_2-t_1)H}\right],\no
\ea
where note that the Hamiltonian $H$ includes interactions between $A$ and $B$.
Finally the expectation value of $\mO_B$ is evaluated to the linear order of $\eta$ as:
\ba
\la \mO_B\lb&=&\mbox{Tr}[\mO_B\rh\mO_B]\no
&\simeq & \mbox{Tr}[\mO_B(t_2)\rho^{(0)}]\no
&&-\rmi\eta \mbox{Tr}\left[[\mO_A(t_1),\mO_B(t_2)]|\rho^{(0)}\right]+O(\eta^2).\label{causalinf}
\ea
Thus using the relation (\ref{Mil}), we find that there is influence between $A$ and $B$ 
if $\rho_{AB}$ is non-hermitian.

\subsubsection{Causal influences due to post-selection}\label{sec:posts}

\begin{figure}[ttt]
   \centering
   \includegraphics[width=4cm]{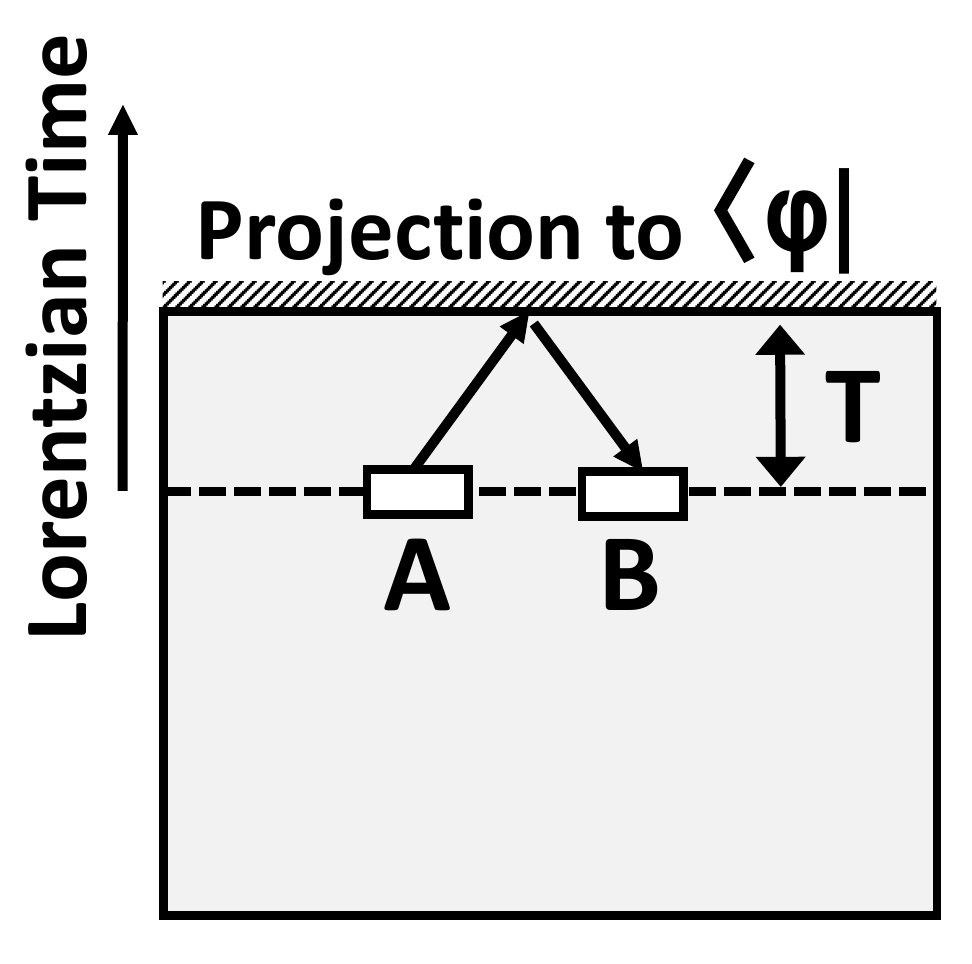}
   \caption{non-hermitian density matrix from a post-selection.}
   \label{fig:Projection}
\end{figure}

Another category of examples where we find causal influences due to the unitary time evolutions is the post-selection. Consider starting with the initial state $|\psi\lb$ and performing the time evolution from $t=0$ to $t=T$. Then finally we project the state to another state $|\vp\lb$. At time $t=0$ we act the unitary transformation $U_A$ on the subregion $A$ and measure the expectation value of the operator $\mO_B$ at the same time (refer to Fig. \ref{fig:Projection}). We consider a relativistic quantum system and choose $A$ and $B$ such that they are space-like separated. The post-selected expectation value is given by 
\ba
\la \mO_B\lb= \frac{\la \vp| e^{-\rmi Ht}U_A \mO_B |\psi\lb}{\la \vp| e^{-\rmi Ht}U_A|\psi\lb}.
\ea
If we take $2T$ to be shorter than the distance $D_{AB}$ between $A$ and $B$, the value of $\la \mO_B\lb$ does not depend on $U_A$. In this case, $A$ and $B$ are causally disconnected. On the other hand, if $2T\geq D_{AB}$, $A$ and $B$ are causally connected due to the reflection at $t=T$ and thus  $\la \mO_B\lb$ depends on $U_A$. Note that this influence phenomenon happens because the generalized density matrix
\ba
\rho_B=\mbox{Tr}_{\bar{B}}[e^{-\rmi Ht}U_A|\psi\lb\la \vp|],
\ea
depends on $U_A$ and still occurs even when $|\psi\lb=|\vp\lb$. A similar mechanism can be found in the influences in non-hermitian systems as we will analyze next.

\subsection{Class 2: non-hermitian systems} \label{sec:classbb}

Now we would like to consider non-hermitian quantum systems where the effective Hamiltonian is non-hermitian.  In such a setup, even though the structure of Hilbert space and its decomposition  ${\cal H}={\cal H}_A\otimes {\cal H}_{\bar{A}}$ are standard, the conjugation which maps a ket state into its bra sate can be different from the standard hermitian conjugation.  
In our argument below, we will make a relatively general argument on how to describe the conjugation and explain why it causes unusual causal influences even in the absence of the direct interactions, peculiar to non-hermitian systems. We will present explicit examples in section \ref{sec:classtwo}.

\subsubsection{non-hermitian deformation}

A simple and fundamental example is a deformation of a Hamiltonian $H_0$ by a perturbation $\lambda H_{\mint}$, where $H_0$ and $H_{\mint}$ are both hermitian operators and $\lambda$ is a complex valued parameter. If we write an eigenstate of the deformed Hamiltonian 
\be
H=H_0+\lambda H_{\mint},  \label{DefH}
\ee
as $|n_+\lb$ with a complex valued eigenvalue $E_n$:
\ba
H|n_+\lb=E_n|n_+\lb,  \label{ega}
\ea
its standard conjugation is given by $|n_+\lb^\dagger=\la n_+|$. This bra state turns out to be an eigenstate of the conjugate Hamiltonian $H^\dagger=H_0+\lambda^* H_{\mint}$ instead of the original one:
\ba
\la n_+|H^\dagger=E^*_n\la n_+|. \label{egb}
\ea
In such a setup, it is also useful to consider the eigenstates of $H^\dagger$, which leads 
\ba
H^\dagger |n_-\lb=E^*_n |n_-\lb,\ \ \ \ \la n_-|H=E_n\la n_-|. \label{egc}
\ea

In the above, we first assume that $\lambda$ is real valued and find the eigenstates as a function of $\lambda$, which can be expressed as $|n(\lambda)\lb$. Then we analytical continue the states to complex values of $\lambda$ and write $|n_+\lb=|n(\lambda)\lb$. Similar we can introduce 
$|n_-\lb=|n(\lambda^*)\lb$. It is clear that they satisfy (\ref{ega}), (\ref{egb}) and (\ref{egc}).

Now consider the analytical continuation from the real valued $\lambda$ to the imaginary valued $\lambda$. Then it is appropriate to introduce the modified conjugation $\ddagger$ such that 
\ba
|n_{\pm}\lb^{\ddagger}=\la n_{\mp}|.
\ea
More generally, we define the conjugation by regarding $\lambda$ as a real valued constant as
\ba
\left(\sum_n C(\lambda)|n_\pm\lb\right)^\ddagger=\sum_n \left(C(\lambda^*)\right)^* \la n_{\mp}|,
\label{mconj}
\ea
for any function $C$ of $\lambda$. Notice that the eigenstates $|n_\pm\lb$ all depends on $\lambda$.

Since we find
\ba
&& \la n_-|H|m_+\lb=E_m\la n_-|m_+\lb=E_n\la n_-|m_+\lb,\no
&& \la n_+|H|m_-\lb=E^*_m\la n_+|m_-\lb=E^*_n\la n_+|m_-\lb,
\ea
the orthogonality property reads
\ba
\la n_+|m_-\lb=\la n_-|m_+\lb=\delta_{n,m},
\ea
and the completeness property looks like
\ba
\sum_{n}|n_+\lb\la n_-|=\sum_{n}|n_-\lb\la n_+|=1.
\ea

In terms of the matrix elements this induces the definition of the new conjugation for an operator $\mO$:
\ba
\la n_+|\mO|m_-\lb^*=\la m_+|\mO^\ddagger|n_-\lb.
\ea
This is a good definition for example because  the following is non-negative:
\ba
\la n_+|\mO^\ddagger \mO|n_-\lb&=&\sum_m \la n_+|\mO^\ddagger|m_-\lb \la m_+|\mO|n_-\lb \no
&=&\sum_m |\la m_+|\mO|n_-\lb|^2\geq 0.
\ea
In particular the Hamiltonian $H=\sum_n E_n|n_+\lb\la n_-|$ is "hermitian" under this new conjugation $H^\ddagger=H$.
In this setup, we regard the correct conjugation as the modified one $^\ddagger$, assuming that the physics is derived by the analytical continuation of $\lambda$, which forces us to regard $\lambda$ as a real valued parameter. If we consider the ket state $|n_+\lb$, then the density matrix is given by $\rho=|n_+\lb\la n_{-}|$. This is not non-hermitian under the standard conjugation $\dagger$ i.e. 
$\rho^\dagger\neq \rho$, but it is "hermitian" under the new one $\ddagger$ i.e. $\rho^\ddagger=\rho$.

This procedure can be formally done for any non-hermitian Hamiltonian. Suppose we have a non-hermitian Hamiltonian $H$. We can always decompose $H= H_{0} + \rmi H_{\mint}$ with Hermitian hamiltonian $H_0=(H+H^\dag)/2$ and $H_{\mint}=(H-H^\dag)/2$.  Then we introduce an parameter $\lambda$ and consider $H(\lambda) = H_0 +\rmi \lambda H_{\mint}$. We denote the right eigenvectors and left eigenvectors of $H=H(1)$ as $\ket{n_+}=\ket{n_{+}(1)},\bra{n_-}=\bra{n_-(1)}$. Then we can consider an operation for $\ket{n_{\pm}}$ by analytical continuation. We first extend $\lambda$ to real value and denote this $\ket{n_+(\lambda)}$. Now we analytically continue $\lambda$ to $\pm \rmi$ and obtain some hermitian Hamiltonian $H(\pm \rmi)$. Under this hermitian situation we take usual dagger conjugation $(\ket{n_+(\pm \rmi)})^\dag $. Then we undo the analytic continuation and obtain some bra state $\bra{\tilde{n}_+}$. In some examples we can confirm that this state is actually a left eigenstates even though we do not have nice structure like pseudo hermitian.
As we will discuss later, a similar non-hermitian system for thermofield double states naturally arises in holographic model of traversable wormholes \cite{Kawamoto:2025oko}
and the above non-hermitian property of the density matrix is crucial to the traversability.

\subsubsection{Influences in non-hermitian systems}\label{sec:infnonh}

Next we would like to study how two subsystems, called $A$ and $B$, influence with each other in such a non-hermitian system. We will show below that even if there is no interaction between $A$ and $B$, we can send a signal from $A$ to $B$ and vice versa if we assume that the modified conjugation (\ref{mconj}) is relevant. The explicit example will be given in \ref{sec:classtwo}, which is a non-hermitian deformation of thermofield double states dual to traversable wormholes.

We again act the unitary $U_A=e^{\rmi\eta \mO_A}$ for a hermitian operator $\mO_A$ on the region $A$ at $t=t_1$ with an infinitesimally small parameter $\eta$.  
The generalized density matrix for $B$ at time $t_2$ looks like
\be
\rho_B=\frac{\mbox{Tr}_{\bar{B}}\left[e^{-\rmi (t_2-t_1)H}U_A e^{-\rmi t_1 H}\rho^{(0)}e^{\rmi t_1H}U^\ddagger_A e^{i(t_2-t_1)H}\right]}{\mbox{Tr}\left[U_A e^{-\rmi t_1 H}\rho^{(0)}e^{\rmi t_1H}U^\ddagger_A \right]},\no
\ee
where note that both the initial state $\rho^{(0)}$ and the Hamiltonian $H$ are not hermitian $(\rho^{(0)})^\dagger \neq \rho^{(0)}$ and 
$H^\dagger\neq H$, while it is self-ajoint under the modified one: $(\rho^{(0)})^\ddagger = \rho^{(0)}$ and 
$H^\ddagger= H$.

The expectation value of a hermitian operator $\mO_B$ at the region $B$ is evaluated to the linear order of $\eta$ as:
\ba
\la \mO_B\lb&=&\mbox{Tr}[\mO_B\rho_B]\no
&\simeq & \la\mO_B(t_2)\lb^{(0)}\no
&&+\rmi\eta \mbox{Tr}\left[\left(\mO_B(t_2)- \la\mO_B(t_2)\lb^{(0)}\right)\mO_A(t_1)\right]\no
&&-\rmi\eta
\mbox{Tr}\left[O^{\ddagger}_A(t_1)\left(\mO_B(t_2)- \la\mO_B(t_2)\lb^{(0)}\right)\rho^{(0)}\right]\no
&& +O(\eta^2).  \label{infja}
\ea
Since in general we have $\mO_{A,B}^\ddagger\neq \mO_{A,B}$, though $\mO_{A,B}^\dagger= \mO_{A,B}$, the expectation value $\la \mO_B\lb$ depends on $\eta$ and thus there is an influence from $A$ to $B$ or vice versa. As the simplest candidate of such a hermitian operator, we can choose $H_{\mint}$ as $\mO_A$ for example.

We can summarize the essential mechanism of the above argument as follows. Consider the bra state $|\psi\lb$ and ket one $\la\vp|$ as in  (\ref{PEden}). We perform different unitary transformations on the former and the latter state at the region $A$, expressed by $U_A^{(\psi)}$ and $U_A^{(\vp)}$, respectively and measure the expectation value $\mO_B$. Then the expectation value looks like
\be
\la \mO_B\lb=\frac{\la\vp|U_A^{(\vp)-1}\mO_B U_A^{(\psi)}|\psi\lb}{\la\vp|U_A^{(\vp)-1}U_A^{(\psi)}|\psi\lb}.
\ee
This expectation value depends on the unitary action at $A$ if  $U_A^{(\psi)}\neq U_A^{(\vp)}$, even when $A$ and $B$ are space-like separated. We can also view the mechanism of the causal influence by the post-selection discussed in (\ref{sec:posts}) as an example of this.

\subsubsection{Modified conjugation in pseudo Hermitian systems}
A modified conjugation has been discussed in pseudo Hermitian systems \cite{Ashida:2020dkc} and we would like to compare it with our previous one (\ref{mconj}). We say that a non-hermitian Hamiltonian $H$ is pseudo hermitian if there exists an invertible operator $W$ such that
\begin{equation}
    H^\dag = W H W^{-1}.
\end{equation}
Then by employing an anti-linear operator $V= WK$ where $K$ is an anti-linear operator which takes complex conjugation $K(a\ket{\psi})=a^* K\ket{\psi}$ for c-number $a$. Then we see the right eigenvector is related to left one by 
\begin{equation}
    V\ket{n_+} =\ket{n_-}.
\end{equation}
Thus in this case, the modified conjugation is defined by  
\begin{equation}
    (\ket{n_\pm})^{\#} = (V\ket{n_{\mp}})^\dag.
\end{equation}
Similar discussion is done in \cite{Castro-Alvaredo:2017udm}. However, to confirm the relation $H^\# = H$ in this formalism, we need real energy spectrum $E_n^*=E_n$. Conversely, when the energy spectrum is real, this coincides with ours i.e. $\#=\ddag$.
Our definition of $\ddagger$ via the analytic continuation method can be used for any non-hermitian systems, while the previous one $\#$ is not well defined for complex valued spectra.
\subsubsection{Example of modified conjugation}
As a simple example, we consider the following two-site tight-binding model:
\begin{equation}
    H = -W_0 \sigma_x + \rmi \lambda \sigma_z 
    = 
    \begin{pmatrix}
        \rmi \lambda & -W_0 \\
        -W_0 & -\rmi \lambda
    \end{pmatrix}.
\end{equation}
This is the simplest $PT$-symmetric model. The energy eigenvalues are given by 
\begin{equation}
    E_{\pm} = \pm \sqrt{W_0^2 - \lambda^2}.
\end{equation}
Hence, for $\abs{W_0} > \abs{\lambda}$, the system has a real energy spectrum and the $PT$ symmetry is unbroken.  
However, for $\abs{W_0} < \abs{\lambda}$, the spectrum becomes purely imaginary and the $PT$ symmetry is broken.  
The left eigenvectors differ between the $PT$-symmetric and $PT$-broken phases, so we analyze each situation separately. 
This system is also pseudo-Hermitian, satisfying
\begin{equation}
    \sigma_x H \sigma_x = H^\dag.
\end{equation}

\par
\noindent\textbf{$PT$-unbroken phase}\par
The right energy eigenvectors are given by
\begin{equation}
    \begin{split}
        \ket{n^{(\pm)}_{+}} &\propto
        \begin{pmatrix}
            W_0 \\
            \rmi \lambda \mp \sqrt{W_0^2 - \lambda^2}
        \end{pmatrix}
        \propto
        \begin{pmatrix}
            -\rmi \lambda \mp \sqrt{W_0^2 - \lambda^2} \\
            W_0
        \end{pmatrix}.
    \end{split}
\end{equation}
The corresponding left eigenvectors are given by
\begin{equation}
    \begin{split}
        \bra{n^{(\pm)}_{-}} &\propto
        (W_0,\rmi \lambda \mp \sqrt{W_0^2 - \lambda^2}) \\
        &\propto
        (-\rmi \lambda \mp \sqrt{W_0^2 - \lambda^2}, W_0).
    \end{split}
\end{equation}

Under the analytic continuation procedure, we obtain
\begin{equation}
    \begin{split}
        &\ket{n^{(\pm)}_{+}} \propto
        \begin{pmatrix}
            W_0 \\
            \rmi \lambda \mp \sqrt{W_0^2 - \lambda^2}
        \end{pmatrix} \\
        &\to
        \ket{n^{(\pm)}_{+}(\lambda=-\rmi\Tilde{\lambda})} \propto
        \begin{pmatrix}
            W_0 \\
            \Tilde{\lambda} \mp \sqrt{W_0^2 + \Tilde{\lambda}^2}
        \end{pmatrix} \\
        &\to
        \bra{n^{(\pm)}_{+}(\lambda=-\rmi\Tilde{\lambda})}
        := \qty(\ket{n^{(\pm)}_{+}(\lambda=-\rmi\Tilde{\lambda})})^\dag \\
        &\qquad\qquad\qquad\qquad
        \propto (W_0,\Tilde{\lambda} \mp \sqrt{W_0^2 + \Tilde{\lambda}^2}) \\
        &\to
        \bra{\Tilde{n}^{(\pm)}_{+}}
        \propto (W_0,+\rmi \lambda \mp \sqrt{W_0^2 - \lambda^2}).
    \end{split}
\end{equation}
The last expression is proportional to the left eigenvectors $\bra{n^{(\pm)}_{-}}$. 
Notice that although we recover the left eigenvectors, the bi-orthogonal normalization is not preserved, \textit{i.e.}
$\braket{\Tilde{n}^{(\pm)}_{+}}{n^{(\pm)}_{+}} \neq 1$ but with $\braket{\Tilde{n}^{(\mp)}_{+}}{n^{(\pm)}_{+}} =0$.

\par
\noindent\textbf{$PT$-broken phase}\par
The right and left energy eigenvectors are given by
\begin{equation}
    \begin{split}
        \ket{n^{(\pm)}_{+}} &\propto
        \begin{pmatrix}
            W_0 \\
            \rmi \lambda \mp \rmi \sqrt{\lambda^2 - W_0^2}
        \end{pmatrix}
        \propto
        \begin{pmatrix}
            -\rmi \lambda \mp \rmi \sqrt{\lambda^2 - W_0^2} \\
            W_0
        \end{pmatrix}, \\
        \bra{n^{(\pm)}_{-}} &\propto
        (W_0,\rmi \lambda \mp \rmi \sqrt{\lambda^2 - W_0^2}) \\
        &\propto
        (-\rmi \lambda \mp \rmi \sqrt{\lambda^2 - W_0^2}, W_0).
    \end{split}
\end{equation}
Under the analytic continuation procedure, we obtain
\begin{equation}
    \begin{split}
        &\ket{n^{(\pm)}_{+}} \propto
        \begin{pmatrix}
            W_0 \\
            \rmi \lambda \mp \rmi \sqrt{\lambda^2 - W_0^2}
        \end{pmatrix} \\
        &\to
        \ket{n^{(\pm)}_{+}(\lambda=-\rmi\Tilde{\lambda})} \propto
        \begin{pmatrix}
            W_0 \\
            \Tilde{\lambda} \mp \sqrt{W_0^2 + \Tilde{\lambda}^2}
        \end{pmatrix} \\
        &\to
        \bra{n^{(\pm)}_{+}(\lambda=-\rmi\Tilde{\lambda})}
        := \qty(\ket{n^{(\pm)}_{+}(\lambda=-\rmi\Tilde{\lambda})})^\dag \\
        &\qquad\qquad\qquad\qquad
        \propto (W_0,\Tilde{\lambda} \mp \sqrt{W_0^2 + \Tilde{\lambda}^2}) \\
        &\to
        \bra{\Tilde{n}^{(\pm)}_{+}}
        \propto (W_0,+\rmi \lambda \mp \sqrt{\lambda^2 - W_0^2}).
    \end{split}
    \label{eq:eigenvec}
\end{equation}
The last expression is proportional to the left eigenvectors $\bra{n^{(\pm)}_{-}}$. 
The choice of branch is fixed so that the analytically continued eigenvectors are proportional to those of the analytically continued Hamiltonian $H(\lambda=-\rmi\Tilde{\lambda})$.

\subsection{Gravity dual of generalized density matrices}

For later use, here we would like to briefly explain the gravity dual of the generalized density matrix via the AdS/CFT \cite{Maldacena:1997re}. The AdS/CFT argues that a gravitational theory on $d+1$ dimensional anti de-Sitter space (AdS) is equivalent to a $d$ dimensional conformal field theory (CFT), which lives on the boundary of the AdS. Under this duality, the partition function of the CFT becomes identical to that of the gravity on the AdS with an appropriate boundary condition at the boundary \cite{Gubser:1998bc,Witten:1998qj}.

As argued in \cite{Nakata:2021ubr}, the generalized density matrix (\ref{PEden}), which is in general non-hermitian, is given by an Euclidean asymptotically AdS background which depends on the imaginary time $\tau$. The full partition function of the gravity gives the inner product $\la \vp|\psi\lb$, where the geometry $\tau<0$ and $\tau>0$ describe $|\psi\lb$ and 
$\la \vp|$, respectively, which are glued at $\tau=0$. In the example of local operator excitations depicted in Fig.\ref{fig:LO}, the dual geometry is realized by inserting the dual shock waves which are emitted from the AdS boundary in appropriate locations \cite{Horowitz:1999gf,Nozaki:2013wia}. The difference from the usual hermitian density matrices is that the geometry is not symmetric under the flip of the sign of imaginary time $\tau\to -\tau$, which simply leads to the non-hermitian density matrix.

Next we consider its real time evolution by an analytical continuation to Lorentzian geometry which is dual to the generalized density matrix:   
\ba
\rho(t)=\frac{e^{-\rmi Ht}|\psi\lb\la \vp|e^{\rmi Ht}}{\la \vp|\psi\lb}. \label{PETden}
\ea
Its dual spacetime can be found by analytically continuing the imaginary time $\tau$ to the real one $t$ by setting $\tau=it$. Note that in general the metric becomes complex valued after the continuation due to the lack of the symmetry $\tau\to -\tau$. The area of the extremal surface $\Gamma_A$, which is anchored at the boundary of $A$ i.e. $\de \Gamma_A=\de A$, is identified with the pseudo entropy (\ref{ENT}) for the generalized reduced density matrix $\rho_A(t)=\mbox{Tr}_{\bar{A}}\rho(t)$ via the geometric formula \cite{Nakata:2021ubr}:
\ba
S_A=\mbox{Ext}\left[\frac{A(\Gamma_A)}{4G_N}\right].
\label{extfor}
\ea
This entropy gets complex valued and this agrees with the general property of pseudo entropy. This generalizes the holographic entanglement entropy \cite{Ryu:2006bv,Ryu:2006ef,Hubeny:2007xt}.
We will discuss explicit examples of AdS traversable wormholes in section \ref{sec:doublewh} and section \ref{sec:classtwo}.

\section{non-hermitian density matrices from local excitations}\label{sec:LO}

As the first simple example of non-hermitian density matrices in quantum field theory we would like to consider  the two dimensional free scalar CFT with a local operator excitation (refer to Fig.\ref{fig:LO}) \cite{Nozaki:2014hna,Nozaki:2014uaa,He:2014mwa}. 
We prepare the states
\begin{align}
  \ket{\psi} =& \frac{1}{\mathcal{N}_1} e^{- \epsilon_1 H} \mathcal{O}(x_1) \ket{0}, \\
  \ket{\varphi} =& \frac{1}{\mathcal{N}_2} e^{- \epsilon_2 H} \mathcal{O}(x_2) \ket{0}. 
\end{align}
Here, $\mathcal{O} = e^{\rmi \ap \phi} + e^{-\rmi\ap \phi}$ is a local operator with conformal dimension $h = \bar{h} = \ap^2/8$, which produces a Bell pair \cite{Nozaki:2014hna, Nozaki:2014uaa}. We set $\ap=\frac{1}{2}$ below for actual numerical plots, though we obtain essnetially same results for any $\ap$.
$x_{1,2}$ is the location of the operator insertion, and $\epsilon_{1,2}$ is a Euclidean time evolution that plays the role of UV cutoff. 
$\mathcal{N}_{1,2}$ is a normalization constant.
We take an interval subsystem $A = [x_l, x_r]$ at time $t = 0$, whose length we denote as $L = x_r - x_l$, 
and consider the reduced density matrix (or transition matrix) 
\begin{equation}
  \rho_A = \frac{\Tr_{\bar{A}} \ket{\psi}\la\vp|}{\la \vp|\psi\lb} \!\! 
\end{equation}
We will now show that this reduced density matrix exhibits expected non-hermitian behaviors. 

\begin{figure}[t]
  \centering
  \includegraphics[scale = 0.25]{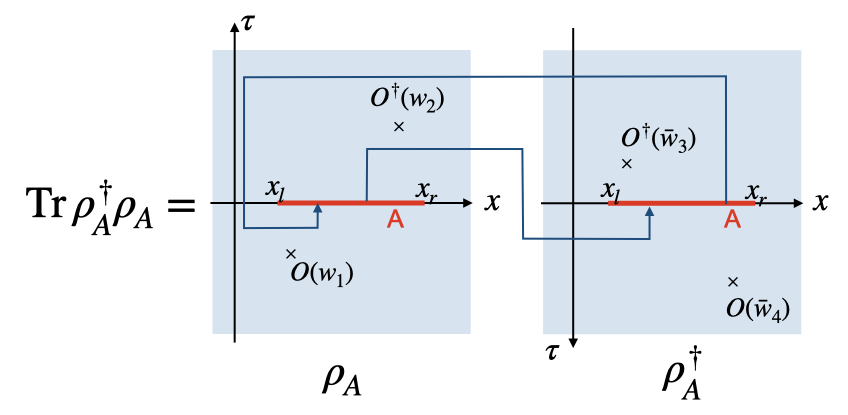}
  \caption{The manifold $\Sigma_2$ for $\mathrm{Tr}(\rho^{\dagger} \rho$). The left copy represent a reduced density matrix $\rho_A$, whereas the right copy is $\rho_A^{\dagger}$. The Hermitian conjugate is achieved by flipping Euclidean time direction. }
  \label{querhd}
\end{figure}

In the path integral formalism, $\rho_A$ is represented as in the left copy of Fig.\ref{querhd}. 
The quantity $\Tr \rho_A^2$ can be computed using the usual replica trick \cite{Calabrese:2004eu,Nozaki:2014hna,Nozaki:2014uaa} and this gives the second Renyi entropy $S^{(2)}_A=-\log\Tr \rho_A^2$.
On the other hand, the other quantity $\Tr \rho^{\dagger}_A \rho_A$ can be evaluated in a similar way with a slight modification, as shown in Fig.\ref{querhd}.  
Explicitly, taking complex coordinate $w_j = x_j + \rmi  \epsilon_j$, these quantities are evaluated as a four point function on a replica manifold;
{\footnotesize
\begin{align}
  \frac{\bTr{\rho_A^2}}{\bTr{\rho_0^2}} &= \frac{\ev{\mO(w_1, \bar{w}_1) \mO^\dagger(w_2, \bar{w}_2) \mO(w_3, \bar{w}_3) \mO^\dagger(w_4, \bar{w}_4)}_{\Sigma_2}}{\ev{\mO(w_1, \bar{w}_1) \mO^\dagger(w_2, \bar{w}_2)}_{\Sigma_1}^2}, \label{rhoptf} \\
  \frac{\bTr{\rho_A^\dagger \rho_A}}{\bTr{\rho_0^2}} &= \frac{\ev{\mO(w_1, \bar{w}_1) \mO^\dagger(w_2, \bar{w}_2) \mO(\bar{w}_4, w_4) \mO^\dagger(\bar{w}_3, w_3)}_{\Sigma_2}}{\ev{\mO(w_1, \bar{w}_1) \mO^\dagger(w_2, \bar{w}_2)}_{\Sigma_1} \ev{\mO(\bar{w}_2, w_2) \mO^\dagger(\bar{w}_1, w_1)}_{\Sigma_1}}. \label{rdrptf}
\end{align}
}

Here $\rho_0$ is a reduced density matrix of the vacuum $\rho_0 = \Tr_{\bar{A}} \ket{0} \!\! \bra{0}$. 
We will use a conformal transformation $\frac{w - x_l}{w - x_r} = z^2$, which maps $\Sigma_2$ to $\Sigma_1$. 
Under this map, the four insertion points are related to each other as $z_3 = -z_1$ and $z_4 = - z_2$ for $\Tr \rho_A^2$, whereas $z_3 = - \bar{z}_2$ and $z_4 = - \bar{z}_1$ for $\Tr \rho_A^{\dagger} \rho_A$. 
Using them, we can represent two quantities as
\begin{align}
  \frac{\bTr{\rho_A^2}}{\bTr{\rho_0^2}} =& \frac{1}{2} \left(1 + \left|\frac{z_{12}^2}{4 z_1 z_2} \right| + \left|\frac{(z_1 + z_2)^2}{4 z_1 z_2} \right| \right) , \label{rhosqu} \\
  \frac{\bTr{\rho_A^\dagger \rho_A}}{\bTr{\rho_0^2}} =& 
    \frac{1}{8} \sqrt{ \frac{|z_2 + z_1|^2 |z_1 + \bar{z}_2|^2}{|z_1| |z_2| \mathrm{Re}(z_1) \mathrm{Re}(z_2)}} \notag \\ 
    & \left(1 + \left| \frac{z_{12}}{z_1 + \bar{z}_2} \right|^2 + \frac{|4 \mathrm{Re}(z_1) \mathrm{Re}(z_2)|}{|z_1 + \bar{z}_2|^2} \right), \label{rhodag}
\end{align}
where we defined $z_{ij} = z_i - z_j$. 
We mention that $\bTr{\rho_0^2}$ only depends on the length of subsystem $A$, as $\bTr{\rho_0^2} = C \left(L / \epsilon \right)^{-\frac{1}{4}}$ \cite{Calabrese:2004eu}, 
which means that this is just a normalization and does not affect the non-hermitian nature of $\rho_A$. 
Finally, we can calculate imagitivity $|| \rho_A^{\dagger} - \rho_A ||_2$ defined by (\ref{imgt}),
up to normalization. 

\subsubsection{Results for Euclidean setup}

\begin{figure}[h]
  \centering
  \includegraphics[width=\linewidth]{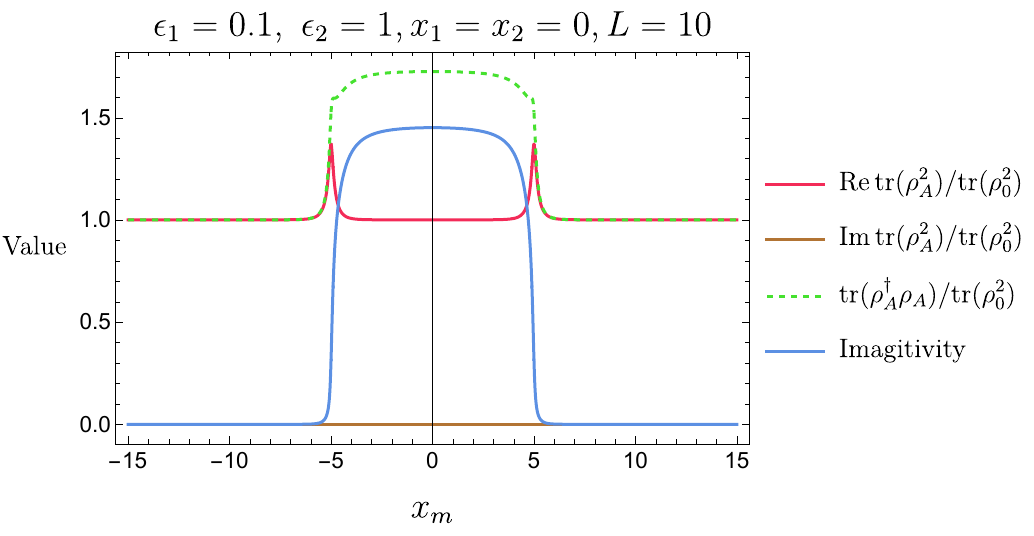}
  \caption{$\bTr{\rho_A^2}$, $\bTr{\rho_A^\dagger \rho_A}$ 
  , and imagitivity for the case with different cutoffs as a function of $x_m=\frac{x_l+x_r}{2}$. }
  \label{qcase1}
\end{figure}

\begin{figure}[h]
  \centering
  \includegraphics[width=\linewidth]{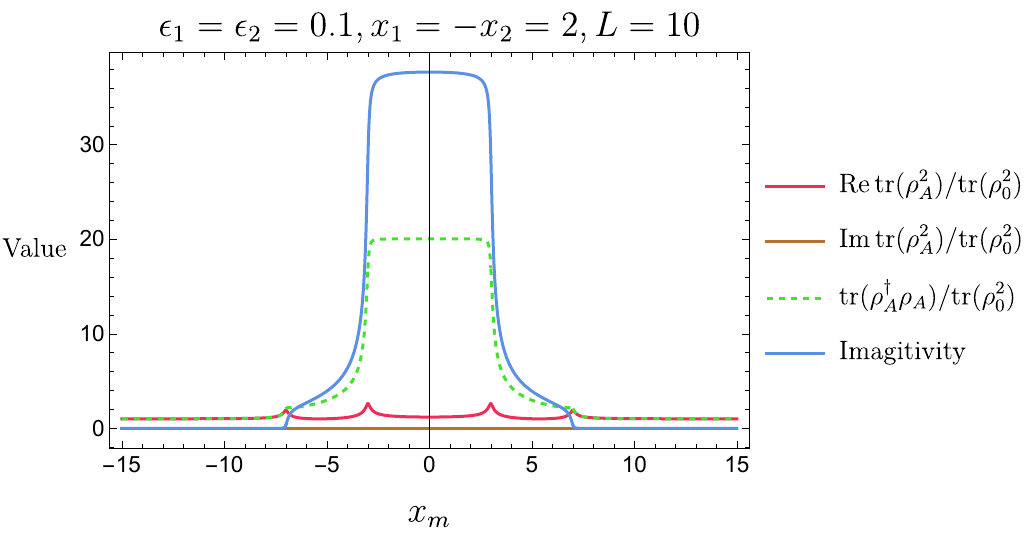}
  \caption{$\bTr{\rho_A^2}$, $\bTr{\rho_A^\dagger \rho_A}$, and imagitivity for the case with different insertion positions, as a function of $x_m=\frac{x_l+x_r}{2}$.}
  \label{qcase2}
\end{figure}

In Fig.\ref{qcase1} and \ref{qcase2}, we will show the behavior of $\bTr{\rho_A^2}$, $\bTr{\rho_A^\dagger \rho_A}$, and imagitivity as a function of the location $x_m$ (i.e. the center of the interval $A$) of the subsystem, where we set $x_m=\frac{x_l+x_r}{2}$.
The first one shows the result for the same insertion points but with different cutoffs ($x_1 = x_2$, $\epsilon_1 \neq \epsilon_2$). 
We can see that imagitivity becomes non-zero only when two insertion points $x_{1,2}$ lie within the interval of subsystem $A=[x_l, x_r]$. 
Since the non-hermitian nature originates solely from the insertion of local operators, 
it is therefore quite natural that the subsystem captures information about non-hermiticity only when it includes the operator insertion points. 
We can also see that $\bTr{\rho_A^2}$ exhibits a sharp increase 
when insertion points get closer to the edge of the interval. This is explained by the fact that the insertion of local operator creates a Bell pair and  is consistent with the behavior of pseudo entropy observed in the previous work \cite{Nakata:2021ubr}. 

The plots in Fig. \ref{qcase2} present the result for the identical cutoff but varying insertion positions ($x_1 \neq x_2$, $\epsilon_1 = \epsilon_2$). 
In this case, the density matrix becomes non-hermitian when both of or either one of two insertion points $x_{1,2}$ is included in $[x_l, x_r]$ (Figure \ref{qcase3}, \ref{qcase4}). 
In addition, the imagitivity takes relatively large values, which is simply due to the small cutoffs. 
We also note that $\bTr{\rho_A^2}$ increases sharply
when one of insertion points becomes closer to the edge of the interval, which is again consistent with the expected behavior of pseudo entropy.

\subsubsection{Lorentzian time evolution}

\begin{figure}[h]
  \centering
  \includegraphics[scale = 0.3]{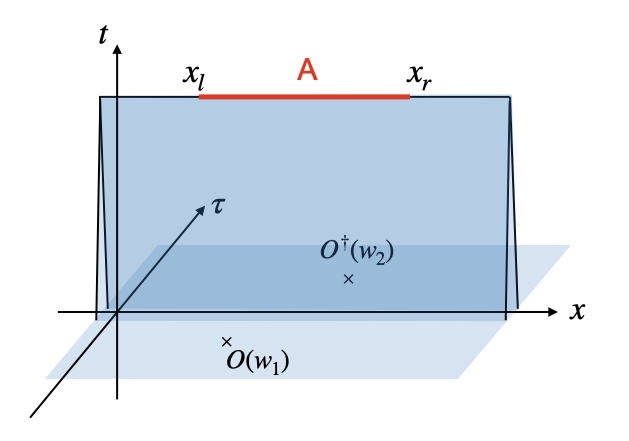}
  \caption{The schematic picture of Lorentzian time evolution of subsystem. The location of subsystem can be obtained by the analytic continuation. }
  \label{lotiev}
\end{figure}

We will now consider the Lorentzian time evolution of the subsystem. 
The schematic picture is shown in Fig.\ref{lotiev}. 
This procedure can be achieved by substituting
\begin{equation}
  x_{l/r} \rightarrow x_{l/r} - t, \quad \bar{x}_{l/r} \rightarrow x_{l/r} + t, 
\end{equation}
which corresponds to the analytic continuation from Euclidean time to Lorentzian time $\tau \rightarrow i t$. 
We note that a Bell pair created by the operator insertion propagates under time evolution, with the two excitations moving away from each other at speed of light. 
Therefore, the time dependence of quantities such as the imagitivity can be understood by tracking the propagation of these excitations.

\begin{figure}[h]
  \centering
  \includegraphics[width=\linewidth]{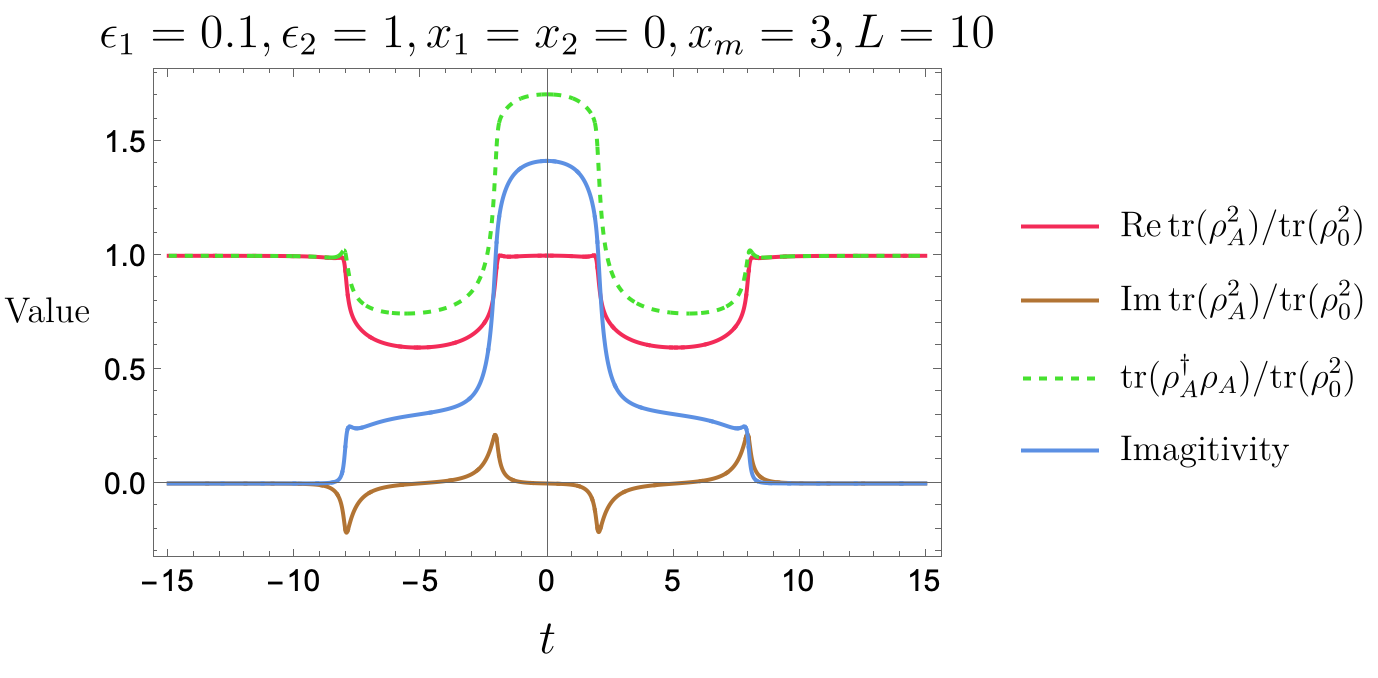}
  \caption{Lorentzian time evolution of $\bTr{\rho_A^2}$, $\bTr{\rho_A^\dagger \rho_A}$, and imagitivity for the case with different cutoffs. }
  \label{qcase3}
\end{figure}

\begin{figure}[h]
  \centering
  \includegraphics[width=\linewidth]{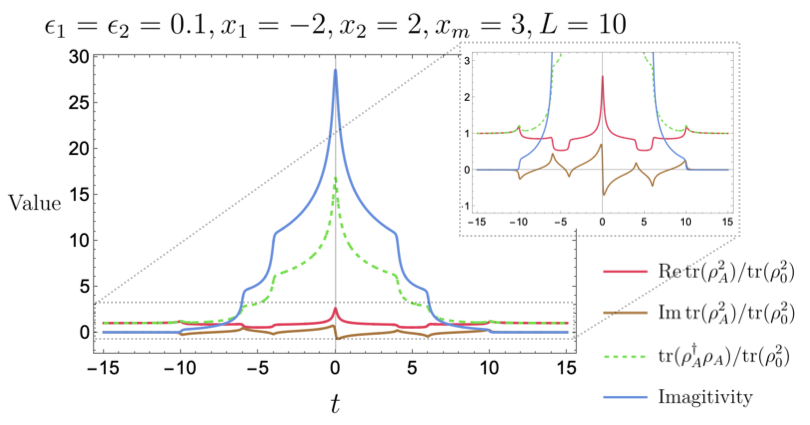}
  \caption{Lorentzian time evolution of $\bTr{\rho_A^2}$, $\bTr{\rho_A^\dagger \rho_A}$, and imagitivity for the case with different insertion positions.}
  \label{qcase4}
\end{figure}

In Fig.\ref{qcase3} and \ref{qcase4}, we will show the time evolution of $\bTr{\rho_A^2}$, $\bTr{\rho_A^\dagger \rho_A}$, and imagitivity. 
The first one shows the result for the case with different cutoffs. 
We can see that $\bTr{\rho_A^\dagger \rho_A}$ and imagitivity have nontrivial values when one or both of excitations are included in the interval. 
Thus, the capacity to detect non-hermiticity are governed by the existence of excitaions inside the subsystem. 
On the other hand, real part of $\bTr{\rho_A^2}$ vary from one when only the one half of Bell pair is included in the subsystem, which detects the entanglement between $A$ and $\bar{A}$. 
We also note that not real part but imaginary part of $\bTr{\rho_A^2}$ shows the sharp behavior 
when any one of excitations get close to the edge of the interval. 

Fig.\ref{qcase4} shows the result for varying insertion points case. 
The basic properties are same to the previous one, although three quantities exhibit more complicated time evolutions because of the shift of operator positions.

\section{non-hermitian density matrices from 
causal influences under unitary evolutions (Class 1)}\label{sec:classone}

Now we move on to the examples of non-hermitian density matrices induced by causal influences under unitary time-evolutions. 
As we explained generally in section \ref{sec:classaa}, the unitary time evolution generates the time-like entanglement, making the generalized density matrices non-hermitian. In this section, first we provide a simple model of coupled harmonic oscillators. Then we analyze the time-like entanglement in its field theory counterpart, namely a two dimensional CFT with a double interval subsystem.

\subsection{Coupled harmonic oscillators} \label{sec:harmoc}

\begin{figure}[ttt]
   \centering
   \includegraphics[width=3.5cm]{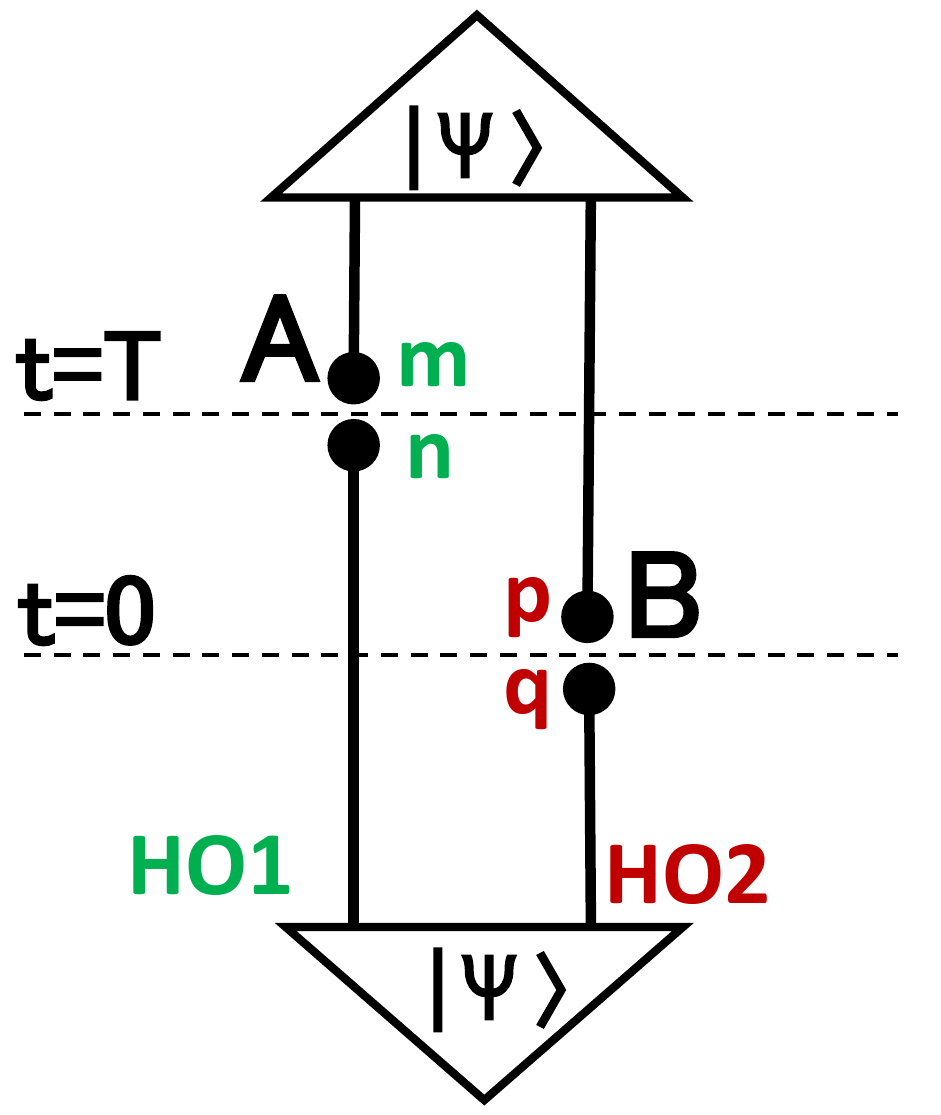}
   \caption{non-hermitian density matrix in coupled harmonic oscillators.}
   \label{fig:HO}
\end{figure}

As a simple example, consider two coupled harmonic oscillators $A$ and $B$ (see also \cite{Kawamoto:2025oko} for an earlier analysis). It is defined by the Hamiltonian 
\begin{align} 
H=\frac{1}{\s{1-\lambda^2}}\left[a^\dagger a+b^\dagger b+\lambda(a^\dagger b^\dagger +ab)+1-\s{1-\lambda^2}\right].
\label{Hamosw}
\end{align}
Via the Bogoliubov transformation, setting $\lambda=\tanh2\theta$, we introduce
\begin{align}
& \tilde{a}=\cosh\theta\ a+\sinh\theta\ b^\dagger,\ \ \tilde{b}=\sinh\theta\ a^\dagger +\cosh\theta\ b,  \label{bgla}
\end{align}
the Hamiltonian gets diagonalized as $H=\tilde{a}^\dagger \tilde{a}+\tilde{b}^\dagger \tilde{b}.$
Thus the ground state $|\Psi\lb$ is found to be
\begin{align}
|\Psi\lb_{AB}=|\tilde{0}\lb_{AB}&=\frac{1}{\cosh\theta}e^{-\tanh\theta a^\dagger b^\dagger}|0\lb_A|0\lb_B \nonumber\\
&=\frac{1}{\cosh\theta}\sum_{k=0}^\infty (-\tanh\theta)^k |k\lb_A |k\lb_B,
\label{gsthdf}
\end{align}
where we introduce the number states $|n\lb_A=\frac{(a^{\dagger})^n}{\s{n!}}|0\lb_A$ and $|m\lb_B=\frac{(b^\dagger)^m}{\s{m!}}|0\lb_B$ as usual. 

Consider the generalized density matrix $\rho_{AB}$ by focusing on at $t=T$ for the harmonic oscillator $A$ and at $t=0$ for $B$ as described, as depicted in Fig.\ref{fig:HO}. This is 
explicitly given by 
\begin{align}
 \left[\rho_{AB}\right]^{mp}_{nq} =&\la \Psi||m\lb_A\la n|e^{-\rmi HT}|p\lb_B\la q||\Psi\lb \nonumber \\
=& \frac{1}{\cosh^2\theta}(-\tanh\theta)^{m+q}\la n|_A\la m|_B
e^{-\rmi HT}|q\lb_A|p\lb_B. \label{rabb}
\end{align}
This is non-vanishing only when $n+p=m+q$. Also it is useful to note the property:
\ba
\la n|_A\la m|_B
e^{-\rmi HT}|q\lb_A|p\lb_B&=&\la q|_A\la p|_B
e^{-\rmi HT}|n\lb_A|m\lb_B \no
&=&\la p|_A\la q|_B
e^{-\rmi HT}|m\lb_A|n\lb_B.\no
\ea

It is straightforward to see that $\rho_{AB}$ is non-hermitian  when 
$T\notin \pi {\mathbb{Z}}$.
We can also easily confirm 
$\mbox{Tr}\rho_{AB}=\la \Psi|e^{-\rmi HT}|\Psi\lb=1.$

The reduced transition matrix $\rho_A$ is estimated as 
\begin{align}
[\rho_A]^m_n
=&\delta_{n,m}\frac{\tanh^{2m}\theta}{\cosh^2\theta}, \label{rhoa}
\end{align}
which is manifestly hermitian $||\rho_A-\rho_A^\dagger|| = 0$ and time-independent as usual. Its second Renyi entropy is computed as 
\begin{align}
  S_A^{(2)} = -\log \mbox{Tr}\rho_A^2=\log \qty[\cosh^4\theta -\sinh^4\theta],
\end{align}
which is due to the quantum entanglement between $A$ and $B$ in the presence of the interaction.

On the other hand, the pseudo entropy and imagitivity of the total system $AB$ get more non-trivial. After explicit calculations presented in appendix \ref{ap:CHO}, we obtain
\begin{align}
    \Tr \rho_{AB}^2 &= \frac{2}{1+e^{-2\rmi T}+(1-e^{-2\rmi T})\cosh 4\theta}, \nonumber\\
    \Tr\rho_{AB}\rho_{AB}^\dagger &= \frac{2}{1+\cos^2 T+\sin^2 T \cosh (4 \theta)}.
\end{align}
This reproduces the second Renyi pseudo entropy computed in \cite{Kawamoto:2025oko}:
\begin{align}
S^{(2)}_{AB}=\log\left[\frac{1+e^{-2\rmi T}+(1-e^{-2\rmi T})\cosh 4\theta}{2}\right]. \label{PEHOQ}
\end{align}
We find this is vanishing either when $\theta=0$ (no coupling) or $T\in\pi{\mathbb{Z}}$.
In general, this entropy takes complex values under the time evolution. This implies the presence of the time-like entanglement, which fits nicely with the fact that $A$ and $B$ are causally related due to the interactions between the two harmonic oscillators.

\begin{figure}[h]
\centering
    \begin{minipage}{0.7\columnwidth}
        \centering
        \includegraphics[width=\columnwidth]{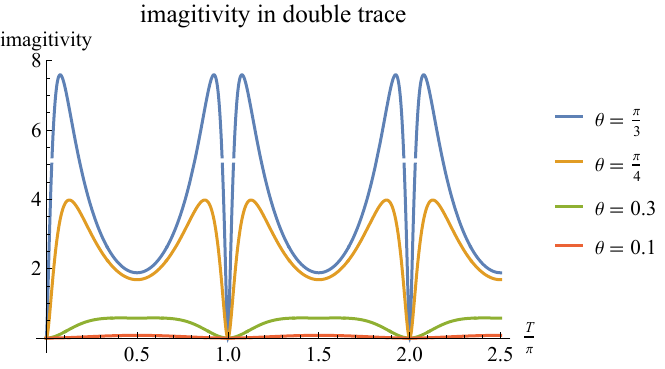}
    \end{minipage}
    \begin{minipage}{0.6\columnwidth}
        \centering
        \includegraphics[width=\columnwidth]{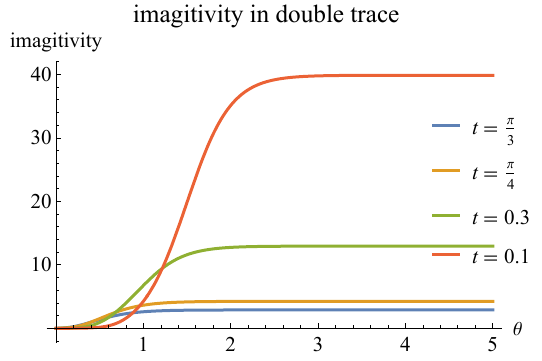}
    \end{minipage}
    \caption{The upper graph is the graph of normalized imagitivity with $T$, and the lower graph is the graph of normalized imagitivity with $\theta$.} 
    \label{fig:Dt_imag_n}
\end{figure}

For the imagitivity (\ref{imgt}), it is useful to normalize as 
\begin{align}
  \frac{||\rho^\dagger_{AB}-\rho_{AB}||_2}{\abs{\Tr[\rho_{AB}^2]}} = \frac{2 \Tr[\rho_{AB}^\dagger \rho_{AB}] - 2 \mathrm{Re} \Tr[\rho_{AB}^2]}{\abs{\Tr[\rho_{AB}^2]}},\label{raitoim}
\end{align}
as the imagitivity itself approaches zero in the limit of  large $\theta$. We plot this normalized imagitivity in Fig.\ref{fig:Dt_imag_n}. This quantity is periodic with respect to time $T$ due to the basic property of harmonic oscillators. At large theta, at first it is rapidly increasing and later decreasing until $T=\pi/2$.
Moreover, it is monotonically increasing about deformation parameter $\theta$. As a function of $\theta$, the normalized imagitivty (\ref{raitoim}) is a monotonically increasing function, which saturates at large $\theta$. This clearly shows that the interaction between $A$ and $B$ generates the time-like entanglement and makes $\rho_{AB}$ non-hermitian. 

\subsection{Time-like entanglement in 2D CFTs}\label{sec:TEE}

As a field theoretic example of time-like entanglement, we would like to consider two dimensional conformal field theories (2D CFTs), where two subsystems $A$ and $B$ are chosen so that they are causally connected. We choose $A$ and $B$ to be two parallel intervals with the same length $2p$, separated by the distance $2q$ from each other as in the right panel of Fig.\ref{fig:Double}.  
Then we expect $\rho^\dagger_{AB}\neq \rho_{AB}$ and the time-like entanglement is expected to be present because $A$ and $B$ are causally connected. As we mentioned in section \ref{sec:tlef}, we expect a similar situation even when $A$ and $B$ are separated in the Euclidean time direction as $A$ and $B$ are influenced by each other via the Euclidean path-integral. 

Below, we first analyze a setup in the Euclidean space and next we discuss the analytical continuation to the Lorentzian signature. As we explained we expect non-hermitian density matrices in both cases. As the tractable examples of 2D CFTs, we especially consider the holographic CFT and the free massless Dirac fermion CFT. The holographic CFT is dual to a classical gravity on a three dimensional anti de-Sitter space (AdS$_3$) via the AdS/CFT. It has a large central charge $c\gg 1$ and is an extremely strongly coupled theory \cite{Maldacena:1997re,Heemskerk:2009pn,Hartman:2014oaa}.

\begin{figure}[H]
   \centering
   \includegraphics[width=7cm]{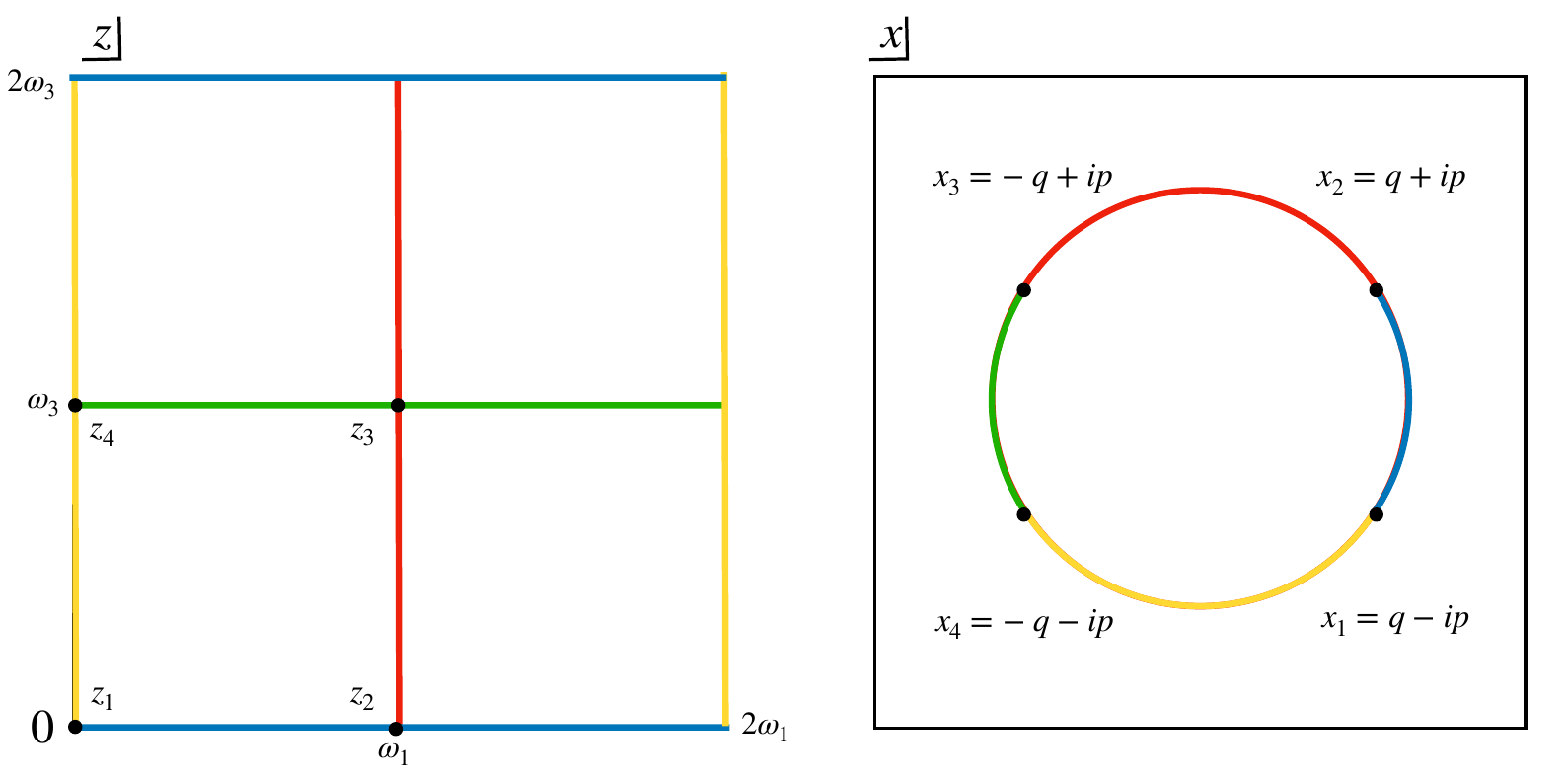}
   \caption{The conformal map from a torus to a branched cover with two sheets used for the calculation of $\mbox{Tr}\rho_{AB}^2$. The first sheet is the image of the region between $[0,\omega_1]$. In the $z$-plane the opposite edges are identified to form the torus. There are three critical points and one pole all of order two $\{z_i\}$ corresponding to the preimages of the branch points $\{x_i\}$. The map is given uniquely up to mobius transformation by the Weierstrass elliptical function $\wp(z,\tau)$. We use this freedom to additionally map the real axis to a circle passing through the points $\pm q\pm ip$. }
   \label{fig:setupeuclideaDI}
\end{figure}

\subsubsection{Euclidean setup}

As mentioned, when $A$ and $B$ are (Euclidean) time-like separated, we expect $\rho^\dagger_{AB}\neq \rho_{AB}$ thus, the imagitivity should be non-trivial. Below we compute both $\mbox{Tr}[(\rho_{AB})^2]$ and $\Tr\rho_{AB}\rho_{AB}^\dagger$ from which we work out the imagitivity \footnote{Throughout we use the unnormalized density matrix $\Tr\rho\neq1$.}.

 The quantity $\mbox{Tr}[(\rho_{AB})^2]$ was computed in \cite{Lunin:2000yv,Headrick:2010zt} and can be understood as the partition function on a torus along with the contribution from the conformal anomaly due to an additional conformal map given uniquely up to mobius transformations by the Weierstrass elliptical function $\wp(z,\tau)$. Using this freedom we can take the four branch points to be $\{x_i\}$ so that $\mbox{Tr}[(\rho_{AB})^2]$  given by 
\be\label{eq:p2arbitary}
\mbox{Tr}(\rho_{AB})^2=2^{-\frac{2c}{3}}|x_{12}x_{13}x_{14}x_{23}x_{24}x_{34}|^{-\frac{c}{12}} Z_{\mathrm{torus}}(\tau,\bar{\tau}),
\ee
where we set $x_{ij}=x_i-x_j$. The subsystems are chosen to be the intervals $A=[x_1,x_2]$ and $B=[x_3,x_4]$.
The moduli $\tau$ of the torus is related to the cross ratio $\eta$ via
\ba\label{eq:lambdafunction}
\eta=\frac{x_{12}x_{34}}{x_{13}x_{24}}=\left[\frac{\theta_2(\tau)}{\theta_3(\tau)}\right]^4,\ \ \ \ 
\frac{x_{23}x_{41}}{x_{13}x_{24}}=\left[\frac{\theta_4(\tau)}{\theta_3(\tau)}\right]^4.
\ea
This can be inverted to give
\be
\tau=\rmi\frac{K(1-\eta)}{K(\eta)}
\ee
where $K$ is the complete elliptical integral of the first kind.
For our set up we will make the choice
\be
\begin{split}
x_1&=q-\rmi p\\
x_2&=q+\rmi p\\
x_3&=-q+\rmi p\\
x_4&=-q-\rmi p\\
\end{split}
\ee
in which case the cross ratio is given by
\be\label{eq:crossratio}
\eta=\frac{x_{12}x_{34}}{x_{13}x_{24}}=\frac{1}{1+\left(\frac{q}{p}\right)^2}.
\ee
Note this only depends on the ratio $\frac{q}{p}$ so we can choose to fix either $q$ or $p$ constant.
It follows that
\be\label{eq:cr_qeta}
|x_{12}x_{13}x_{14}x_{23}x_{24}x_{34}|=64\left|p^6\frac{(1-\eta)}{\eta^2}\right|=64\left|q^6\frac{\eta}{(1-\eta)^2}\right|.
\ee
Working with the later form for the Euclidean calculation we choose to fix $q=1$.

For holographic theories we can evaluate the torus partition function by taking the leading semiclassical approximation \cite{1998JHEP...12..005M,2020JHEP...02..170D}
\be\label{eq:holopf}
\begin{split}
S_{\text{min}}(\tau)=\min_{a,b,c,d\in\mathbb{Z},
\;ad-bc=1}\left[\frac{\rmi\pi c}{12}\left(\frac{a\tau+b}{c\tau+d}-\frac{a\overline{\tau}+b}{c\overline{\tau}+d}\right)\right]\\ \text{ with }Z_{\text{torus}}(\tau)=e^{-S_{\text{min}}(\tau)}.
\end{split}
\ee
here for $\eta\in[0,1]$ there are two contributing phases corresponding to the identity $e$ and $S$ modular transformations of the the modular parameter with the phase transition occurring at $\tau=\rmi$ or equivalently $\eta=\frac{1}{2}$.

We will also consider the Dirac fermion where the partition function is given explicitly \cite{Takayanagi:2022xpv} in terms of Jacobi theta functions as 
\ba\label{eq:diracpf}
Z_{torus}&=&\frac{|\theta_3(\tau)|^2+|\theta_2(\tau)|^2+|\theta_4(\tau)|^2}{2|\eta(\tau)|^2}\no
&=&2^{\frac{2}{3}}\left|\frac{x_{13}^{\frac{1}{3}}
x_{24}^{\frac{1}{3}}}
{x_{12}^{\frac{1}{6}}x_{34}^{\frac{1}{6}}x_{23}^{\frac{1}{6}}x_{41}^{\frac{1}{6}}}\right|.
\ea
By plugging this into (\ref{eq:p2arbitary}), we find the following simple expression of Tr$\rho_{AB}^2$ for the Dirac fermion CFT:
\ba
\mbox{Tr}(\rho_{AB})^2_{D}\propto\left|\frac{p^2+q^2}{p^2q^2}\right|^{\frac{1}{4}}.
\label{diracren}
\ea

The situation for $\Tr\rho_{AB}\rho_{AB}^\dagger$ is far less studied though the method for calculation proceeds similarly. We have relegated many of the details to appendix \ref{app:2cftdets} and here summarize the main results. The required conformal transformation from the torus to the complex plane is given by
\be
f(z)=\frac{\rmi q}{\pi}\left(\zeta(z)+\zeta\left(z+\omega_3\right)-4\eta_1 z-\eta_3\right)
\ee
where $\zeta$ is the Weierstrass zeta function and $\omega_i,\eta_i$ are the half-periods and half-quasiperiods corresponding to the modular parameter $\tau$. This conformal transformation was first considered in the context of 2d CFTs in \cite{Numasawa:2016emc,2019JHEP...09..018C} and appears more generally in the context of conformal mappings of multiply connected domains see e.g. \cite{CrowdyMarshall2006,Crowdy2012}. The effect of this mapping is shown in Fig. \ref{fig:setupDI}:

\begin{figure}[H]
   \centering
   \includegraphics[width=7cm]{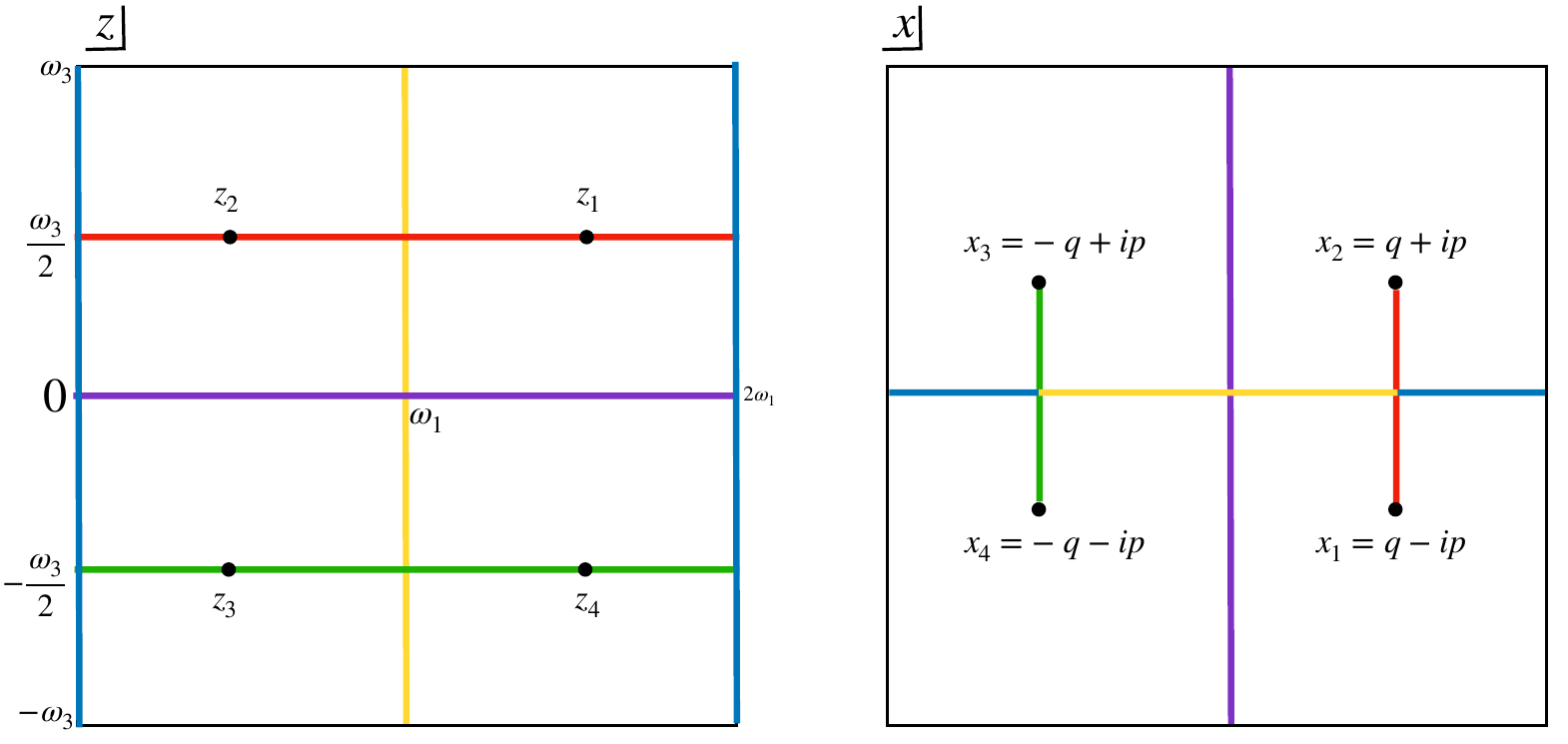}
   \caption{Conformal map $x=f(z)$ from a rectangular torus to a branched cover with two sheets. The first sheet is the image of the region between $\pm\frac{\omega_3}{2}$. In the z-plane the opposite edges are identified to form the torus. There are four critical points $\{z_i\}$ corresponding to the preimages of the branch points $\{x_i\}$. There are also two poles (accounting for the side identifications of the torus) at $0,\omega_3$.}
   \label{fig:setupDI}
\end{figure}

\begin{figure}[H]
   \centering
   \includegraphics[width=7cm]{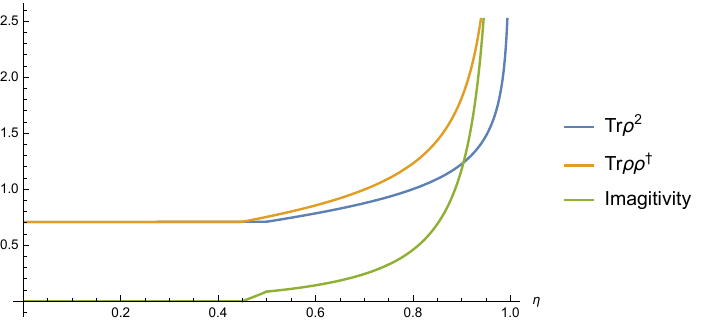}
   \caption{The imagitivity (green) as a function of cross ratio $\eta$ for holographic theories with torus partition function \eqref{eq:holopf}. }
   \label{fig:euclidean_all}
\end{figure}

\begin{figure}[H]
   \centering
   \includegraphics[width=7cm]{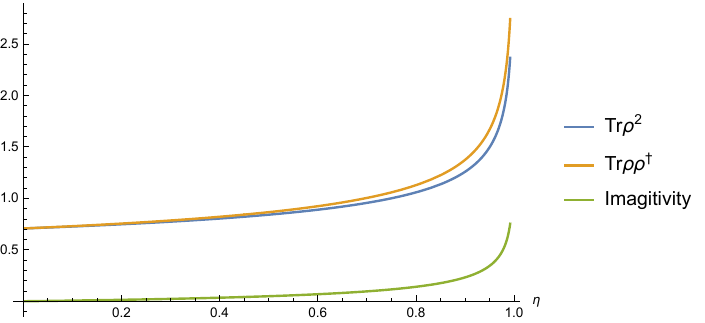}
   \caption{The imagitivity (green) as a function of cross ratio $\eta$ for the Dirac fermion with torus partition function \eqref{eq:diracpf}.   }
   \label{fig:diracfermion}
\end{figure}

While we are able to determine the final answer equation \eqref{eq:trrhorhodeta} exactly as a function of the modular parameter $\tau$ it is necessary for us to numerically determine the relation between $\eta$ and $\tau$ (see figure \ref{fig:etataunumeric}) which can then be related to $\frac{q}{p}$ using equation \eqref{eq:crossratio}. The final imagitivity as a function of cross-ratio $\eta$ is shown in Fig. \ref{fig:euclidean_all} for holographic theories and in Fig. \ref{fig:diracfermion} for the Dirac fermion.

In both CFTs, $\mbox{Tr}[(\rho_{AB})^2]$  and $\Tr\rho_{AB}\rho_{AB}^\dagger$ are monotonically increasing functions of $\eta$. The former shows that the second Renyi entropy $S^{(2)}_{AB}$ decreases and the Renyi mutual information $I^{(2)}_{AB}=S^{(2)}_A+S^{(2)}_B-S^{(2)}_{AB}$ increases  
as the intervals gets closer, which agrees with our expectation. 
The imagitivity, obtained from the difference between the former and latter, also turns out to be a monotonically increasing function of $\eta$. This is also quite sensible as we expect that as $A$ and $B$ gets closer, it gets more non-hermitian. When the ratio $\frac{q}{p}$ becomes infinite the imagitivity should be vanishing, while in the limit $\frac{q}{p}\to 0$, it gets divergent. In the holographic CFT, due to the large $c$ phase transition, for a finite but small values of $\eta$, the imagitivity becomes zero, while in the Dirac fermion CFTs, it is nonzero for any nonzero $\eta$.

\subsubsection{Lorentzian setup}

\begin{figure}[H]
   \centering
   \includegraphics[width=5cm]{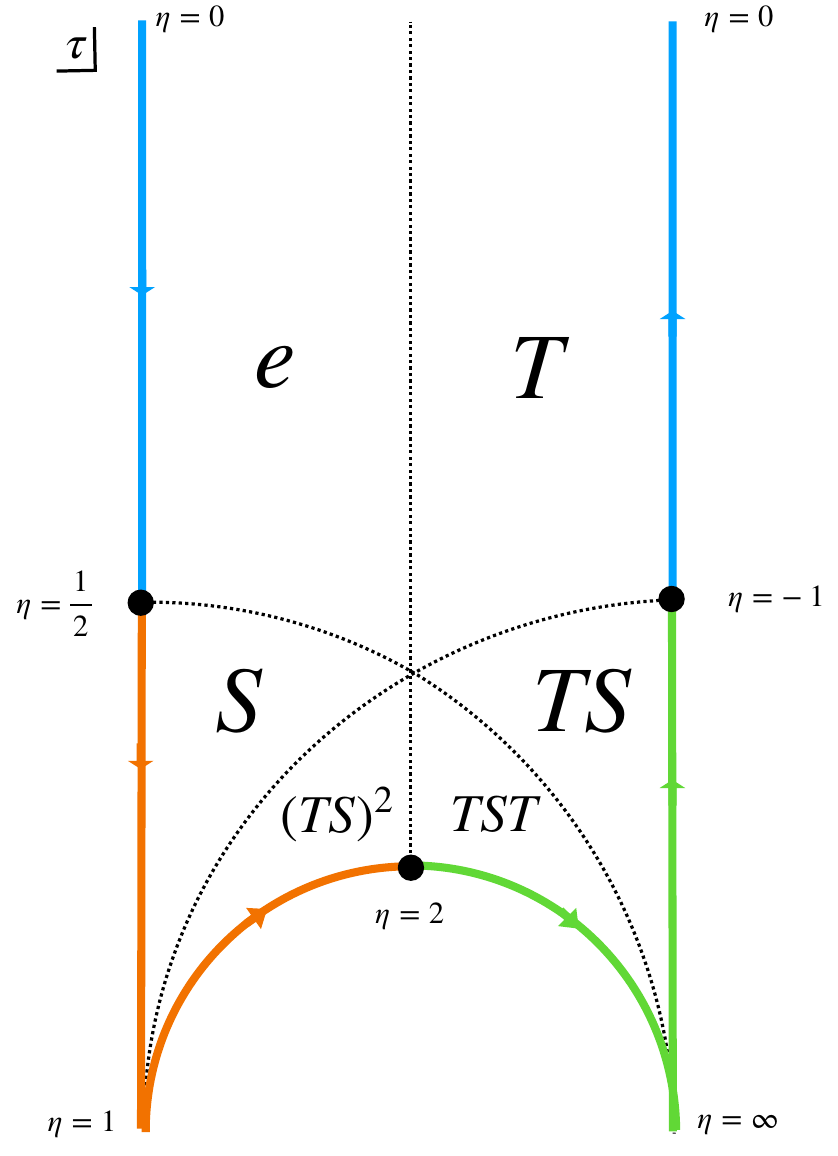}
   \caption{A contour for real $\eta$ in the $\tau$ upper half plane between the points $\tau: \infty\rightarrow0\rightarrow1\rightarrow\infty$. There are three contributing phases corresponding to the modular transformations $e,S,(TS)^2$. The phase transitions occur at $\eta=\frac{1}{2},2,-1$ and the contour has been colored to match the dominate phase as shown in Fig. \ref{fig:TP_phase}. Note that the dominate phases are symmetric in the sense that the transitions occur exactly at the halfway points (using the natural hyperbolic metric on the upper half plane) from the three cusp points at $\tau=0,1,\infty$ along the contour.  }
   \label{fig:toruscontour}
\end{figure}

\begin{figure}[H]
   \centering
   \includegraphics[width=7cm]{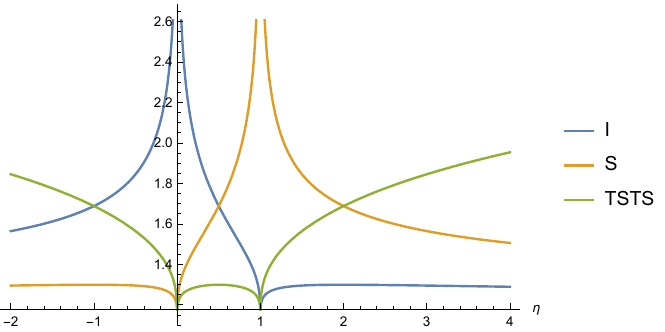}
   \caption{The holographic torus partition function \eqref{eq:holopf} as a function of $\eta\in\mathbb{R}$. Shown are the possible contributing phases with the largest being the dominate contribution. The region $\eta\in[0,1]$ correspond to the usual euclidean evolution with  phase transition between the identity and $S$ phase which occurs at $\eta=\frac{1}{2}$ or precisely when the torus is square with modular parameter $\tau=i$. When we consider the analytic continuation $q\rightarrow it$ $\eta$ now takes the values $\eta\in[1,\infty]\cup[-\infty,0]$. In particular this gives rise to a new dominate phase corresponding to the $(TS)^2$ modular transformation of the modular parameter with phase transitions occurring at $\eta=2,-1$. the colors for the phases have been coordinated with the contour shown in Fig. \ref{fig:toruscontour}.  }
   \label{fig:TP_phase}
\end{figure}

Here we consider the analytic continuation to Lorentzian signature by taking $q\longrightarrow it$ and fixing $p=1$. In particular the cross ratio is now given by
\be
\eta=\frac{1}{1-\left(\frac{t}{p}\right)^2}
\ee
with $\eta\in(-\infty,0)\cup(1,\infty)$. For  $\mbox{Tr} \rho_{AB}^2$ we can use the relation equation \eqref{eq:lambdafunction} to determine a contour in the $\tau$ plane for which $\eta$ is real\footnote{The function~\eqref{eq:lambdafunction} is automorphic with respect to the
congruence subgroup $\Gamma(2)\subset PSL(2,\mathbb{Z})$, which has index $6$ in the full modular group. Consequently, a fundamental domain for $\Gamma(2)$ may be taken as the union of the standard fundamental domain for
$PSL(2,\mathbb{Z})$ together with its five additional images under coset
representatives of $\Gamma(2)\backslash PSL(2,\mathbb{Z})$.

Since \eqref{eq:lambdafunction} maps this entire fundamental region onto a
single copy of the Riemann sphere, identifying the image of the real axis
requires distinguishing between the upper and lower half-planes. To achieve this, one adjoins the reflection about the imaginary axis to the group, forming the $\mathbb{Z}_{2}$-extension generated by this reflection together with $\Gamma(2)$. The contour of interest is then the boundary of the fundamental region for this extended group.}. For the holographic CFT, this is shown in Fig. \ref{fig:toruscontour} and the holographic partition function along this contour is shown in Fig. \ref{fig:TP_phase}.

\begin{figure}[H]
   \centering
   \includegraphics[width=7cm]{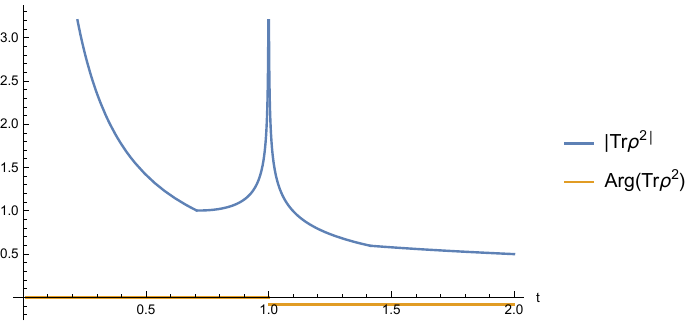}
   \caption{The magnitude and argument of the purity $|\mbox{Tr} \rho_{AB}^2|$ as a function of Lorentzian time $t$ for the holographic CFT. Here we have chosen $p=1$ with the lightcone singularity occurring at $t=p=1$. The two phase transitions of the torus partition function occur at $\frac{p}{\sqrt{2}}$ and $\sqrt{2}p$.  }
   \label{fig:rho2L}
\end{figure}

Using these it is possible to determine $\mbox{Tr} \rho_{AB}^2$ for holographic theories as a function of Lorentzian time $t$ as shown in Fig. \ref{fig:rho2L}. Note that \eqref{eq:cr_qeta} reduces to
\be
64|t^8(1-t^2)|
\ee
and then from \eqref{eq:p2arbitary} we conclude the second Renyi pseudo entropy with have a constant imaginary piece $\frac{\pi c}{12}i$ for $|t|> p$.

In this way, for the holographic CFT, we find that the Renyi pseudo entropy $S^{(2)}_{AB}=-\log\mbox{Tr}\rho^2_{AB}$ is mostly a increasing function of the time separation. However, it shows a sharp dip when the size $2p$ of the intervals coincides with the time separation $2t$. At this point, an end point of $A$ is null separated by an end point of $B$. If we remember the holographic pseudo entropy, this is natural as the geodesic which connects these boundary points become null at this time. 

On the other hand, in the Dirac fermions CFT, we simply find the following analytical expression via the Euclidean result (\ref{diracren}):
\ba
\mbox{Tr}(\rho_{AB})^2_{D}\propto\left|\frac{t^2-p^2}{p^2t^2}\right|^{\frac{1}{4}}.
\label{diracrenn}
\ea
Moreover, as we can see from (\ref{diracrenn}), we obtain the imaginary part $\frac{\pi c}{4}i$, which is characteristic to the time-like entanglement entropy, for the second Renyi pseudo entropy for $|t|<p$
\ba
S^{(2)}_{AB}= \frac{c}{4}\log \frac{p^2t^2}{(p^2-t^2)\ep^2}+\frac{\pi c}{4}\rmi,
\ea
where $\ep$ is the UV cut off. Notice that this positively diverges at $t=p$, as opposed to the holographic CFT result (depicted in Fig.\ref{fig:rho2L}), though in the limit $t\to 0$ and $t\to \infty$, both of them behave similarly.

While it would certainly be interesting to consider the analytic continuation for $\mbox{Tr} \rho_{AB}\rho^{\dagger}_{AB}$ and compute the imagitivity as a function of Lorentzian time $t$ there is a significant challenge in performing the analytic continuation due to the lack of a specific functional form of $\eta(\tau)$ which in particular must be valid for all $\tau$ (not just pure imaginary $\tau=is$ as was done in our Euclidean analysis). We leave further consideration of this to future work.

However in order for the imagitivity to be positive we anticipate that $\Tr\rho_{AB}\rho^\dagger_{AB}$ will follow a similar structure to $\Tr\rho^2_{AB}$. Like the Euclidean case the same three phases should contribute but with the phase transitions will occur at different values of $\eta$. We also expect the imagitivity also behaves similar, which will be decreasing under the time evolution except intermediate phases as in Fig.\ref{fig:rho2L}.

\subsection{Traversable wormholes and time-like entanglement}\label{sec:doublewh}

The holographic duality equivalently relates the dynamics of a holographic CFT to that of gravity in an  asymptotically AdS. The traversable AdS wormholes provide an interesting class of non-hermitian density matrices, where the time-like entanglement and causal influences play the crucial role to make the wormhole traversable as first argued in \cite{Kawamoto:2025oko}. 

Consider two identical CFTs, called CFT$_{(1)}$ and  CFT$_{(2)}$ and assume a thermofield double (TFD) state $|\mbox{TFD}(\beta)\lb$ defined by 
\ba
|\mbox{TFD}(\beta)\lb=\sum_n e^{-\frac{\beta}{4}(H_1+H_2)}|n\lb_1|n\lb_2, \label{TFD}
\ea
where $|n\lb_{1,2}$ are the energy eigenstates in the two CFTs. 
If we trace out one of the two CFTs, then we obtain the density matrices of the canonical distributions $\rho_{1,2}\propto e^{-\beta H_{1,2}}$. This TFD state is well-known to be dual to the eternal AdS black hole \cite{Maldacena:2001kr} via the AdS/CFT. This is a two sided black hole with two asymptotically AdS regions, which are connected through the wormhole so called the Einstein-Rosen bridge. This wormhole is not traversable due to the black hole horizon. 
We cannot send a physical signal from one of the two AdS boundaries to the other. This is consistent with the fact that there are not interactions or causal influences between the two CFTs, though they are entangled.  
Moreover, we can consider the Lorentzian time evolution in each CFT as 
\ba
&&|\mbox{TFD}(t_1,t_2,\beta)\lb\no
&&=\sum_n e^{-\rmi t_1H_1-\rmi t_2H_2}e^{-\frac{\beta}{4}(H_1+H_2)}|n\lb_1|n\lb_2,
\label{timeTFD}
\ea
However, as first shown in \cite{Gao:2016bin}, if we add the interactions between two CFTs, so called the double trace interactions $\sim\int d^dx \mO_1(x) \mO_2(x)$, then the wormhole turns to be traversable. From the CFT viewpoint, this is again expected as the two CFTs are now causally connected. We can also realize eternal traversable wormholes by choosing appropriate interactions in CFTs \cite{Maldacena:2018lmt,Harvey:2023oom}. 

Consider the generalized density matrix in this model. If we take the subsystem $A$ to be a region in the CFT$_{(1)}$ at $t=t_1$ and another region $B$ in the CFT$_{(2)}$ at $t=t_2$,
the generalized density matrix $\rho_{AB}$ gets non-hermitian if $|t_1- t_2|$ gets enough large. This is because of the causal influence between the two CFTs due to the double trace interactions \cite{Kawamoto:2025oko}. A toy model which shows the analogous effect is the coupled harmonic oscillators, presented in section \ref{sec:harmoc}.

In the gravity dual, we can confirm that the generalized density matrix becomes the non-hermitian by computing the holographic pseudo entropy (\ref{extfor}). Indeed, since $A$ and $B$ are time-like separated in the AdS traversable wormhole when $|t_1- t_2|$ is larger than a certain value, the extremal surface which connects $\de A$ with $\de B$ (simply we can assume that $A$ and $B$ are semi-infinite) can have time-like part and this makes its area complex valued. The presence of the imaginary part of pseudo entropy shows that the dual density matrix is non-hermitian.

As mentioned, the presence of causal influence in the traversable AdS wormhole is clear from its geometry. We can physically send a signal from one boundary to the other within a finite time. From the CFT viewpoint, this signal propagation can be explicitly confirmed by the relation (\ref{causalinf}), where the commutator of operators $\mO_A$ and $\mO_B$ does not vanish due to the double trance interactions. This shows that $\rho_{AB}$ is non-hermitian owing to (\ref{Mil}). In this way, to realize a traversable wormhole, we need not only the ordinary quantum entanglement but also the time-like entanglement.

\section{non-hermitian density matrices 
from non-unitary evolutions (Class 2)} \label{sec:classtwo}

Here we would like to investigate explicit examples of generalized density matrices in quantum systems with non-hermitian hamiltonian $H^\dagger\neq H$. We focus on the non-hermitian deformation of the originally unitary quantum system as we introduced in (\ref{DefH}). When the parameter $\lambda$ is real, the hamiltonian is hermitian and we employ the standard conjugation $\dagger$ to relate the ket state to bra one. However, when $\lambda$  takes complex values with non-zero imaginary part, we need to employ the modified conjugation $\ddagger$ as we explained in section \ref{sec:classbb}. We will focus on the examples using the thermofield double (TFD) states. After we present a general description of such quantum systems, we will see that this class of non-hermitian systems naturally arise in the context of CFT duals of traversable AdS wormhole, based on Janus deformations. We will provide both the CFT and gravity analysis below. 

\subsection{non-hermitian TFD states and thermal pseudo entropy}\label{nonhtfd}

The main setup we would like to discuss below in explicit examples of CFTs is the thermofield double state for the non-hermitian hamiltonians. Consider two identical quantum systems $Q_{(1)}$ and $Q_{(2)}$ and assume that their Hamiltonians are given by $H$ and $H^\dagger$, respectively. As we introduced in 
(\ref{ega}), (\ref{egb}) and (\ref{egc}), we write their eigenstates as $|n_+\lb$ and $|n_-\lb$, respectively.  

The (unnormalized) thermofield double state is explicitly expressed as follows 
 \begin{equation}
     \begin{split}
          \ket{\TFD(\beta,\lambda)} &=  e^{-\frac{\beta H}{4}}\otimes e^{-\frac{\beta H^\dag}{4}} \sum_n \ket{n_+}_1\ket{n_+}_2. \label{TFDMa}
     \end{split}
 \end{equation}
Since we define the non-hermitian system by the analytical continuation with respect to the parameter $\lambda$, its conjugate state $ \bra{\overline{\TFD}(\beta,\lambda)}$ should be introduced by the modified conjugation $\ddagger$ defined by (\ref{mconj}). This leads to the following expression:
 \begin{equation}
     \begin{split}
     \bra{\overline{\TFD}(\beta,\lambda)} &=\left(\ket{\TFD(\beta,\lambda)}\right)^\ddagger \no
        &=\sum_n \bra{n_-}_1\bra{n_-}_2  e^{-\frac{\beta H}{4}}\otimes e^{-\frac{\beta H^\dag}{4}},  \label{TFDMb}
     \end{split}
 \end{equation}
 where we employed the property $H^\ddagger=H$.
Notice that $(\ket{\TFD(\beta,\lambda)})^\dag \neq \bra{\overline{\TFD}(\beta,\lambda)}$ and thus the generalized density matrix 
\be
\rho_{12}(\beta,\lambda)={\cal N}\ket{\TFD(\beta,\lambda)}\bra{\overline{\TFD}(\beta,\lambda)}
\label{gened}
\ee
is not hermitian when $\lambda$ is complex valued. Here ${\cal N}$ is the overall normalization which is determined by requiring Tr$\rho_{12}(\beta,\lambda)=1$.

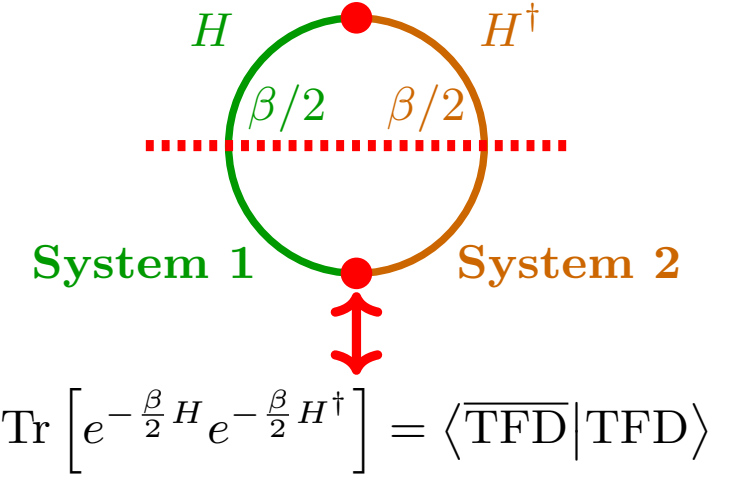
\begin{figure}
    \centering
  \begin{tikzpicture}[scale=1]

\tikzset{
  every node/.style={scale=2.0, font=\bfseries},
}

\def\R{1.7}

\draw[line width=2.8pt, green!60!black]
  (0,\R) arc[start angle=90,end angle=270,radius=\R];

\draw[line width=2.8pt, orange!80!black]
  (0,-\R) arc[start angle=-90,end angle=90,radius=\R];

\draw[red, line width=4.2pt, dashed] (-2.8,0) -- (2.8,0);

\fill[red] (0,\R) circle (0.21);
\fill[red] (0,-\R) circle (0.21);

\node[anchor=south east, text=green!60!black] at (-1.4,1.1) {$H$};
\node[anchor=north east, text=green!60!black] at (-0.17,1.05) {$\beta/2$};
\node[anchor=north east, text=green!60!black] at (-1.1,-1.1) {System 1};

\node[anchor=south west, text=orange!80!black] at (1.4,1.1) {$H^\dag$};
\node[anchor=north west, text=orange!80!black] at (0.17,1.05) {$\beta/2$};
\node[anchor=north west, text=orange!80!black] at (1.1,-1.1) {System 2};

\draw[red, line width=3.5 pt, <->] (0,-1.95) -- (0,-3.05);

\node[align=center] at (0,-3.8)
{$\bTr{e^{-\frac{\beta}{2}H}e^{-\frac{\beta}{2}H^\dag}}
= \braket{\overline{\TFD}}{\TFD}$};

\end{tikzpicture}
    \caption{Sketch of the Euclidean path integral for $\bTr{e^{-\frac{\beta H}{2}}e^{-\frac{\beta H^\dag}{2}}}$. The red points express the defect which connects two different Hamiltonian $H$ and $H^\dag$. The region above and the one below the red line prepare the state $\ket{\TFD}$ and $\bra{\overline{\TFD}}$, respectively. }
    \label{fig:Thermodynamcis}
\end{figure}

 One natural thermodynamic path integral for non-hermitian deformed theory is given by the Euclidean path integral, sketched in Fig. \ref{fig:Thermodynamcis} :
 \begin{equation}
     \bTr{e^{-\frac{\beta H}{2}}e^{-\frac{\beta H^\dag}{2}}} = \sum_{n,m}e^{-\frac{\beta (E_n+E_m^*)}{2}}\braket{n_-}{m_-}\braket{m_+}{n_+}. 
 \end{equation}
 which is clearly real valued. Similar to the standard canonical ensemble case (\ref{TFD}), 
 we have a purification or Choi-map of this mixed states to a pure states in doubled Hilbert space. With short algebra, similar to \cite{Bak:2007qw}, now by using the modified TFD states (\ref{TFDMa}) and (\ref{TFDMb}), we can confirm that this partition function coincides with the inner product of our TFD states:
 \begin{equation}
 \begin{split}
 \braket{\overline{\TFD}(\beta,\lambda)}{\TFD(\beta,\lambda)}=  \bTr{e^{-\frac{\beta H}{2}}e^{-\frac{\beta H^\dag}{2}}}.
 \end{split} \label{tfdpa}
 \end{equation}

Now we consider a partial trace of  $\rho_{12}(\beta,\lambda)$ over $Q_{(1)}$ or $Q_{(2)}$
as follows:
\begin{equation}
\begin{split}
\rho_1  &= \mathrm{Tr}_2{[\rho_{12}]}= e^{-\frac{\beta}{4}H}e^{-\frac{\beta}{2}H^\dag}e^{-\frac{\beta}{4}H},\\
    \rho_2 &= \mathrm{Tr}_1{[\rho_{12}]}= e^{-\frac{\beta}{4}H^\dag}e^{-\frac{\beta}{2}H}e^{-\frac{\beta}{4}H^\dag}.
\end{split}
\end{equation}
which are non-hermitian again. For intuition see Fig.\ref{fig:Thermodynamcis}.
Similar to the standard TFD states, we can consider pseudo entropy for our TFD states, which is analogous to the thermodynamic entropy of each system. Importantly, we find that the Renyi pseudo entropy and von Neumann pseudo entropy are all real valued, in spite that the density matrices are non-hermitian. 
To see this, first write down
\begin{equation}
   \bTr{\rho_1^n}= \bTr{\rho_2^n}= \bTr{e^{-\frac{\beta H}{2}}e^{-\frac{\beta H^\dag }{2}}\cdots e^{-\frac{\beta H}{2}}e^{-\frac{\beta H^\dag }{2}}}  \label{gentrre}
\end{equation}
From the trace cyclicity, this is indeed real valued. The von Neumann pseudo entropy is also real value since
\begin{equation}
    S_{\mathrm{v.N.}}[\rho_1]=S_{\mathrm{v.N.}}[\rho_2] = \lim_{n\to 1}\frac{1}{1-n}\log{ \frac{\bTr{\rho_1^n}}{\bTr{\rho_1}^n}}. \label{entreal}
\end{equation}
Soon later, we will confirm this real-valued property from the holographic computation of the horizon entropy of the AdS traversable wormhole \cite{Kawamoto:2025oko}. Interestingly, the entropy can exceed the value of the eternal AdS black hole, which is expected to be maximal as we will see. This can happen because the entropy (\ref{entreal}) should be regarded as the pseudo entropy instead of entanglement entropy. Even though this is pseudo entropy, its values remains real valued as opposed to the pseudo thermal entropy \cite{Caputa:2024gve}.

Moreover, the above analysis shows that the Euclidean partition function (\ref{tfdpa}) is real valued and non-negative. This implies that our non-hermitian TFD system may actually become well-defined system for at least small non-hermitian deformations.

\subsection{Imaginary Janus deformed free CFT}\label{sec:imjan}

As a CFT example of the non-hermitian TFD state (\ref{TFDMa}), we would like to consider a Janus deformation with the deformation parameter $\lambda$ taken to be imaginary valued, which was first introduced in \cite{Kawamoto:2025oko} based on the solution \cite{Bak:2007jm}. Though we can apply the same procedure to any CFTs, below we would like to focus on the two dimensional CFT described by a $c=1$ free scalar field. This is aiming at a weak coupling limit of AdS$_3/$CFT$_2$. In the next subsection we will give results for a holographic CFT, which is expected to be in strong coupling limit, using the gravity dual. 

Consider two identical CFTs, denoted by  CFT$_{(1)}$  and  CFT$_{(2)}$. The Janus deformation
\cite{Bak:2003jk, Freedman:2003ax, DHoker:2007zhm,Bak:2007jm, Bak:2007qw}
is performed via a perturbation of the CFTs by an exactly marginal primary operator $\mO$ in an asymmetric way, where the original actions $S^{(0)}_1$ and $S^{(0)}_2$ of the CFTs are deformed as follows:
\ba
&& S_1=S^{(0)}_1+\lambda\int d^2x \mO_{(1)}(x), \no
&& S_2=S^{(0)}_2-\lambda\int d^2x \mO_{(2)}(x).\label{Januscftd}
\ea

We consider the specific model where the CFT$_{(i)}$ consists of a massless free scalar field $\phi^{(i)}$ for $i=1,2$.  If we prepare multiple scalar fields and their super partners, then this correspond to the weakly coupled limit of the D1-D5 Janus deformed CFT introduced and studied in the pioneering works \cite{Bak:2007jm, Bak:2007qw}.

We write the compactification radius of each of two free bosons as $R_{i}$ such that the identification looks like $\phi^{(i)}\sim \phi^{(i)}+2\pi R_i$ for $i=1,2$. We start with the common radius $R$ and deform the CFTs such that $R_1\neq R_2$. We parametrize this Janus deformation by the parameter $\theta$ defined by
\ba
R_1=\frac{R}{\s{\tan\theta}},\ \ \ \ R_2=\s{\tan\theta} R.   \label{J:rad}
\ea
In this example, the exact marginal operator $\mO$ is the one which changes the radius or its metric 
as $\mO_{(i)}=\de_a \phi^{(i)}\de^a \phi^{(i)}$. The point $\theta=\frac{\pi}{4}$ corresponds to the TFD state before the Janus deformation and thus the difference 
\ba
\lambda=\theta-\frac{\pi}{4},
\ea
shows the amount of the Janus deformation. When $\lambda$ is real, the Hamiltonian is hermitian as the radii $R_1$ and $R_2$ take real values. However when $\lambda$ is imaginary, which we are interested in, the radii get complex valued and Hamiltonian becomes non-hermitian.

Now we would like to present an explicit description of our CFT. We write the coordinates of the Euclidean time and space as $\tau$ and $\sigma$, respectively. In the description of Fig.\ref{fig:Thermodynamcis} of the inner product of the TFD state, we take $0\leq \tau\leq \beta$ and $0\leq \sigma\leq 2\pi$. A convenient way to deal with such interfaces in CFTs is to employ the folding method \cite{Bachas:2001vj} so that both $\phi^{(1)}$ and $\phi^{(2)}$ live on the same interval $0\leq \tau\leq \frac{\beta}{2}$, where the boundaries $\tau=0$ and $\tau=\frac{\beta}{2}$ are situated at the interface where $\phi^{(1)}$ and $\phi^{(2)}$ are related with each other by a conformal boundary condition, parameterized by $\theta$, whose details can be found in \cite{Bachas:2001vj,Kawamoto:2025oko}. 

The mode expansions of $\phi^{(1)}$ and $\phi^{(2)}$ read
\begin{align}
  \phi^{(i)}_L&=x^{(i)}_L-\rmi p^{(i)}_L(\tau-\rmi \sigma)+\rmi\sum_{m\in \mathbb{Z}}
  \frac{\alpha^{(i)}_m}{m}e^{-m(\tau-\rmi \sigma)},\nonumber\\
  \phi^{(i)}_R&=x^{(i)}_R-\rmi p^{(i)}_R(\tau+\rmi \sigma)+\rmi \sum_{m\in \mathbb{Z}}
  \frac{\tilde{\alpha}^{(i)}_m}{m}e^{-m(\tau+\rmi \sigma)}, \label{modee}
\end{align}
where i=1,2 and the oscillators satisfy
\begin{align*}
  [\alpha^{(i)}_m,\alpha^{(j)}_n]=m\delta_{ij}\delta_{m+n,0} ,\ \ 
  [x^{(i)}_{L,R},p^{(j)}_{L,R}]=\rmi\delta_{ij}.
\end{align*}
By compactifying the scalars on circles with the radius $R_1$ and $R_2$, the quantization of the momenta looks like:
\begin{align}
  P^{(i)}_{L,R}=\frac{n_i}{R_i} \pm \frac{w_iR_i}{2},\label{nw}
\end{align}
where $n_i$ and $w_i$ describe the momenta and winding number and are integer valued. The total Hamiltonian is given by
\begin{align}
  H=\sum_{n=1}^\infty  &( \alpha^{(1)}_{-n} \alpha^{(1)}_n +  \alpha^{(2)}_{-n} \alpha^{(2)}_n +
  \ti{\alpha}^{(1)}_{-n} \ti{\alpha}^{(1)}_n +  \ti{\alpha}^{(2)}_{-n} \ti{\alpha}^{(2)}_n
  ) \nonumber\\
  &+\frac{n_1^2}{R^2_1} + \frac{w^2_1R^2_1}{4} + \frac{n_2^2}{R^2_2} + \frac{w^2_2R^2_2}{4} - \frac{1}{6}.
\end{align}

The folding method allows us to describe the modified TFD state (\ref{TFDMa}) in terms of the boundary state \cite{Bachas:2001vj,Sakai:2008tt,Kawamoto:2025oko}:
\begin{align}
  |\TFD(\beta,\lambda)\lb &= e^{-\frac{\beta}{4}H}|B\lb  
  \label{bsta}
\end{align}
where the boundary state $|B\lb$ is explicitly given by (see appendix \ref{ap:janus}
for more details):
\begin{align}
  |B\lb=\exp&\left[\sum_{m=1}^\infty\frac{1}{m}\left[\cos2\theta(-\alpha^{(1)}_{-m}\tilde{\alpha}^{(1)}_{-m}+\alpha^{(2)}_{-m}\tilde{\alpha}^{(2)}_{-m})\right.\right.\nonumber\\
  &\left.\left.+\sin2\theta(\alpha^{(1)}_{-m}\tilde{\alpha}^{(2)}_{-m}+\tilde{\alpha}^{(1)}_{-m}\alpha^{(2)}_{-m})\right]\right]|\Omega\lb.  \label{bstb}
\end{align}
Here $|\Omega\lb$ is the vacuum of the Fock spaces spanned by the oscillators $\ap^{(i)}_n$ and $\ti{\ap}^{(i)}_n$ and includes the summation over the zero modes with the constraint $w_1+w_2=0$ and $n_1-n_2=0$, set by the conformal boundary condition. Thus we can write it as $|\Omega\lb=\sum_{n,w\in \mathbb{Z}}|n_1=n,w_1=w,n_2=n,w_2=-w\lb$.

By employing (\ref{bsta}) and (\ref{bstb}), we would like to compute the second Renyi pseudo entropy $S^{(2)}_A=-\log \mbox{Tr}\rho_A^2$ and imagitivity for the reduced density matrix $\rho_A$ for $A=$CFT$_{(1)}$. We can compute $\rho_A$ by tracing out the subsystem $B=$CFT$_{(2)}$ from the generalized density matrix (\ref{gened}).  After some algebras presented in appendix \ref{ap:janus}, we obtain
\begin{align}
    S_{A}^{(2)}=& -\log \left[ \frac{2 \sin 2\theta \qty[\eta\qty(\frac{\rmi\beta}{2\pi})]^3}{\vartheta_1(\frac{2\theta}{\pi},\frac{\rmi\beta}{2\pi})}\right.\nonumber\\ 
     &\times\left.\frac{\vartheta_3 \qty(0, \frac{\rmi\beta}{\pi}\qty(\frac{1}{R_1^2}+\frac{1}{R_2^2})) 
    \vartheta_3 \qty(0, \frac{\rmi\beta}{\pi}\frac{R_1^2+R_2^2}{4})}
    {\vartheta_3 \qty(0, \frac{\rmi\beta}{2\pi}\qty(\frac{1}{R_1^2}+\frac{1}{R_2^2}))^2 
    \vartheta_3 \qty(0, \frac{\rmi\beta}{2\pi}\frac{R_1^2+R_2^2}{4})^2}\right] ,  \label{stwojan}
\end{align}
where $\eta(\tau)$ is the eta function and $\theta_{1,3}(\nu,\tau)$ are the theta functions and we follow the convention of \cite{DiFrancesco:1997nk}.

When we consider the imaginary Janus deformation $\theta = \frac{\pi}{4}+\rmi\frac{\delta}{2}$, we have $R_1 = R \sqrt{\frac{1+\rmi\sinh \delta}{\cosh\delta}}=R^*_2$. Thus the Renyi pseudo entropy is expressed as
\begin{align}
  S_A^{(2)}=& -\sum_{n = 1}^{\infty} \log \Biggl[
  \frac{2\cosh\delta\cdot \eta\left(\frac{\rmi\beta}{2\pi}\right)^3}{\vartheta_2 
\left(\frac{i\delta}{\pi}, \frac{\rmi\beta}{2\pi}\right)}\nonumber\\ 
  &\quad\times \frac{\vartheta_3 \qty(0, \frac{2i\beta}{\pi R^2\cosh\delta}) \vartheta_3 \qty(0, \frac{i\beta R^2}{2\pi\cosh\delta})}
  {\vartheta_3 \qty(0, \frac{i\beta}{\pi R^2\cosh\delta})^2\vartheta_3 \qty(0, \frac{\rmi \beta R^2}{4\pi\cosh\delta})^2} \Biggr].\label{deltape}
\end{align}
When $\delta$ is large, by using the modular transformation of theta functions, we obtain the following behavior 
\ba
 S_A^{(2)}\simeq \frac{2\delta^2}{\beta}+\delta.
\ea

\begin{figure}[h]
  \centering
  \begin{minipage}{0.45\columnwidth}
     \centering
     \includegraphics[width=\columnwidth]{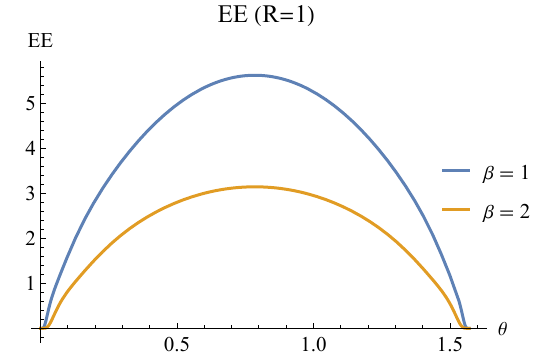}
  \end{minipage}
  \begin{minipage}{0.45\columnwidth}
     \centering
     \includegraphics[width=\columnwidth]{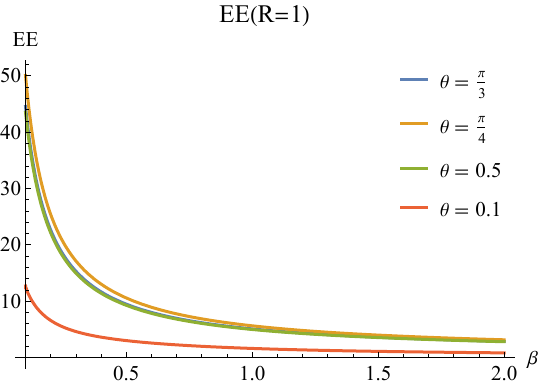}
  \end{minipage}
  \caption{The plots of Renyi entropy $S^{(2)}_A$ at $R=1$ for real values of $\theta$  as a function of $\theta$ (left) and of $\beta$ (right).} 
  \label{fig:Josi_rEE}
\end{figure}

\begin{figure}[h]
  \centering
  \begin{minipage}{0.45\columnwidth}
     \centering
     \includegraphics[width=\columnwidth]{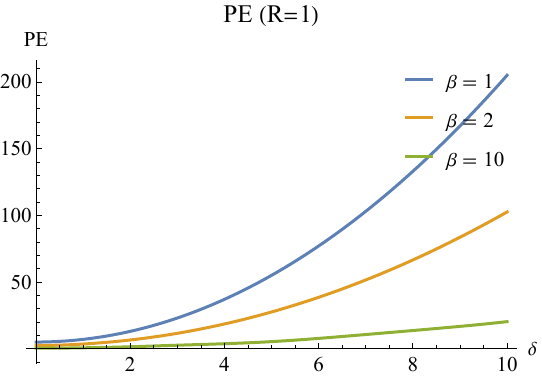}
  \end{minipage}
  \begin{minipage}{0.45\columnwidth}
     \centering
     \includegraphics[width=\columnwidth]{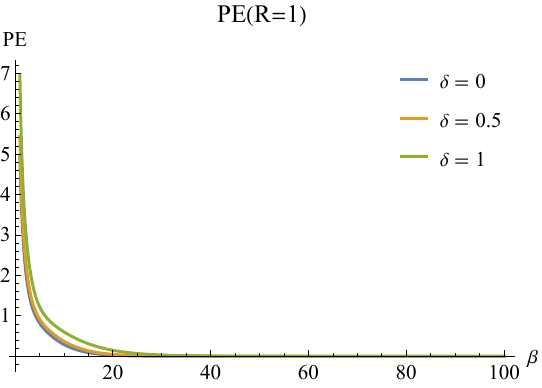}
  \end{minipage}
  \caption{The plots of Renyi pseudo entropy $S^{(2)}_A$ at $R=1$ for $\lambda=\rmi\frac{\delta}{2}$ as a function of $\delta$ (left) and $\beta$ (right).} 
  \label{fig:Josi_iEE}
\end{figure}

We plotted the Renyi entropy in Fig.\ref{fig:Josi_rEE} for real $\theta$. At $\theta=\pi/4$ (i.e. $\lambda=0$), which corresponds to the standard TFD state before the Janus deformation, the second Renyi entropy is maximized. This should be so because the original TFD state is the maximally entangled state when we fix the energy.  When the Janus deformation is imaginary as $\lambda=\frac{i}{2}\delta$ for a real valued parameter $\delta$, we plotted the Renyi pseudo entropy in Fig.\ref{fig:Josi_iEE}. Notice that as we showed generally in (\ref{gentrre}), the pseudo entropy takes real values in spite of the non-hermitian density matrix. Its shows that the pseudo entropy is a monotonically increasing function of $\delta$. Even though $\delta=0$ is the maximally entangled state, the pseudo entropy can go beyond this bound as is known in general \cite{Nakata:2021ubr,Ishiyama:2022odv}. In both cases, we also note that the pseudo entropy is monotonically decreasing function of $\beta$ as expected.

The normalized imagitivity is given by;
\begin{align}
  \frac{||\rho^\dagger_{A}-\rho_{A}||_2^2}{|\Tr [\rho_{A}^2]|} 
  &= \frac{2 \mathrm{Tr} [\rho_{A}^\dagger \rho_{A}]_{\text{non-zero}}-2\mathrm{Re Tr}[\rho_{A}^2]_{\text{non-zero}}}{|\Tr [\rho_{A}^2]|_{\text{non-zero}}}
\end{align}
where, the non-zero mode contributions are computed as
\begin{align}
    &\qty[\mathrm{Tr}\rho_A^2]_{\text{non-zero}}
    = \frac{2 \sin 2\theta \qty[\eta(\frac{\rmi\beta}{2\pi})]^3}{\vartheta_1(\frac{2\theta}{\pi},\frac{\rmi\beta}{2\pi})} \\
    &\qty[\text{Tr}\rho_A\rho_A^\dagger]_{\text{non-zero}} \nonumber\\
    &= \prod_{n=1}^\infty \frac{(1-e^{-m\beta})^2}{\left(1- e^{-m\beta} ((\cos 2\theta + \cos 2\bar{\theta})^2-2) +e^{-2m\beta}\right)}. \label{J_trrd}
\end{align}
Here we have only to consider non-zero mode contributions
because the zero mode contribution to $\qty[\text{Tr}\rho_A\rho_A^\dagger]$ is equal to that to $\qty[\mathrm{Tr}\rho_A^2]$ and this is real valued.

\begin{figure}[h]
\centering
    \begin{minipage}{0.45\columnwidth}
        \centering
        \includegraphics[width=\columnwidth]{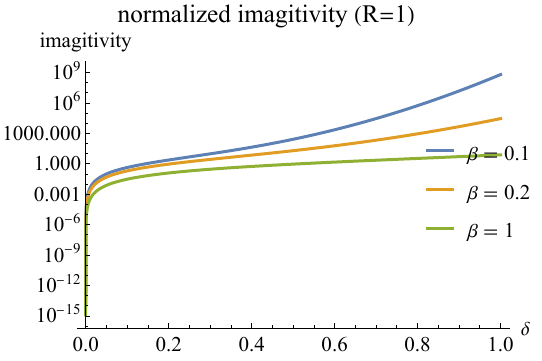}
    \end{minipage}
    \begin{minipage}{0.45\columnwidth}
        \centering
        \includegraphics[width=\columnwidth]{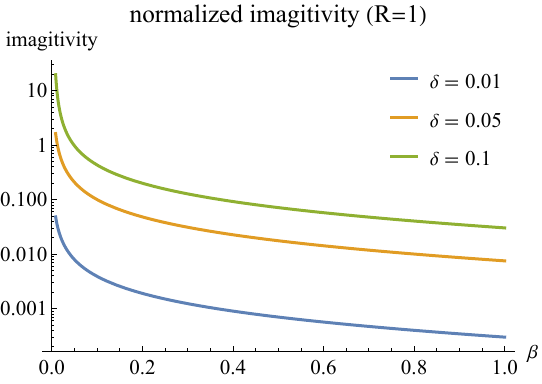}
    \end{minipage}
    \caption{The plot of the normalized imagitivity at $R=1$ as a function of $\delta$ (left) and $\beta$ (right).} 
    \label{fig:J_osi_imag_log}
\end{figure}

We find that the imagitivity is vanishing when $\theta$ is real. For the imaginary deformation $\theta=\frac{\pi}{4}+\rmi\frac{\delta}{2}$, (\ref{J_trrd}) can be represented by the elliptic functions:
\begin{align}
    \qty[\text{Tr} \rho_A\rho_A^\dagger]_{\text{non-zero}}
    =\frac{2 \qty[\eta (\frac{\rmi\beta}{2\pi})]^3} {\vartheta_2 \left(0,\frac{\rmi\beta}{2\pi}\right)}.
\end{align}
This leads to the normalized imagitivity plotted in Fig.\ref{fig:J_osi_imag_log}.
 The imagitivity is a rapidly increasing function of $\delta$. It decreases monotonically as $\beta$ gets larger.


\begin{figure}[h]
\centering
    \begin{minipage}{0.45\columnwidth}
        \centering
        \includegraphics[width=\columnwidth]{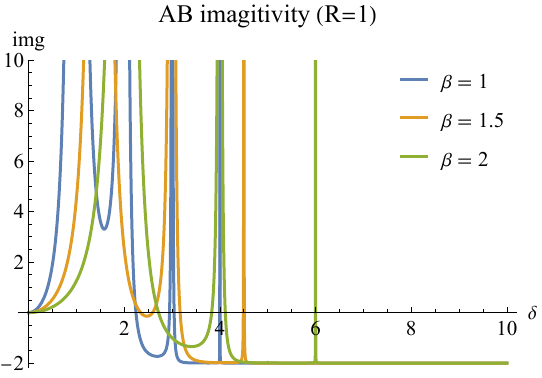}
    \end{minipage}
    \begin{minipage}{0.45\columnwidth}
        \centering
        \includegraphics[width=\columnwidth]{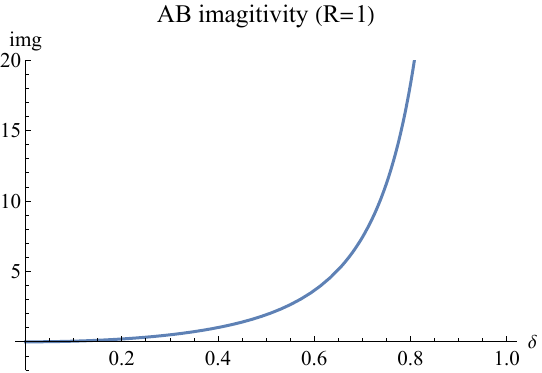}
    \end{minipage}
    \caption{The plots of the imagitivity as a function of $\delta$ at $R=1$. In the right panel we took $\beta=1$ and focused on the small $\delta$ region. }
    \label{fig:Josi_Abs}
\end{figure}

Finally, we consider the full generalized density matrix $\rho_{AB}$.  
Note that the second Renyi pseudo entropy simply vanishes $S_{AB}^{(2)}=0$ as our TFD state is a pure state. The normalized imagitivity is zero when $\theta$ is real. However,  in the presence of imaginary Janus deformation $\theta=\frac{\pi}{4}+\rmi\frac{\delta}{2}$, we obtain
\begin{align}
  &||\rho^\dagger_{AB}-\rho_{AB}||^2 \nonumber\\
  &= 2 \qty[\prod_{m=1}^\infty \qty(\frac{1-2 e^{-m\beta} +e^{-2m\beta}} {1-2 e^{-m\beta} \cosh\delta + e^{-2m\beta}})^2 -1]\label{J_img}\\
  &= 2 \qty[\frac{4 \sin^2 \qty(\frac{\delta}{2}i) \  \qty[\eta\qty(\frac{\rmi\beta}{2\pi})]^6}{\vartheta_2(\frac{\delta}{2\pi}i, \frac{\rmi\beta}{2\pi})^2} -1].
\end{align}
For $\delta<\beta$, the imagitivity is a monotonically increasing function of $\delta$ as depicted in (\ref{fig:Josi_Abs}). However, the imagitivity gets divergent at $\delta=n\beta$ for any positive integer $n$ and even becomes negative for large enough $\delta$ as depicted in (\ref{fig:Josi_Abs}). This implies that our non-hermitian system gets ill-defined for  $\delta<\beta$.
In this sense, it may be appropriate for us to regard our description of a non-hermitian system as the low energy effective description which is valid only for energy below a certain scale.

\subsection{Traversable wormholes from imaginary Janus deformation} \label{janholo}

In this final part of this paper, we will study the three dimensional traversable AdS wormhole solution, which is equivalent to strong coupling limit of imaginary Janus CFTs in two dimensions via the AdS$/$CFT correspondence.

The gravity solution dual to the Janus deformation of a two dimensional holographic CFT was given  in \cite{Bak:2007jm}
as a solution of Einstein equation with a massless scalar (dilation). In particular, we are interested in the finite temperature solution as we have in mind the TFD state. The corresponding solution is given by 
\begin{align}
  ds^2 &= dy^2 + \frac{r_0^2 f(y)}{\cosh^2 r_0 t} \left( - d t^2 + d \theta^2 \right) \label{jansol1} \\
  f(y) &= \frac{1}{2} \left( 1 + \sqrt{1 - 2 \gamma^2} \cosh 2 y \right) \label{jansol2} \\
  \varphi(y) &= \varphi_0 + \frac{1}{\sqrt{2}} \log \left[ \frac{1 + \sqrt{1 - 2 \gamma^2} + \sqrt{2} \gamma \tanh y}{1 + \sqrt{1 - 2 \gamma^2} - \sqrt{2} \gamma \tanh y} \right]  \label{JanusBHS}
\end{align}
where $\gamma$ is a parameter of Janus deformation. 
We expect $\gamma$ is monotonically related to the Janus deformation parameter $\lambda$ in (\ref{Januscftd}).
The metric reduces to that of the eternal BTZ black hole when $\gamma = 0$. For $0\leq |\gamma|\leq \frac{1}{\s{2}}$, the solution (\ref{JanusBHS}) describe the Janus deformed black hole solution, whose properties have been studied in \cite{Bak:2007qw}. The two asymptotic boundaries $y\to \pm\infty$ are not causally connected and thus the wormhole realized on each time slices in this geometry is not traversable.  
Its black hole entropy can be found from the horizon area 
at $y=t=0$ as
\ba
S_{\text{horizon}}=\frac{\pi r_0}{2G_N}\s{\frac{1+\s{1-2\gamma^2}}{2}}.
\label{SBHj}
\ea
\begin{figure}[!ht]
\centering
\resizebox{0.15\textwidth}{!}{
\begin{circuitikz}
\tikzstyle{every node}=[font=\LARGE]
\draw(6.5,9.5)node[above]{\huge{$\gamma^2<0\;(\chi>1)$}};
\draw [short] (8.75,9) -- (8.75,2.25);
\draw [short] (4.25,9) -- (4.25,2.25);
\draw[domain=4.25:8.75,samples=100,smooth] plot (\x,{0.1*sin(10*\x r -10.75 r ) +9});
\draw [dashed] (4.25,4.25) -- (8.75,9);
\draw [dashed] (4.25,9) -- (8.75,4.25);
\draw [dashed] (4.25,2.25) -- (8.75,7);
\draw [dashed] (4.25,7) -- (8.75,2.25);

\draw[domain=4.25:8.75,samples=100,smooth] plot (\x,{0.1*sin(10*\x r -10.75 r ) +2.25});
\draw(3,9.1)node[left]{\huge{$t$}};
\draw [short,->] (3,2.25) -- (3,9);

\draw(9,1)node[below]{\huge{$y$}};
\draw [short,->] (3.8,1.3) -- (9,1.3);

\draw[thick,dotted](4.25,5.625)--(8.75,5.625);
\fill[red] (6.5,5.625) circle (8pt); 
\end{circuitikz}}
\caption{Penrose diagram of the traversable wormhole via imaginary Janus deformation. The red point is a cross section whose are gives horizon entropy \eqref{SBHj}.}
\label{fig:my_label}
\end{figure}
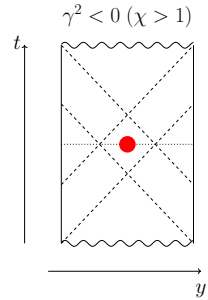
On the other hand, as first noted in \cite{Kawamoto:2025oko}, if we choose $\gamma$ to take imaginary values, we can connect the two asymptotic boundaries by a null geodesic and its wormhole becomes traversable. Note that in this case, although the dilaton takes imaginary value, the metric remains real-valued everywhere in this spacetime. In the D1-D5 CFT example \cite{Bak:2007jm}, the massless dilaton is dual to the exactly marginal perturbation which shifts the compactification  radius of the scalar fields in the CFT. Thus we can regard this gravity dual describes the strongly coupled limit of the example studied in section \ref{sec:imjan}. Notice that in our gravity solution argument, we essentially take an analytical continuation of $\gamma$ to imaginary values and this perfectly matches with our prescription of non-hermitian systems presented in section  
and with our general description of TFD states in section \ref{nonhtfd}.  We would also like to mention that the earlier work \cite{Garcia-Garcia:2020ttf} gave a construction of Euclidean wormhole due to non-hermitian interactions, though this mechanism is different from ours. Refer also to \cite{Arean:2019pom,Xian:2023zgu,Kanda:2023jyi,Kanda:2023zse,Suzuki:2025wlc} for another different approach to a holography for non-hermitian systems. 

At first, it looks surprising that one of the two CFTs is causally influenced by the other even though there are no interactions between the two CFTs as is clear from the deformed action (\ref{Januscftd}). However, this only occurs when the imaginary part of $\lambda$ (or equally $\gamma$) is non-vanishing. In section \ref{sec:infnonh}, we gave an argument using modified conjugation peculiar to non-hermitian systems which shows how the influences occur even without interactions, as manifested in (\ref{infja}). If we identify $\rho^{(0)}$ with (\ref{gened}) in our Janus CFT, then this argument directly explains the presence of the influence.

The holographic (thermal) pseudo entropy which measures the entanglement between CFT$_{(1)}$ and CFT$_{(2)}$ is given by the area of extremal surface (\ref{extfor}). Thus it is still given by (\ref{SBHj}). When $\gamma$ is imaginary, this entropy is a monotonically increasing function of $|\gamma|$, even larger than the BTZ black hole entropy $\gamma=0$. This amplification of entropy qualitatively agrees with our results of pseudo entropy (\ref{stwojan}) in the free scalar Janus CFT, plotted in Fig.\ref{fig:Josi_iEE}. 

\subsubsection{Single sided pseudo entropy}

Now let us examine holographic pseudo entropy when we choose subsystems in CFT$_{(1)}$ and CFT$_{(2)}$. 
First we choose a subsystem A in CFT$_{(1)}$ to be the interval $\theta \in [-\theta_\infty, \theta_\infty]$ at time $t = t_{\infty}$. 
We compute the pseudo entropy $S_A$, which is given by an area of extremal surface that is anchored to the edge of the subsystem  (\ref{extfor}) \cite{Nakata:2021ubr}.
Here we will show the results for a large subsystem $(\theta_\infty \gg r_0^{-1})$ case, at early time $(t_\infty \ll \theta_\infty)$ and late time $(t_\infty \gg \theta_\infty)$ limit. 
The details of the calculation and exact form of the area are given in Appendix \ref{janusent} (see also \cite{Nakaguchi:2014eiu}). 
In early time limit, after removing the UV divergence, the area, which is the geodesic length in our three dimensional geometry, is given by
\begin{equation}
    \begin{aligned}
          A(\Gamma_A) \underset{(t_\infty \ll \theta_\infty)}{\simeq} 2 \kappa_+ r_0 \theta_{\infty} + 2 (1 - \kappa_{+}) \log \cosh r_0 t_{\infty} \\ 
          - 2 \log \left[ \frac{\kappa_+ + \sqrt{\kappa_+^2 - \kappa_-^2}}{2} r_0 \right], \label{earlylim}
    \end{aligned}
\end{equation}
where we defined $ \kappa_{\pm} := \sqrt{\frac{1 \pm \sqrt{1 - 2 \gamma^2}}{2}}$. 
On the other hand, in late time limit, the area becomes constant: 
\begin{equation}
  A \underset{(t_\infty \gg \theta_\infty)}{\simeq} 2 (r_0 \theta_{\infty} - \log r_0). \label{latelim}
\end{equation}
In particular, since this value does not depend on the deformation parameter $\gamma$, it coincides with the entropy of BTZ black hole. 

\begin{figure}[t]
  \centering
  \includegraphics[width=0.8\columnwidth]{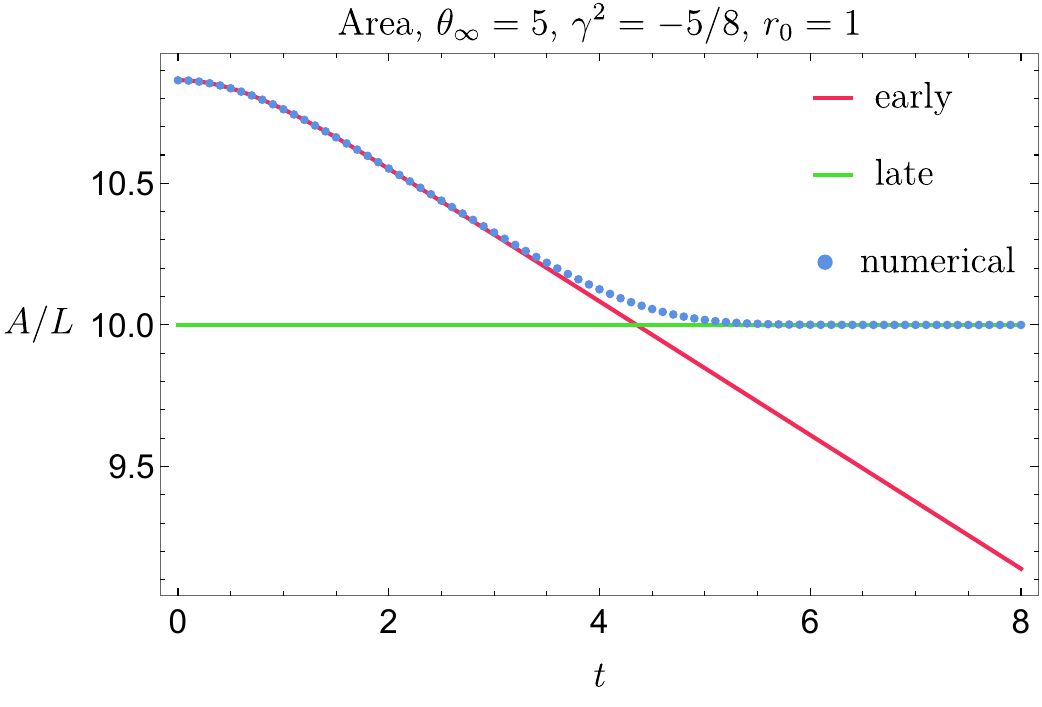}
  \caption{The time evolution of extremal surface area in imaginary Janus deformed spacetime. The red line: early time approximation, green line: late time approximation, blue markers: exact solution. }
  \label{imjaen}
\end{figure}

In general, for $\frac{1}{r_0}\ll t_\infty \ll \theta_\infty$, we find the linear behavior $A(\Gamma_A)\simeq 2(1-\kappa_+)t_\infty+$const.
Note that before the deformation $\gamma=0$, there is no linear growth, which is because a BTZ black hole is static. For real values of $\gamma$ in the range $|\gamma|<\frac{1}{\s{2}}$, the area grows linearly. This is natural in that the initial thermal entropy is smaller than the maximal one for $\gamma>0$ as we already observed in (\ref{SBHj}) and therefore the entanglement entropy tends to increase to reach the common thermal value (\ref{latelim}) of the BTZ black hole.
This is a process that can be regarded as thermalization \cite{Bak:2007qw}. 
This may also look similar to the known linear growth of holographic entanglement entropy in \cite{Hartman:2013qma}, which is a model of global quantum quenches \cite{Calabrese:2005in}.

On the other hand, when $\gamma$ is imaginary valued, we find that $1 - \kappa_+$ is negative and thus the pseudo entropy $S_A$ gets linearly decreasing, as opposed to the usual Janus deformation. This reflects the fact that the initial pseudo entropy is larger than the maximal value as in  (\ref{SBHj}) for imaginary $\gamma$ and the entropy tends to decrease to reach the common value (\ref{latelim}). 
The time evolution of renormalized area is shown in Fig.\ref{imjaen}.

One may think this traversable wormhole result seems to violate the second law of thermodynamics. Indeed this is due to the obvious violation of the achronal averaged null energy condition \cite{Wall:2009wi, Wall:2010jtc} due to the imaginary valued dilaton, which makes the traversable wormhole possible. However, from the CFT viewpoint, we have now quantum information theoretic interpretation. Namely, the violation of the second law is not surprising as we are considering the pseudo entropy instead of standard entropy. For the pseudo entropy, due to the amplification effect \cite{Ishiyama:2022odv,Nakata:2021ubr}, we often observe exotic behaviors as we stressed several times before.

\subsubsection{Double sided pseudo entropy}

Finally we would like to analyze the pseudo entropy for a subsystem $A$ which spans the two CFTs, so that $A=A_1\cup A_2$. It is useful to introduce another time coordinate $\tau$ by $\cos\tau=\frac{1}{\cosh r_0 t}$, where $\tau$ takes the values in the range $-\frac{\pi}{2}<\tau<\frac{\pi}{2}$. We choose $A_1$ to the region $\theta>0$ at $\tau=-\tau_\infty$ in the CFT$_{(1)}$, while $A_2$ is the region $\theta>0$ at $\tau=\tau_\infty$ in the CFT$_{(2)}$.

Consider the geodesic $\tau=\tau(y)$ at $\theta=0$ in the imaginary Janus solution (\ref{JanusBHS}). The geodesic extends in the time interval $-\tau_\infty<\tau<\tau_\infty$ and $-y_{\infty}<y<y_{\infty}$. This gives the holographic pseudo entropy $S_{A_1A_2}$. After some algebras, we obtain
\ba
\tau_\infty &=& \int^{y_{\infty}}_0 \frac{\s{C} dy}{f(y)\s{C+f(y)}}\no
&=&\frac{2\s{C}}{\s{(\chi-1)(1+2C+\chi)}}\no
&& \times F\left[\arcsin\s{\f{\chi-1}{2\chi}},\frac{4C\chi}{(2C+1+\chi)(\chi-1)}\right],\no \label{geoqwe}
\ea
where $C$ is an integration constant and we introduced the parameter $\chi=\s{1-2\gamma^2}$.
In general we find in the limit $C\to\infty$ and $C\to 0$:
\ba
&&\tau_\infty\simeq\tau_{0}-\s{\frac{1}{C}},\ \ \ (C\to \infty),\label{wedfwr} \\ 
&&\tau_\infty\to 0, \ \ \ (C\to 0), 
\ea
where $\tau_0=\frac{1}{\s{\chi}}K\left(\frac{\chi-1}{2\chi}\right)$.
Since $\tau_\infty$ is monotonically increasing function of $C$, we have $0<\tau_\infty<\tau_0$.
When $\chi=\s{1-2\gamma^2}<1$ (i.e. $\gamma$ is real), we find $\tau_0>\frac{\pi}{2}$. Therefore $\tau_\infty$ can cover all time period.

However , when $\chi=\s{1-2\gamma^2}>1$ (i.e. $\gamma$ is imaginary), we find $\tau_0<\frac{\pi}{2}$. In this imaginary Janus case, the geodesic becomes null at $C=\infty$. We cannot find a geodesic which has $\tau_0<\tau_\infty<\frac{\pi}{2}$ from the above calculations. We expect the actual geodesic which connects $(-\tau_\infty,-y_\infty)$ to $(\tau_\infty,y_\infty)$ does not pass through the origin $(0,0)$, such that it extends in the complex direction. Therefore we cannot use the formula (\ref{geoqwe}). 

However, we can still perform an analytical continuation of the geodesic length and trajectory of geodesic. The geodesic length $L$ is originally given by 
\ba
A(\Gamma_A)=\int^{y_{\infty}}_0 \frac{dy}{\s{1+\frac{C}{f(y)}}}.
\ea
Since we are interested in the geodesic which is close the null geodesic, we assume $C\gg 1$.  The integral is suppressed around $|y|\leq \frac{1}{2}\log C$. Therefore we can estimate 
\ba
L\simeq 2\left(y_{\infty}-\log \s{C}\right)+\mbox{const}.
\ea
On the other hand we find from (\ref{wedfwr}) the behavior 
$\frac{1}{\s{C}}\simeq \tau_0-\tau_\infty$. Therefore we find the holographic PE
\ba
S_A=\frac{c}{3}\log \frac{1}{\ep\s{C}}\simeq \frac{c}{3}\log \frac{\tau_0-\tau_\infty}{\ep}, \label{emagrpngqkf}
\ea
where $\ep$ is the UV cut off of the CFT and is related to $y_\infty$ as 
$\ep=e^{-y_\infty}$. Though the above estimation (\ref{emagrpngqkf}) is obtained by assuming $\tau_0-\tau_\infty>0$, we expect it is still true for the other side. Then we find for $\tau_0-\tau_\infty<0$
\ba
S_A=\frac{c}{3}\log \frac{1}{\ep\s{C}}\simeq \frac{c}{3}\log \frac{\tau_\infty-\tau_0}{\ep}+\frac{\pi c}{3}\rmi, \label{emagrpngqkfaa}
\ea
The imaginary part implies that the geodesic consists of a union of the time-like part and space-like one \cite{Doi:2022iyj,Doi:2023zaf}, or extends into complex valued coordinate directions  \cite{Heller:2024whi,Heller:2025kvp,Nunez:2025ppd,Fujiki:2025rtx}.
as in the time-like entanglement entropy. Indeed the presence of the imaginary entropy which is an integer multiple of $\frac{\pi c}{6}\rmi$ is characteristic feature of the time-like entanglement entropy in two dimensional CFTs.
This complex valued pseudo entropy manifestly shows that the CFT dual of the generalize density matrix is not hermitian and this is directly related to the fact that the two asymptotics are causally connected as the bulk is traversable wormhole.

\section{Conclusions} \label{sec:concl}

In this paper, we provided extensive studies of non-hermitian generalized density matrices in diverse quantum systems. We fully employ the time-like entanglement entropy \cite{Doi:2022iyj,Doi:2023zaf} and imagitivity \cite{Milekhin:2025ycm} as useful probes of generalized density matrices to explore physical properties. It would be also intriguing to employ other probes such as the SVD entropy \cite{Parzygnat:2023avh}, which is left for future work. At the same time it would be an important future problem to better understand precise quantum information theoretic meanings of these quantities, such as the imaginary part of the time-like entanglement entropy.

In the first part of this paper, we classified the possible setups where we encounter non-hermitian density matrices, into two classes: class 1 and class 2. The class 1 occurs when we consider time-like entanglement in unitary quantum systems. When we consider a reduced density matrix $\rho_A$ for a subsystem $A$, the time-like entanglement implies the causal influences between two points inside $A$. This effect is equivalent to the non-hermitian properties of $\rho_A$, which can be probed by the imaginary part of time-like entanglement entropy and the no-vanishing imagitivity. A unitary operation at a point influences the expectation value of an operator at another point if $\rho_A$ is not hermitian. We presented a simple model of time-like entanglement in class 1 by considering coupled harmonic oscillators. We computed the second Renyi entropy and imagitivity and showed that they show expected behaviors. 

As a field theoretic example, we also analyzed these quantities in the two dimensional free Dirac fermion and holographic CFT for a subsystem $AB$ consists of equal size intervals $A$ and $B$, which are causally connected. We worked out the full calculations for the case where the two intervals are separated in the Euclidean time direction. In both the free and holographic CFT, the imagitivity gets monotonically increasing as $A$ and $B$ get closer. The difference between them is that in the latter case there is a phase transition when the distance between the intervals become smaller than a specific value. The imagitivity does vanish in the leading order of large $c$ limit until the distance reaches the critical value in the holographic CFT. On the other hand, in the free fermion CFT, 
the imagitivity is always positive.

The Renyi pseudo entropy in these CFTs is monotonically decreasing as the separation gets smaller, which makes the mutual information monotonically increasing. When $A$ and $B$ are separated in the Lorentzian time direction, due to complications of the Wick rotating the Euclidean result, we obtain a full result only for the second Renyi entropy. Though this entropy is mostly a increasing function of the time separation, we found a crucial difference around the time when the subsystem size coincides with the time separation. In the holographic CFT, we find that the entropy gets negatively divergent, while in the free fermion CFT it becomes positively divergent. 
This corresponds to the time when an end point of the interval is null separated by an end point of another interval. From the viewpoint of holographic pseudo entropy, this is natural as the geodesic which connects these boundary points become light-like. The analysis of Lorentzian time evolution of imagitivity seems to require a careful treatment of phase transitions, which we leave for a future problem.

The time-like entanglement and non-hermitian density matrix also arise when we perform a post-selection, where the bra state is different from the ket state. As such an example, we computed the second Renyi pseudo entropy and imagitivity for an interval subsystem $A$ in a two dimensional free CFT where the bra and ket state are generated by local excitations at different locations. We found that the results perfectly agree with a quasi-particle picture.
In particular, the imagitivity detects the non-hermitian property only when the local excitation can influence the subsystem $A$.  

For the class 2, we explained how we can construct generalized density matrices in non-hermitian systems by introducing the modified conjugation operation and how we compute physical quantities. From this formulation, we showed the presence of influences between two quantum systems even without interactions between them. As an explicit example, we studied a non-hermitian deformation of a thermofield double (TFD) state, called imaginary Janus deformation. After we give a general prescription, we showed that the pseudo entropy which describes the entanglement between two quantum systems remain real valued even if the Hamiltonian is deformed to be non-hermitian. Then we analyzed an example of the Janus deformation of a pair of free scalar CFTs and computed the second Renyi entropy and imagitivity. We obtained analytical expressions of them. This shows that the Renyi pseudo entropy between the two CFTs grows as the non-hermitian deformation is increased. It gets larger than the entropy for the undeformed TFD state. Since the TFD state is expected to be maximally entangled when we fix the temperature, one may think this is surprising. However, this kind of amplification phenomena \cite{Ishiyama:2022odv,Nakata:2021ubr} was already known for pseudo entropy. We observed that the imagitivity is also monotonically increasing as the system gets more non-hermitian as expected. The real-valuedness of the entropy and partition function (\ref{tfdpa}) may suggest that our non-hermitian system can be stable for a certain parameter region. In the dual gravity dual, we expect that the partition function should be computed by assuming the scalar field fluctuation should be real valued, in spite that the expectation value of scalar field gets complex valued. It would be intriguing to explore this in a future work. 

Finally, we examined the gravity dual of this imaginary Janus deformation, which is given by a traversable AdS wormhole solution \cite{Kawamoto:2025oko}. We evaluated the holographic pseudo entropy and showed the results qualitatively agree with the ones from the dual CFT. The horizon area calculation confirms the mentioned amplification phenomena of pseudo entropy. We also computed the holographic pseudo entropy $S_A$ when we choose the subsystem $A$ to be one of the two CFTs. Interestingly, we find that it gets linearly decreasing under the time evolution in the presence of non-hermitian deformation as opposed to the ordinary intuition of the second law of thermodynamics. In this way, our traversable AdS wormhole provides a useful example of class 2 systems with its CFT dual, where we can find essential features of non-hermitian quantum system. 
Therefore further studies and applications of this holographic setup will be an interesting future direction. At the same time, it would be also important to explore other examples in different holographic contexts from the viewpoint of generalized density matrices. In particular, this includes the dS/CFT correspondence \cite{Strominger:2001pn}, where non-hermitian CFTs are expected to play important roles \cite{Anninos:2011ui,Hikida:2021ese,Hikida:2022ltr}.

\vspace{5mm}
{\it Acknowledgments:}
We are grateful to Dongsu Bak, Michal Heller, Rene Meyer, Rob Myers, Erik Tonni, Okuto Morikawa for valuable discussions. We are grateful to the workshops 
"Quantum Gravity and Information in Expanding Universe" (YITP-W-24-19) and "Extreme Universe 2025"  (YITP-T-25-01), both held at YITP, Kyoto U., where parts of this work were developed and presented.
This work is supported by MEXT KAKENHI Grant-in-Aid for Transformative Research Areas (A) through the ``Extreme Universe'' collaboration: Grant Number 21H05187. TT is also supported by JSPS Grant-in-Aid for Scientific Research (B) No.~25K01000. 
TK is supported by Grant-in-Aid for JSPS Fellows No.~25KJ1455.
Part of the numerical computations in this work were performed on Yukawa-21 at Yukawa Institute for Theoretical Physics, Kyoto University.

\appendix

\section{Details of the calculation for 2d CFTs}\label{app:2cftdets}
\subsection{Elliptic functions}
First we define two useful functions the the \emph{Weierstrass elliptic function} and \emph{Weierstrass zeta function}. Our conventions follow from from \cite{Sansone1969-fm}.

First we consider the \emph{Weierstrass elliptic function} $\wp(z,\tau)$ which is the solution to the differential equation
\be
\begin{split}
\wp'(z)^2&=4\wp^3(z)-g_2\wp(z)-g_3\\
&=4(\wp(z)-e_1)(\wp(z)-e_2)(\wp(z)-e_3)
\end{split}
\ee
It is doubly periodic with respect to a lattice $\Lambda$ with periods $2\omega_1$ and $2\omega_3$:
\be
\begin{split}
\wp(z+2\omega_1)=\wp(z)\\
\wp(z+2\omega_3)=\wp(z)
\end{split}
\ee
the modular parameter is given by $\tau=\frac{\omega_3}{\omega_1}$.

At the half periods $\omega_{i}$ the Weierstrass elliptic function takes the special values
\be
\wp(0)=\infty,\quad \wp\left(\omega_1\right)=e_1, \quad \wp\left(\omega_3\right)=e_3, \quad \wp\left(\omega_2\right)=e_2
\ee
where $\omega_2=\omega_1+\omega_3$.
The elliptical invariants $g_2$ and $g_3$ are then given by
\be\label{eq:g_e}
g_2=-4(e_1e_2+e_2e_3+e_1e_3),\quad g_3=4e_1e_2e_3,\quad e_1+e_2+e_3=0.
\ee
The Weierstrass elliptic function obeys the following addition theorem
\be
\wp(a+b)+\wp(a)+\wp(b)=\frac{1}{4}\left(\frac{\wp'(a)-\wp'(b)}{\wp(a)-\wp(b)}\right)^2.
\ee
We will also make use of the \emph{Weierstrass zeta function} $\zeta(z,\tau)$ which is defined by
\be
\zeta'(z)=-\wp(z)
\ee
A consequence is that $\zeta(z)$ is quasi-periodic
\be
\zeta(z+2\omega_i)=\zeta(z)+2\eta_i
\ee
where $\eta_i$ are called the half quasi-periods. We then have that
\be
\zeta(\omega_i)=\eta_i.
\ee
The half periods and half quasi-periods are related by
\be
\eta_a\omega_b-\eta_b\omega_a=\frac{1}{2}\pi i, \quad a<b
\ee
The addition theorem is given by
\be
\zeta(a+b)-\zeta(a)-\zeta(b)=\frac{1}{2}\frac{\wp'(a)-\wp'(b)}{\wp(a)-\wp(b)}
\ee
this can also be put in the form
\be
\zeta(z-a)+\zeta(z+a)-2\zeta(z)=\frac{\wp'(z)}{\wp(z)-\wp(a)}.
\ee
\subsection{Calculation via the Liouville action}
Given the partition function of a 2d CFT on a Riemann surface $\Sigma$ it is well understood how this quantity changes under a conformal transformation $\Gamma(z)$. There is a conformal anomaly which is captured by the Liouville action $S_L$
\be
\begin{split}
e^{S_L}\frac{Z_{\Sigma}}{Z^s_{\mathbb{CP}}},\quad S_L&=\frac{c}{96\pi}\int_{\Sigma}d^2z\sqrt{g}\left[\partial_\mu\phi\partial_\nu\phi g^{\mu\nu}+2R\phi\right]\\
\phi(z)&=2\log\left(|\partial_z\Gamma(z)|\right).
\end{split}
\ee
here $R$ is the Ricci scalar of $\Sigma$ and $Z_\mathbb{CP}$ is the partition function of the Riemann sphere. Because the Liouville action  diverges one has to be careful to correctly regulate it. In particular those values of $z$ for which $\phi(z)$ diverges e.g. the critical points $\partial \Gamma(z)=0$ and poles $\Gamma(z)=\infty$  play an important role in its evaluation. We consider the expansion around each of these types of points:
\be
\Gamma(z)\approx x_i+a_i(z-z_i)^{\omega_i}, \quad
\Gamma(z)\approx b_j(z-z_j)^{-q_j}
\ee

For the calculation we will consider we have that $\Sigma$ is a torus with genus $g=1$ and around all such points $\omega_i=2$ and $q_j=1$. With these specifications one can show that the final form of $\log(e^{S_L}\frac{Z_{\Sigma}}{Z^s_{\mathbb{CP}}})$ taking into account the regularization is given by \cite{Lunin:2000yv,2010PhDT.......134A}
\be\label{eq:mastertorus}
\begin{split}
&-\frac{c}{12}\Biggr\{
\frac{1}{2}\log\left(\prod_i |a_i|\right)+2\log \left(\prod_j|b_j|\right)+12\log(2)
\Biggr\}\\
&+\log(Z_{\text{torus}}).
\end{split}
\ee

\subsection{Calculation of $\Tr(\rho_{AB}^{\dagger}\rho_{AB})$}
Following \cite{Numasawa:2016emc,2019JHEP...09..018C} we start with
\be
K(z)=\partial_{z'} \log{\theta_{1}(z',is)}|_{z'=z}
\ee
where the Jacobi theta function is defined as
\be
\begin{split}
&\theta_{1}(\nu,\tau)=2e^{\pi i \frac{\tau}{4}}\sin(\pi\nu)*\\&\prod_{k=1}^{\infty}(1-e^{2\pi i k \tau})(1-e^{2\pi i \nu}e^{2\pi i k \tau})(1-e^{-2\pi i \nu}e^{2\pi i k \tau}).
\end{split}
\ee
and then make use of the identity
\be
\partial_z\left(\log{\theta_{1}(z)}\right)=2\omega_1\zeta(2\omega_1 z)-4\eta_1\omega_1 z.
\ee
 The conformal transformation to consider is given by
\be
\begin{split}
&f(z)=A\left(\zeta(z)+\zeta\left(z+\omega_3\right)-4\eta_1 z+B\right)\\
&=A\left(2\zeta(z)+\frac{1}{2}\frac{\wp'(z)}{\wp(z)-e_3}-4\eta_1 z+\eta_3+B\right)\\
&=A\left(2\zeta(z+\frac{1}{2}\omega_3)+\frac{1}{2}\frac{\wp'(z+\frac{1}{2}\omega_3)}{\wp(z+\frac{1}{2}\omega_3)-\wp(\frac{1}{2}\omega_3)}-4\eta_1 z+B\right)
\end{split}
\ee
with $\omega_1=\frac{1}{2}$ , $\omega_3=\frac{is}{2}$ and $s$ real and positive. In the second and third lines we have used the addition theorem. Making the choice $f(\omega_1)=0$ sets $B=-\eta_3$.

To determine $A$ we further require that $f(\omega_1+\frac{1}{2}\omega_3)=q$. This is easiest to accomplish using the third line and $\wp'(\omega_1+\omega_3)=0$ from which we find
\be
\begin{split}
A(2(\eta_1+\eta_3)-4\eta_1(\omega_1+\frac{1}{2}\omega_3)-\eta_3)&=q\\
-A(\eta_1\omega_3-\eta_3\omega_1)&=q\omega_1\\
-i\pi A&=q,
\end{split}
\ee

Thus,
\be
f(z)=\frac{iq}{\pi}\left(\zeta(z)+\zeta\left(z+\omega_3\right)-4\eta_1 z-\eta_3\right).
\ee

At the critical points we have $\partial_zf(z)=0$ which is given by
\be
\begin{split}
-\wp(z)-\wp(z+\omega_3)-4\eta_1=0\\
\wp(z)+e_3+\frac{(e_1-e_3)(e_2-e_3)}{\wp(z)-e_3}+4\eta_1=0\\
\wp^2(z)+e_3\eta_1\wp(z)+(e_1-e_3)(e_2-e_3)-e_3^2-4e_3\eta_1=0
\end{split}
\ee
so that
\be\label{eq:wpalpha}
\begin{split}
\wp(z_i)&=-2\eta_1\pm \sqrt{4\eta_1(\eta_1+ e_3)+e_3^2-(e_1-e_3)(e_2-e_3)}\\
&=\alpha_{\pm}
\end{split}
\ee
There are four critical points of order two to account for. They have expansions
\be
\begin{split}
x_1&=q-ip+a_-(z-z_1)^2+\cdots\\
x_2&=q+ip+a_+(z-z_2)^2+\cdots\\
x_3&=-q+ip+a_+(z-z_3)^2+\cdots\\
x_4&=-q-ip+a_-(z-z_4)^2+\cdots\\
\end{split}
\ee
To determine $a_{\pm}$ we examine the second derivative
\be
\begin{split}
|\partial^2f(z)|=&\frac{q}{\pi}\wp'(z)\left(1-\frac{(e_1-e_3)(e_2-e_3)}{(\wp(z)-e_3)^2}\right)\\
=&\frac{2q}{\pi}\sqrt{(\wp(z)-e_1)(\wp(z)-e_2)(\wp(z)-e_3)}*\\ &\left(1-\frac{(e_1-e_3)(e_2-e_3)}{(\wp(z)-e_3)^2}\right)
\end{split}
\ee
Thus,
\be
|a_\pm|=\frac{1}{2}|\partial^2f(z)||_{\wp(z)=\alpha_\pm}.
\ee
The full contribution from all critical points is 
\be
 \frac{1}{2}\log\left( |a_{+}|^2|a_{-}|^2\right)
\ee
which after some algebra is given by
\be
\frac{1}{2}\!\log\left(\!\frac{16q^4}{\pi^4}|4\eta_1\!-\!e_3|^2\!|4\eta_1\!(e_3\!+\!\eta_1)\!+\!e_2e_3\!+\!e_1e_3\!-\!e_1e_2|^2\!\right).
\ee

The two poles are located at $z=0,\omega_3$ coming from the first order poles of the zeta functions. We have the expansions
\be
\begin{split}
&\frac{iq}{\pi}\frac{1}{z}\cdots\\
&\frac{iq}{\pi}\frac{1}{z-\omega_3}\cdots
\end{split}
\ee
and the contribution
\be
2\log\left(\frac{q^2}{\pi^2}\right)
\ee

So now gathering all the contributions we find
\be\label{eq:trrhorhodeta}
\begin{split}
&\log\left(\Tr(\rho_{AB}^{\dagger}\rho_{AB})\right)=
\\&\!-\!\frac{c}{12}\!\log\left(\!\frac{2^{14}q^6}{\pi^6}\!|4\eta_1\!-\!e_3||4\eta_1\!(e_3\!+\!\eta_1\!)\!+\!e_2e_3\!+\!e_1e_3\!-\!e_1e_2|\!\right)\\&+\log(Z_{\text{torus}}).
\end{split}
\ee

We will make use of the cross ratio
\be\label{eq:crossratioapp}
\eta=\frac{x_{12}x_{34}}{x_{13}x_{24}}=\frac{1}{1+\left(\frac{q}{p}\right)^2}
\ee
In order to determine $\frac{q}{p}$ we consider $f(z)=qw(z)$ then
\be
\begin{split}
x_3\!-\!x_4\!=\!2ip\!=\!q(w(z_3)\!-\!w(z_4))
\longrightarrow \frac{q}{p}\!=\!\frac{2i}{(w(z_3)-w(z_4))}
\end{split}
\ee
and using equation \eqref{eq:wpalpha} we have
\be
z_{3,4}(\tau)=\wp^{-1}(\alpha_{\pm}(\tau))
\ee
which allows us to directly relate $\frac{q}{p}$ and $\tau$.
We can use the constraint
\be
x_3+x_4=-2q=q(w(x_3)+w(z_4))
\ee
 to choose among the the possible values of $\wp^{-1}$ which are related by translations of the image by the lattice periods. This along with equation \eqref{eq:crossratioapp}  allows us to numerically determine the relationship between $\eta$ and $\tau$:
\begin{figure}[H]
   \centering
   \includegraphics[width=7cm]{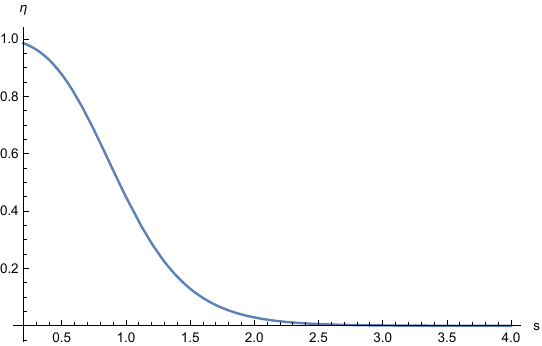}
   \caption{\label{fig:etataunumeric} $\eta(\tau)$ for $\tau=is$ pure imaginary. For holographic theories the phase transition in the partition function occurs when the torus is square $s=1$. This corresponds to when $\eta\approx.45$. }
   \label{fig:etatau}
\end{figure}

\section{Detailed analysis of harmonic oscillators}\label{ap:CHO}
Here we would like to present the detailed calculations of the coupled harmonic oscillators presented in section \ref{sec:harmoc}.

\subsection{Useful formulas}
First we summarize useful formulas. For the number state 
$|n\lb=\frac{1}{\s{n\!}}(a^\dagger)^n|0\lb$, we have 
\begin{align}
  \sum_{n=0}^\infty \ket{n} \bra{n}=\int \frac{dzd\bar{z}}{\pi}e^{-z\bar{z}}e^{za^\dagger}\ket{0}\bra{0}e^{\bar{z}a}, \label{sumz}
\end{align}
where $dzd\bar{z}=dxdy$ with $z=x+iy$. In the above we employed the gaussian integral
\begin{align}
  \int \frac{dzd\bar{z}}{\pi}e^{-z\bar{z}}z^n\bar{z}^m=\delta_{n,m} \int^\infty_0 2r^{2n+1}e^{-r^2}dr=\delta_{n,m}n!.
\end{align}

Moreover, we often use following formulae for the four independent creation and annihilation operators  $(a^\dagger,\tilde{a}^\dagger,b^\dagger,\tilde{b}^\dagger)$ and  
$(a,\tilde{a},b,\tilde{b})$
one is
\begin{align}
  & \bra{0}e^{Aa\tilde{a}}e^{p_La+p_R\tilde{a}}e^{q_La^\dagger+q_R\tilde{a}^\dagger}e^{Ba^\dagger \tilde{a}^\dagger}\ket{0}  \nonumber\\
  &=\frac{1}{1-AB}\cdot e^{\frac{1}{1-AB}(p_Lq_L+p_Rq_R+Bp_Lp_R+Aq_Lq_R)}.
\label{formu}
\end{align}

\subsection{Calculating pseudo entropy and imagitivity}

To calculate pseudo entropy and imagitivity, the following generating function is helpful: 
\begin{align}
&F(x,y,z,w)\equiv  \bra{0}e^{xa+yb}e^{-\rmi Ht}e^{za^\dagger+wb^\dagger}\ket{0}\nonumber \\
&=\sum_{m,n,p,q=0}^\infty \frac{1}{\sqrt{m!n!p!q!}}x^n y^m z^q w^p
\bra{n}_A\bra{m}_B e^{-\rmi Ht} \ket{q}_A \ket{p}_B,
\end{align}
where the Hamiltonian $H$ is given by (\ref{Hamosw}).
From this function $F$, we can read off $\bra{n}_A\bra{m}_B e^{-\rmi Ht}\ket{q}_A\ket{p}_B$ from its series expansion.

By the direct calculation using (\ref{formu}), we obtain
\begin{align}
F(x,y,z,w)=\frac{e^{A(xy+zw)+B(xz+yw)}}{\cosh^2\theta-\sinh^2\theta e^{-2\rmi T}},\label{Ffor}
\end{align}
where 
\ba
&& A=\frac{\tanh\theta(e^{-2\rmi T}-1)}{1-\tanh\theta^2 e^{-2\rmi T}},\no
&& B=\frac{e^{-iT}}{\cosh^2\theta-\sinh^2\theta e^{-2\rmi T}}.
\ea

Now we would like to calculate $\Tr \rho\rho^\dagger$. (We employ the simple notation: $\ket{n,m} = \ket{n}_A\ket{m}_B$, and $c=$cosh$\theta$, $s=$sinh$\theta$, $t=$tanh$\theta$.)
\begin{align}
  &\frac{F(x,y,z,w)F(x',y',z',w')}{c^4} \no
 & = \frac{1}{c^4} \sum_{mnpq,m'n'p'q'} \frac{1}{\sqrt{m!n!p!q!}} \frac{1}{\sqrt{m'!n'!p'!q'!}} \nonumber\\
 &\quad\times x^m y^n z^p w^q \bar{x}^{m'} \bar{y}^{n'} \bar{z}^{p'} \bar{w}^{q'} \nonumber\\
  &\quad\times \bra{n,m} e^{-\rmi Ht_1} \ket{q,p} \bra{n',m'} e^{-\rmi Ht_2} \ket{p',q'} \nonumber\\
  &= \frac{e^{A_1t(xy+zw)+B_1t(xz+yw)}}{c^2-s^2 e^{-2\rmi t_1}} \frac{e^{A_2(\bar{x}\bar{y}+\bar{z}\bar{w})+B_2(\bar{x}\bar{z}+\bar{y}\bar{w})}}{c^2-s^2e^{-2\rmi t_2}},
\end{align}
where
\begin{align}
  A_i = \frac{t(e^{-2\rmi t_i}-1)}{1-t^2e^{-2\rmi t_i}},\quad
  B_i = \frac{e^{-\rmi t_i}}{c^2-s^2 e^{-2\rmi t_i}}.
\end{align}
Then we replace as $y\rightarrow -yt,\ z\rightarrow -zt,\ \bar{y} \rightarrow -\bar{y}t,\ \bar{z} \rightarrow -\bar{z}t$, and perform the integration
\ba
\int\frac{dxd\bar{x}}{\pi} \frac{dyd\bar{y}}{\pi} \frac{dzd\bar{z}}{\pi} \frac{dwd\bar{w}}{\pi} e^{-(x\bar{z}+y\bar{w}+z\bar{x}+w\bar{y})},
\ea
 to both side, and use \eqref{Ffor}. This leas to 
\begin{align}
  & \sum_{mnpq} \frac{(-t)^{m+q+p+n}}{c^4} \bra{nm} e^{-iHt_1} \ket{qp} \bra{qp} e^{-iHt_2} \ket{nm} \nonumber\\
  & =\frac{1}{c^4(c^2-s^2 e^{-2\rmi t_1})(c^2-s^2 e^{-2\rmi t_2})} \nonumber\\ 
  &\quad\times \int\frac{dxd\bar{x}}{\pi} \frac{dyd\bar{y}}{\pi} \frac{dzd\bar{z}}{\pi} \frac{dwd\bar{w}}{\pi}
  e^{-(x\bar{z}+y\bar{w}+z\bar{x}+w\bar{y})}\nonumber\\ 
  &\quad\times
  e^{A_1t(xy+zw)+B_1t(xz+yw)} e^{A_2 (\bar{x}\bar{y} +\bar{z}\bar{w}) +B_2 (\bar{x}\bar{z} +\bar{y}\bar{w})} \nonumber\\
  &= -\frac{1}{c^4(c^2-s^2 e^{-2\rmi t_1}) (c^2-s^2 e^{-2\rmi t_2})} \times \nonumber\\ 
  & \frac{4\left(\text{sech}^2(\theta )+e^{2 i t_1}-1\right) \left(\text{sech}^2(\theta )+e^{2 i t_2}-1\right)}
   {\text{sech}^8(\theta ) \left(\sinh ^2(2 \theta ) \left(e^{i t_1}+e^{i t_2}\right)^2-4 \cosh ^2(2 \theta ) e^{2 i (t_1+t_2)}\right)},
\end{align}
where the last extra minus comes from $\sqrt{x^2} = -x$ when $x$ is negative. If we put $t_1 = t_2 = T$, we have
\begin{align}
  \Tr\rho_{AB}^2 = \frac{2 e^{2 i T}}{-\cosh (4 \theta )+e^{2 i T} (\cosh (4 \theta )+1)+1}.
\end{align}
This matches the previous result in \cite{Kawamoto:2025proceedings}. On the other hand, if we set $t_1 = -t_2 = T$, we obtain
\begin{align}
  \Tr\rho_{AB}\rho_{AB}^\dagger 
 \! =\! \frac{8 e^{2 i T}}{2 e^{2 i T} (\cos (2 T)+3)\!-\!\left(-1+e^{2 i T}\right)^2 \cosh (4 \theta)}.
\end{align}


\section{Detailed Analysis of Janus scalar CFT}
\label{ap:janus}
Here we present the detailed calculations of Janus deformed CFT which was fully employed in section \ref{sec:imjan}.

\subsection{Description of generalized density matrix}

The Janus deformation of $c=1$ free scalar CFT was described by the following quantum state (\ref{bsta}) (see also \cite{Kawamoto:2025oko}):
\begin{align}
\ket{\Psi} =e^{-\rmi Ht}e^{-\frac{\beta}{4}H}\ket{B},
\end{align}
up to the normalization factor. Here $\ket{B}$ is the boundary state. This is explicitly written as 
\begin{align}
&\ket{\Psi} \nonumber\\
&=\exp\left[\sum_{m=1}^\infty\frac{1}{m}e^{-2m\left(it+\frac{\beta}{4}\right)}
\left[\cos 2\theta(-\alpha^{(1)}_{-m}\tilde{\alpha}^{(1)}_{-m}+\alpha^{(2)}_{-m}\tilde{\alpha}^{(2)}_{-m})\right.\right. \nonumber\\ 
  &\quad\quad\quad \left.\left.
+\sin2\theta(\alpha^{(1)}_{-m}\tilde{\alpha}^{(2)}_{-m}+\alpha^{(2)}_{-m}\tilde{\alpha}^{(1)}_{-m})\right]\right]\ket{0}\nonumber\\
&\ \ \ \ \ \ \otimes\ e^{-\left(it+\frac{\beta}{4}\right)E_{n,w}}\sum_{n,w\in Z}\ket{n,w}_A\ket{n,-w}_B,
\end{align}
where $E_{n,w}$ is the energy of zero modes and is found to be
\begin{align}
E_{n,w}=\frac{1}{\sin 2\theta}\left(\frac{2n^2}{R^2}+\frac{w^2}{2}R^2\right).
\end{align}
We regard $(\alpha^{(1)}_m,\tilde{\alpha}^{(1)}_m)$ as the degrees of freedom in CFT$_1$ and $(\alpha^{(2)}_m,\tilde{\alpha}^{(2)}_m)$ as that of CFT$_{2}$. 
At $\theta=\frac{\pi}{4}$, the state $\ket{\Psi} $ is the maximally entangled state between the two CFTs. 
On the other hand at $\theta=0$, the state $\ket{\Psi} $ has no entanglement. 
The imaginary Janus deformation is described by setting $\theta=\frac{\pi}{4}+\rmi\frac{\delta}{2}$.

It is useful to introduce the normalized creation and annihilation operator as
\begin{align}
& a_m=\frac{1}{\sqrt{m}}\alpha^{(1)}_m, \ 
a^\dagger_m=\frac{1}{\sqrt{m}}\alpha^{(1)}_{-m},\nonumber\\
&\tilde{a}_m=\frac{1}{\sqrt{m}}\tilde{\alpha}^{(1)}_m, \ \tilde{a}^\dagger_m=\frac{1}{\sqrt{n}}\tilde{\alpha}^{(1)}_{-m},\nonumber\\
& b_m=\frac{1}{\sqrt{m}}\alpha^{(2)}_m,  \ 
b^\dagger_m=\frac{1}{\sqrt{m}}\alpha^{(2)}_{-m},\nonumber\\ 
&\tilde{b}_n=\frac{1}{\sqrt{m}}\tilde{\alpha}^{(2)}_m,
\ \tilde{b}^\dagger_m=\frac{1}{\sqrt{m}}\tilde{\alpha}^{(2)}_{-m}.  \label{relaq}
\end{align}
Then the quantum state looks like (we only show the non-zero mode parts):
\begin{align}
\ket{\Psi} &=\exp\left[\sum^\infty_{m=1}\lambda_m\cos2\theta(-a^\dagger_n \tilde{a}^\dagger_n+b^\dagger_n \tilde{b}^\dagger_n)\right.\nonumber\\ 
  &\quad\left. +\lambda_m\sin 2\theta
(a^\dagger_n \tilde{b}^\dagger_n+b^\dagger_n \tilde{a}^\dagger_n)\right]\ket{0},
\end{align}
where $\lambda_m=e^{-2m\left(it+\frac{\beta}{4}\right)}$.

\subsection{Normalization}

The analysis of non-zero modes can be simply decomposed into each $m$ sector.
Thus we calculate the Renyi entropy for the following state $\ket{\psi} $ in a coupled harmonic oscillators of $(a^\dagger,\tilde{a}^\dagger,b^\dagger,\tilde{b}^\dagger)$:
\begin{align}
\ket{\psi} &={\cal N}\exp\left[\lambda\cos2\theta(-a^\dagger \tilde{a}^\dagger+b^\dagger \tilde{b}^\dagger)\right.\nonumber\\ 
  &\quad\left.+\lambda\sin 2\theta
(a^\dagger \tilde{b}^\dagger+b^\dagger \tilde{a}^\dagger)\right]\ket{0},
\end{align}
where ${\cal N}$ is the normalization factor such that $\bra{\psi}\ket{\psi} =1$.
We can diagonalize this state by introducing 
\begin{align}
& \alpha=\cos\theta~ a-\sin\theta~ b,
\ \ \ \alpha^\dagger=\cos\theta~ a^\dagger-\sin\theta~ b^\dagger,\nonumber\\
& \tilde{\alpha}=\cos\theta~\tilde{a}-\sin\theta~\tilde{b},
\ \ \ \tilde{\alpha}^\dagger=\cos\theta~\tilde{a}^{\dagger}-\sin\theta~\tilde{b}^{\dagger},\nonumber\\
& \beta=\sin\theta~ a +\cos\theta~ b,
\ \ \ \beta^\dagger=\sin\theta~ a^{\dagger}+\cos\theta~ b^{\dagger},\nonumber\\
& \tilde{\beta}=\sin\theta~\tilde{a}+\cos\theta~\tilde{b},
\ \ \ \tilde{\beta}^\dagger= \sin\theta~ \tilde{a}^{\dagger}+\cos\theta~ \tilde{b}^{\dagger}.  \label{relaqaq}
\end{align}
The quantum state can be simply rewritten as
\begin{align}
\ket{\psi} ={\cal N}e^{\lambda\left(-\alpha^\dagger \tilde{\alpha}^\dagger+\beta^\dagger \tilde{\beta}^\dagger\right)}\ket{0}.
\end{align}
It is straightforward to compute its norm
\begin{align}
\bra{\psi}\ket{\psi} ={\cal N}^2\cdot \frac{1}{(1-|\lambda|^2)^2}.
\end{align}
Thus we find 
\begin{align}
{\cal N}=1-|\lambda|^2.  \label{norn}
\end{align}

\subsection{Generalized reduced density matrix}
Now we would like to trace out $(b^\dagger,\tilde{b}^\dagger)$, i.e.  CFT$_{(2)}$ to obtain the reduced density matrix $\rho_A$. 
By employing (\ref{sumz}) we can write it as follows (we write $c\equiv\cos 2\theta$ and $s\equiv\sin 2\theta$)
\begin{align}
\rho_A&=\mbox{Tr}_B\ket{\psi} \bra{\psi} \nonumber\\
&=\sum_{n,n'}\bra{n}_{b}\bra{n'}_{\tilde{b}}\left(\ket{\psi} \bra{\psi}\right)\ket{n} _{b}\ket{n'}_{\tilde{b}}\nonumber\\
&={\cal N}^2\int\frac{dzd\bar{z}}{\pi}\int\frac{dwd\bar{w}}{\pi}
e^{-|z|^2-|w|^2}\nonumber\\
&\quad\times \bra{0}_B e^{\bar{z}b+\bar{w}\tilde{b}}e^{\lambda c(a^\dagger \tilde{a}^\dagger+b^\dagger \tilde{b}^\dagger)
+\lambda s(a^\dagger \tilde{b}^\dagger+b^\dagger \tilde{a}^\dagger)}\ket{0}_{A}\ket{0}_{B}\nonumber\\
&\ \ \  \times \bra{0}_{A}\bra{0}_B e^{\bar{\lambda}c(-a\tilde{a}+b\tilde{b})+\bar{\lambda}s(a\tilde{b}+b\tilde{a})}e^{zb^\dagger+w\tilde{b}^\dagger}
\ket{0}_B.
\end{align}
By applying the formula (\ref{formu}) we can evaluate the above matrix element for the $(b^\dagger,\tilde{b}^\dagger)$ modes. This leads to the expression
\begin{align}
\rho_A&={\cal N}^2\int\frac{dzd\bar{z}}{\pi}\int\frac{dwd\bar{w}}{\pi}e^{-|z|^2-|w|^2}\nonumber\\
& \ \ \ \ e^{\lambda (\!-c a^\dagger \tilde{a}^\dagger\!+\!s\bar{z}\tilde{a}^\dagger\!+\!s\bar{w}a^\dagger\!+\!c\bar{z}\bar{w})}\ket{0}_A 
\bra{0} e^{\bar{\lambda} (\!-\!c a\tilde{a}\!+\!sz\tilde{a}\!+\!sw a\!+\!c zw)}.
\label{rhoone}
\end{align}

\subsection{Second Renyi pseudo entropy for subsystem $A$}
Now we would like to the second Renyi pseudo entropy
\begin{align}
  S^{(2)}_A=-\log\mbox{Tr}\rho_{A}^2.
\end{align}
For this let us evaluate $\mbox{Tr}(\rho_A)^2$. This can be done by multiplying (\ref{rhoone}):
\begin{align}
&\mbox{Tr}(\rho_A)^2  \nonumber\\
&={\cal N}^4\int\frac{dzd\bar{z}}{\pi}\int\frac{dwd\bar{w}}{\pi}\int\frac{dxd\bar{x}}{\pi}\int\frac{dyd\bar{y}}{\pi} 
e^{-|z|^2-|w|^2-|x|^2-|y|^2}\nonumber\\
&\ \ \ \times \bra{0}e^{\bar{\lambda} (-c a\tilde{a}+sx\tilde{a}+sy a+c xy)}
\cdot e^{\lambda (-c a^\dagger \tilde{a}^\dagger+s\bar{z}\tilde{a}^\dagger+s\bar{w}a^\dagger+c\bar{z}\bar{w})}\ket{0}\nonumber\\
&\ \ \  \times \bra{0}e^{\bar{\lambda} (-c a\tilde{a}+sz\tilde{a}+sw a+c zw)}
\cdot e^{\lambda (-c a^\dagger \tilde{a}^\dagger+s\bar{x}\tilde{a}^\dagger+s\bar{y}a^\dagger+c\bar{x}\bar{y})}\ket{0}.
\end{align}
Again we evaluate the matrix elements by using (\ref{formu}) and obtain
\begin{align}
& \mbox{Tr}(\rho_A)^2  \nonumber\\
&=\int\frac{dzd\bar{z}}{\pi} \int \frac{dwd\bar{w}}{\pi} \int \frac{dxd\bar{x}}{\pi} \int \frac{dyd\bar{y}}{\pi}
e^{-|z|^2 -|w|^2 -|x|^2 -|y|^2}\nonumber\\
&\quad\times \frac{(1-|\lambda|^2)^4}{(1-|\lambda|^2c^2)^2} 
\cdot e^{\lambda c(\bar{z}\bar{w} +\bar{x}\bar{y}) +\bar{\lambda} c(zw+xy)} \nonumber\\
&\ \ \ \ \  \times \exp\left[\frac{1}{1-|\lambda|^2c^2}\left(|\lambda|^2 s^2(y\bar{w}+\bar{y}w+x\bar{z}+z\bar{x})\right.\right.\nonumber\\
&\quad\left.\left.
-|\lambda|^2\lambda cs^2(\bar{z}\bar{w} +\bar{x}\bar{y})
-|\lambda|^2 \bar{\lambda} cs^2(zw+xy)\right)\right].
\end{align}
Finally by performing the gaussian integral of $(z,w,x,y)$,
we obtain
\begin{align}
  \bTr{\rho_A}^2 
  &=\frac{(1-|\lambda|^2)^2} {\left(-2 |\lambda|^2 \cos (4 \theta)+|\lambda|^4+1\right)},
\end{align}
where we employed the integration formula 
\begin{align}
   \int dzd\bar{z} e^{-\frac{1}{2}x^T A x} = \frac{\pi}{\sqrt{\det A}}.
\end{align}

The above analysis corresponds to the non-zero modes of the free scalars
by regarding $(\alpha^{(1)}_n,\tilde{\alpha}^{(1)}_n)$ and 
$(\alpha^{(2)}_n,\tilde{\alpha}^{(2)}_n)$ 
as the creation and annihilation operators of the harmonic oscillators  $(a^\dagger,a)$ and $(b^\dagger,b)$
as in (\ref{relaq}). From this relation, compute the pseudo entropy for the non-zero modes. 
We set $\lambda_m = e^{-2m(it+\beta/4)}$, so $|\lambda_m|^2=e^{-m\beta}$
\begin{align}
  \qty[\mathrm{Tr}\rho_A^2]_m = \frac{(1-e^{-m\beta})^2}{\left(-2 e^{-m\beta} \cos (4 \theta)+e^{-2m\beta}+1\right)}.
\end{align}
By taking the product of these over $m$, we obtain the non-zero model contributions:
\begin{align}
    \qty[\mathrm{Tr}\rho_A^2]_{\text{non-zero}}=& \prod_{n = 1}^{\infty} \frac{(1-e^{-m\beta})^2}{\left(-2 e^{-m\beta} \cos (4 \theta)+e^{-2m\beta}+1\right)} \label{rtrr_p}\\
    =& \frac{2 \sin 2\theta \qty[\eta(\frac{\rmi\beta}{2\pi})]^3}{\vartheta_1(\frac{2\theta}{\pi},\frac{\rmi\beta}{2\pi})}.
\end{align}
by using
\begin{align}
  &\eta(\tau) = e^{\frac{\pi i \tau}{12}} \prod_{n=1}^{\infty} (1-e^{2\pi i n \tau}), \\
  &\vartheta_1(\nu,\tau) \\
  &= 2 e^{\frac{\pi i\tau}{4}} \sin (\pi\nu) \prod_{n=1}^{\infty} (1-e^{2\pi i\nu}e^{2\pi in\tau }) (1- e^{-2\pi i\nu}e^{2\pi in\tau }) \nonumber\\
  &\ge 2 e^{\frac{\pi i\tau}{4}} \sin (\pi\nu)  \prod_{n=1}^{\infty} (1-e^{2\pi in\tau })^2 \geq 2 \sin u\ \qty[\eta(\tau)]^3 \label{ellip_ge}.
\end{align}
Using \eqref{ellip_ge} for a real value of $\theta$, we find
\begin{align}
  \qty[\mathrm{Tr}\rho^2]_{\text{non-zero}}\le 1,
\end{align}
as expected.

Finally we would like to add the contributions from the zero modes. The density of state for the zero mode looks like
\begin{align}
  \rho^{(0)}_{AB}=\tilde{\mathcal{N}}^2\cdot e^{-\frac{\beta}{4}H_0}\ket{B^{(0)}}\bra{B^{(0)}}e^{-\frac{\beta}{4}H_0},
\end{align}
where 
\begin{align}
  \ket{B^{(0)}}=\sum_{n,w=-\infty}^\infty  \ket{n,w}_{A}\ket{n,-w}_{B},
\end{align}
and $H_0=\frac{n^2}{R_1^2}+\frac{n^2}{R_2^2}+\frac{w^2 R_1^2}{4}+\frac{w^2 R_2^2}{4}$. 
Note that the overall normalization ${\cal \tilde{N}}$ is chosen so that $\mbox{Tr}[\rho_{AB}]=1$. 
We get it,
\begin{align}
  \frac{1}{\tilde{\mathcal{{N}}}^2} &= \sum_{n,w} \exp[-\frac{\beta+\bar{\beta}}{4}\qty(n^2\qty(\frac{1}{R_1^2}+\frac{1}{R_2^2})+ \frac{w^2(R_1^2+R_2^2)}{4})] \nonumber\\
  &= \vartheta_3 \qty(0, \frac{\rmi\beta}{2\pi}\qty(\frac{1}{R_1^2}+\frac{1}{R_2^2})) \vartheta_3 \qty(0, \frac{\rmi\beta}{2\pi}\frac{R_1^2+R_2^2}{4}) .
\end{align}
Thus the zero mode contribution is found to be
\begin{align}
  \qty[\mathrm{Tr}\rho_A^2]_{\text{zero}} 
  &= \frac{\vartheta_3 \qty(0, \frac{\rmi \beta}{\pi}\qty(\frac{1}{R_1^2}+\frac{1}{R_2^2})) \vartheta_3 \qty(0, \frac{\rmi \beta}{\pi}\frac{R_1^2+R_2^2}{4})}
  {\vartheta_3 \qty(0, \frac{\rmi\beta}{2\pi}\qty(\frac{1}{R_1^2}+\frac{1}{R_2^2}))^2 \vartheta_3 \qty(0, \frac{\rmi\beta}{2\pi}\frac{R_1^2+R_2^2}{4})^2} .
\end{align}
where we use $\vartheta_3 (0,\tau) = \sum_{n=-\infty}^{\infty} e^{\pi in^2\tau}$. Finally, by combining the result of non-zero and zero mode, we obtain the complete expression given by (\ref{stwojan}).

When we set $\theta = \frac{\pi}{4}+\rmi\frac{\delta}{2}$, we have $R_1 = R \sqrt{\frac{1+i\sinh \delta}{\cosh\delta}}$ and $R_2 = R\sqrt{\frac{\cosh\delta}{1+i\sinh\delta}}$. Thus we note $R_2^*=R_1$ and $R_1^*=R_2$. This leads to the expression (\ref{deltape}).

\subsection{Imagitivity for subsystem $A$}

Next we compute the imagitivity $||\rho^\dagger_{A}-\rho_{A}||^2$ in this free scalar Janus model. We can write
\begin{align}
&\mbox{Tr} \rho_A\rho_A^\dagger \nonumber\\
&={\cal N}^4\int\frac{dzd\bar{z}}{\pi}\int\frac{dwd\bar{w}}{\pi}\int\frac{dxd\bar{x}}{\pi}\int\frac{dyd\bar{y}}{\pi}
e^{-|z|^2-|w|^2-|x|^2-|y|^2}\nonumber\\
&\ \ \ \times \bra{0}e^{\bar{\lambda} (-c^* a\tilde{a} +s^*x\tilde{a} +s^*y a +c^* xy)}
\cdot e^{\lambda (-c a^\dagger \tilde{a}^\dagger+s\bar{z}\tilde{a}^\dagger+s\bar{w}a^\dagger+c\bar{z}\bar{w})}\ket{0}\nonumber\\
&\ \ \  \times \bra{0}e^{\bar{\lambda} (-c a\tilde{a}+sz\tilde{a}+sw a+c zw)}
\cdot e^{\lambda (-c^* a^\dagger \tilde{a}^\dagger +s^*\bar{x}\tilde{a}^\dagger +s^*\bar{y}a^\dagger +c\bar{x}\bar{y})}\ket{0}.
\end{align}
We can perform the gaussian integral in the above explicitly.
By setting $c_n=\cos n\theta$ and $\bar{c}_n=\cos n\bar{\theta}$,
the final result looks like
\begin{align}
  &\mathrm{Tr}\rho_A\rho_A^\dagger \nonumber\\
  =& \frac{2(1-|\lambda|^2)^2} {(-|\lambda|^2 (4 c_2 \bar{c}_2+c_4+\bar{c}_4)+2 |\lambda|^2 (|\lambda|^2+1)+2)}\nonumber\\
  =&  \frac{(1-e^{-m\beta})^2} {1 - e^{-m\beta} ((\cos 2\theta + \cos 2\bar{\theta})^2-2) + e^{-2m\beta}}.
\end{align}
Thus for the non-zero mode contributions of the free scalar CFT we obtain 
\begin{align}
  &\qty[\text{Tr}\rho_A\rho_A^\dagger]_{\text{non-zero}} \nonumber\\
  =& \prod_{n=1}^\infty \frac{(1-e^{-m\beta})^2}{\left(1- e^{-m\beta} ((\cos 2\theta + \cos 2\bar{\theta})^2-2) +e^{-2m\beta}\right)}. \nonumber\\
\end{align}
We can calculate the zero mode contribution as
\begin{align}
  \qty[\Tr \rho_A \rho_A^\dagger]_{\text{zero}}
  = &\frac{\vartheta_3 \qty(\frac{\rmi\beta}{2\pi}\qty(\frac{1}{R_1^2}+\frac{1}{R_2^2}+\frac{1}{R_1^{*2}}+\frac{1}{R_2^{*2}}))}
  {\vartheta_3 \qty( \frac{\rmi\beta}{2\pi}\qty(\frac{1}{R_1^2}+\frac{1}{R_2^2})) \vartheta_3 \qty(\frac{\rmi\beta}{2\pi}\qty(\frac{1}{R_1^{*2}}+\frac{1}{R_2^{*2}}))} \nonumber\\
  &\quad\times
  \frac{\vartheta_3 \qty(\frac{\rmi\beta}{2\pi}\frac{R_1^2+R_2^2+R_1^{*2}+R_2^{*2}}{4})}{\vartheta_3 \qty(\frac{\rmi\beta}{2\pi}\frac{R_1^2+R_2^2}{4}) \vartheta_3 \qty(\frac{\rmi\beta}{2\pi}\frac{R_1^{*2}+R_2^{*2}}{4})}.
\end{align}

We focus on the imaginary Janus deformation $\theta = \frac{\pi}{4}+\rmi\frac{\delta}{2}$. Then since $R^*_1=R_2$ and $R^*_2=R_1$, we note that $\qty[\Tr \rho_A \rho_A^\dagger]_{\text{zero}}= \qty[\Tr \rho_A^2]_{\text{zero}}$.


In this case, the non-zero mode contributions get simplified to
\begin{align}
  &\mathrm{Tr} [\rho_{A}^\dagger \rho_{A}]_{\mathrm{non-zero}}\nonumber\\
  =& \prod_{n=1}^\infty \frac{2 (1-e^{-m\beta})^2} {2-e^{-m\beta} (4 \sinh^2 \delta - 2\cosh (2 \delta))+2 e^{-m\beta} (e^{-m\beta}+1)}\nonumber\\
  =& \prod_{n=1}^\infty \qty(\frac{1-e^{-m\beta}}{1+e^{-m\beta}})^2 = \frac{2\eta(\frac{\rmi\beta}{2\pi})^3}{\vartheta_2(0,\frac{\rmi\beta}{2\pi})}.
\end{align}

Finally the normalized imagitivity is found to be
\begin{align}
   \frac{2 \Tr[\rho_{A}^\dagger \rho_{A}] - 2 \mathrm{Re} \Tr[\rho_{A}^2]} {\abs{\Tr[\rho_{A}^2]}}
   = 2\qty[\frac{\vartheta_1(\frac{2\theta}{\pi}, \frac{\rmi\beta}{2\pi})}{\vartheta_2(0,\frac{\rmi\beta}{2\pi}) \sin 2\theta}-1],
\end{align}
where the zero mode contributions simply cancels in the ratio. 
We plotted this in Fig.\ref{fig:J_osi_imag_log}.

\subsection{Imagitivity for the total system}

Finally,we calculate the imagitivity for the total system $\rho_{AB}$. It is obvious to have $\Tr \rho^2_{AB} = 1$ as the state is pure. Then we would like to evaluate:
\be
\mbox{Tr}\rho_{AB}\rho_{AB}^\dagger 
    = \Tr\ket{\psi}\bra{\psi} \ket{\bar{\psi}}\bra{\bar{\psi}}=
    |\bra{\psi} \ket{\bar{\psi}}|^2.
\ee
    The relevant inner product can be done as in our previous computations:
\begin{align}
    &\bra{\psi}\ket{\bar{\psi}} \nonumber\\
    &= (1-|\lambda|^2)^2 \bra{0} 
    e^{\bar{\lambda} \cos\bar{\theta} (-a\tilde{a} + b\tilde{b}) + \bar{\lambda} \sin\bar{\theta}(a\tilde{b} + b\tilde{a})}\nonumber\\
    &\quad e^{\lambda \cos\theta (-a^\dagger \tilde{a}^\dagger + b^\dagger \tilde{b}^\dagger) 
    + \lambda \sin\theta (a^\dagger \tilde{b}^\dagger + b^\dagger \tilde{a}^\dagger)}\ket{0} \\
    &= (1-|\lambda|^2)^2 \int \frac{dxd\bar{x}}{\pi} \int \frac{dyd\bar{y}}{\pi} \int \frac{dzd\bar{z}}{\pi} \int \frac{dwd\bar{w}}{\pi} \nonumber\\
    &\quad\times e^{-x\bar{x} -y\bar{y} -z\bar{z} -w\bar{w}} \nonumber\\
    &\ \ \ \times \bra{0} e^{\bar{\lambda} \cos\bar{\theta} (-a\tilde{a} + b\tilde{b}) + \bar{\lambda} \sin\bar{\theta}(a\tilde{b} + b\tilde{a})}
    e^{a^\dagger x+\tilde{a}^\dagger y +b^\dagger z+\tilde{b}^\dagger w} \ket{0} \nonumber\\
    &\ \ \ \times \bra{0} e^{a \bar{x}+\tilde{a}\bar{y} +b\bar{z}+\tilde{b}\bar{w}}
    e^{\lambda \cos\theta (-a^\dagger \tilde{a}^\dagger + b^\dagger \tilde{b}^\dagger) 
    + \lambda \sin\theta (a^\dagger \tilde{b}^\dagger + b^\dagger \tilde{a}^\dagger)}\ket{0} \\
    &= (1-|\lambda|^2)^2 \int \frac{dxd\bar{x}}{\pi} \int \frac{dyd\bar{y}}{\pi} \int \frac{dzd\bar{z}}{\pi} \int \frac{dwd\bar{w}}{\pi} \nonumber\\
    &\quad\times
    e^{-x\bar{x}-y\bar{y}-z\bar{z}-w\bar{w}} \nonumber\\
    &\ \ \ \times e^{\bar{\lambda} \cos\bar{\theta} (-x y + z w) + \bar{\lambda} \sin\bar{\theta}(x z + y w)}\nonumber\\
    &\quad\times
    e^{\lambda \cos\theta (-\bar{x}\bar{y} + \bar{z}\bar{w}) + \lambda \sin\theta(\bar{x}\bar{z} + \bar{y}\bar{w})} \\
    &= \frac{(1-|\lambda|^2)^2}{1-2|\lambda|^2 \cos(\theta - \bar{\theta}) + |\lambda|^4}.
\end{align}
For the real valued Janus deformation $\theta = \bar{\theta}$, we have $\bra{\psi} \ket{\bar{\psi}}=1$. Thus the imagitivity is vanishing. For the imaginary Janus deformation $\theta=\frac{\pi}{4}+\rmi\frac{\delta}{2}$, we obtain the non-zero mode contributions
\begin{align}
  &\qty[\Tr \rho_{AB}\rho_{AB}^\dagger]_{\text{non-zero}} \nonumber\\
  &= \prod_{m=1}^\infty \qty(\frac{1-2 e^{-m\beta} +e^{-2m\beta}} {1-2 e^{-m\beta} \cosh\delta + e^{-2m\beta}})^2 \\
  &= \frac{4 \sin^2 \qty(\frac{\delta}{2}i)\  \eta(\frac{\rmi\beta}{2\pi})^6}{\vartheta_1(\frac{\delta}{2\pi}i, \frac{\rmi\beta}{2\pi})^2}.
\end{align}
Moreover, the zero mode contribution reads 
\begin{align}
  &\qty[\Tr \rho_{AB}\rho_{AB}^\dagger]_{\text{zero}} \nonumber\\ 
  &= \frac{\vartheta_3 \qty(0, \frac{\rmi\beta}{2\pi}\qty(\frac{1}{R_1^2}+\frac{1}{R_2^2}+\frac{1}{\bar{R}_1^2}+\frac{1}{\bar{R}_2^2})) 
    \vartheta_3 \qty(0, \frac{\rmi\beta}{2\pi}\frac{R_1^2+R_2^2+\bar{R}_1^2+\bar{R}_2^2}{4})}
    {\vartheta_3 \qty(0, \frac{\rmi\beta}{2\pi}\qty(\frac{1}{R_1^2}+\frac{1}{R_2^2}))^2 
    \vartheta_3 \qty(0, \frac{\rmi\beta}{2\pi}\frac{R_1^2+R_2^2}{4})^2} \nonumber\\
    &=1.
\end{align}
Thus, the imagitivity is
\begin{align}
  ||\rho^\dagger_{AB}-\rho_{AB}||^2
  =2\qty[\frac{4 \sin^2 \qty(\frac{\delta}{2}i)\  \eta(\frac{\rmi\beta}{2\pi})^6}{\vartheta_2(\frac{\delta}{2\pi}i, e^{-\beta/2})^2} -1].
\end{align}

\section{Detail calculations of entanglement/pseudo entropy in Janus wormhole} \label{janusent}

\begin{figure}[H]
  \centering
  \includegraphics[width=0.8\columnwidth]{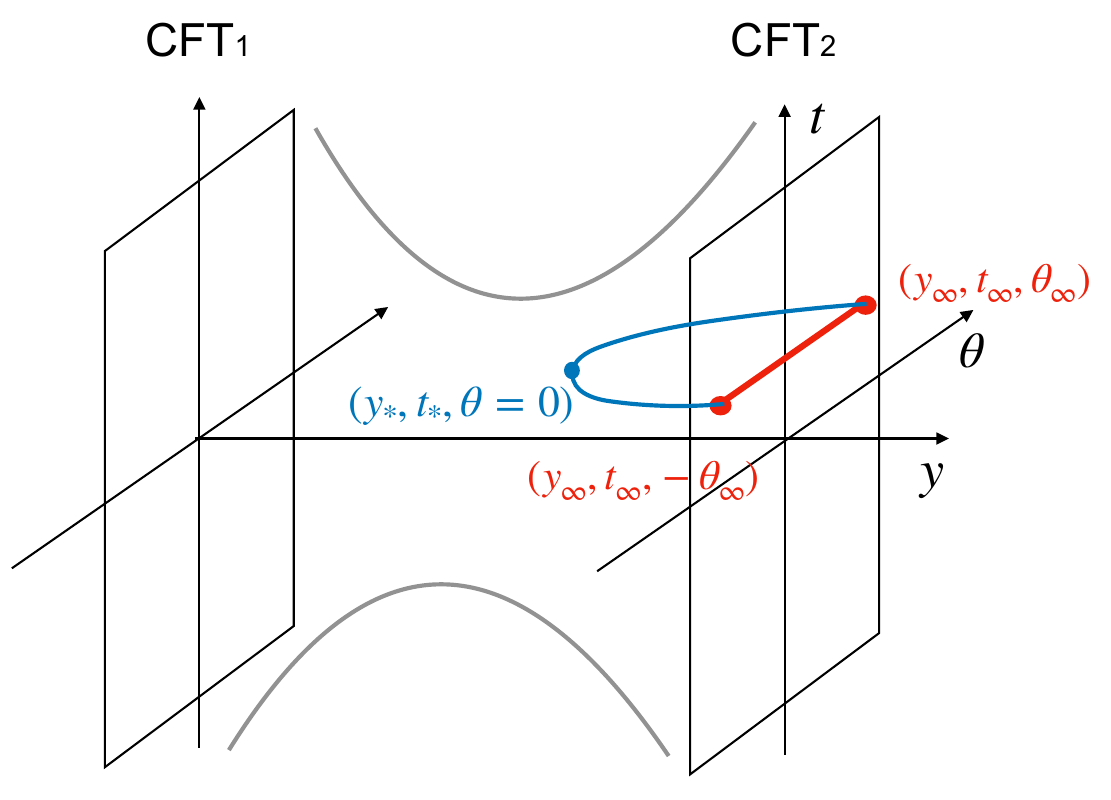}
  \caption{Subsystem $A$ and extremal surface which is related to holographic entanglement/pseudo entropy between two CFTs. }
  \label{extsur}
\end{figure}

Here we provide the explicit calculation of holographic entanglement/pseudo entropy in Janus wormhole discussed in Sec.\ref{janholo}. 
The earlier calculation was given in \cite{Nakaguchi:2014eiu}
\footnote{They used a function $\tilde{g}(y)$ in the metric, which corresponds to $f(y) = 1 / \tilde{g}(y)^2$ in our notation. Moreover, we set AdS radius $L = 1$ here. }
, where they discussed for the usual Janus black hole, where $\gamma$ is real valued. Our main focus is to extend this to the solution for imaginary $\gamma$.
We start from the metric given in Eq.\eqref{jansol1} \eqref{jansol2}
\begin{align}
  ds^2 &= dy^2 + \frac{r_0^2 f(y)}{\cosh^2 r_0 t} \left( - d t^2 + d \theta^2 \right), \tag{\ref{jansol1}} \\
  f(y) &= \frac{1}{2} \left( 1 + \sqrt{1 - 2 \gamma^2} \cosh 2 y \right). \tag{\ref{jansol2}}
\end{align}
Here $r_0$ is a constant corresponding to horizon radius of BTZ black hole when we take $\gamma = 0$. 
In this geometry, the AdS boundary is located at $(y, t) = (y_\infty, t_\infty)$ which is related to the UV cutoff of the dual CFT $\epsilon_{\mathrm{CFT}}$ via
\begin{equation}
  \epsilon_{\mathrm{CFT}} = \frac{2}{\sqrt[4]{1 - 2 \gamma^2} r_0} e^{- y_{\infty}} \cosh r_0 t_\infty.
\end{equation}
The surface area (geodesic length) functional is given by
\begin{equation}
  A[t(y), \theta(y)] = \int \dd{y} \sqrt{1 +  \frac{r_0^2 f(y)}{\cosh^2 r_0 t} (-\dot{t}^2 + \dot{\theta}^2)},  \label{areafn} 
\end{equation}
where dot denotes the derivative of $y$. 
To get the extremal surface area, we have to minimize this functional under the boundary condition that $t(y_{\infty}) = t_{\infty}$ and $\theta(y_\infty) = \pm \theta_\infty$, that gives the Euler-Lagrange equations. 
More explicitly, one equation is
\begin{equation}
  \begin{aligned}
    & \dv{y} \frac{\delta A}{\delta \dot{\theta}} = 0, \\
    \Leftrightarrow ~ & J = \frac{f(y)}{\cosh r_0 t} \frac{r_0^2 \dot{\theta}}{\sqrt{\cosh^2 r_0 t + r_0^2 f(y) (- \dot{t}^2 + \dot{\theta}^2)}}, \label{eqtheta}
  \end{aligned}
\end{equation}
where $J$ is a integral constant. 
The other equation is 
\begin{equation}
  \begin{aligned}
    \dv{y} \frac{\delta A}{\delta \dot{t}} - \frac{\delta A}{\delta t} = 0 
    ~ \Leftrightarrow ~ & \dv{y} \left( J \frac{\dot{t}}{\dot{\theta}} \right) = J r_0 \frac{- \dot{t}^2 + \dot{\theta}^2}{\dot{\theta}} \tanh r_0 t \\
    ~ \Leftrightarrow ~ & \dv[2]{t}{\theta} = r_0 \left[ 1 - \left( \dv{t}{\theta} \right)^2 \right] \tanh r_0 t. 
  \end{aligned}
\end{equation}
Using the condition $\theta|_{y = y_*} = 0$ and $\dd{t} / \dd{\theta}|_{y = y_*} = 0$ to determine integral constants, the solution is given by
\begin{equation}
  \sinh r_0 t = \sinh r_0 t_* \cosh r_0 \theta. \label{ttheta}
\end{equation}
Here, $t_*$ and $y_*$ are a location of returning point (Fig. \ref{extsur}). 
Substituting this into Eq.\eqref{eqtheta}, we obtain 
\ba
  \dot{t} &=& \frac{\cosh r_0 t}{r_0 \cosh r_0 t_*} \sqrt{\cosh^2 r_0 t - \cosh^2 r_0 t_*} \no
 &&\ \ \  \times\sqrt{\frac{f(y_*)}{f(y) (f(y) - f(y_*))}}, \label{tdoteq}
\ea
whose solution is given by
\begin{equation}
  \begin{aligned}
      \sqrt{1 - \frac{\sinh^2 r_0 t_*}{\sinh^2 r_0 t}} &= \tanh r_0 \theta \\ 
      &= \cosh r_0 t_* \\
      &\times \tanh \left[ \int_{y_*}^{y_\infty} dy \sqrt{\frac{f(y_*)}{f(y) (f(y) - f(y_*))}}\right]. 
  \end{aligned}
\end{equation}
Now we can determine the returning point $(y_*, t_*)$ as a function of boundary condition by
\begin{align}
  \frac{\sinh r_0 t_\infty}{\cosh r_0 \theta_\infty} &= \sinh r_0 t_*, \label{eqcond1} \\
  \frac{\sinh r_0 \theta_\infty}{\cosh r_0 t_\infty} &= \sinh \left[\int_{y_*}^{y_\infty} dy \sqrt{\frac{f(y_*)}{f(y) (f(y) - f(y_*))}} \right]. \label{eqcond2}
\end{align}
Substituting these solutions into eq.\eqref{areafn}, we get the exact form of area
\begin{equation}
  A (t_\infty, \theta_\infty) = 2 \int_{y_*}^{y_\infty} dy \sqrt{\frac{f(y)}{f(y) - f(y_*)}}. \label{areasol}
\end{equation}
This area has a UV divergence: at $y \sim y_\infty$ region, $f(y_{\infty}) \gg f(y_*)$ implies
\begin{equation}
  A \rightarrow 2 \int^{y_{\infty}} \dd{y} \sim 2 y_{\infty} = 2 \log \frac{2 \cosh r_0 t_{\infty}}{\sqrt[4]{1 - 2 \gamma^2} r_0 \epsilon_{\mathrm{CFT}}}. 
\end{equation}
Therefore, we define the renormalized area
\begin{equation}
  \begin{aligned}
    A^{\mathrm{(ren)}} :=& A + 2 \log \epsilon_{\mathrm{CFT}} \\
    =& A - \frac{1}{2} \log(1 - 2 \gamma^2) + 2 \log \frac{2 \cosh r_0 t_\infty}{r_0} - 2 y_\infty. 
  \end{aligned}
\end{equation}

Let us consider large subsystem $\theta_{\infty} \gg r_0^{-1}$ and early time limit $(t_\infty \ll \theta_\infty)$. In this case, Eq.\eqref{eqcond1} is much smaller than $1$; 
\begin{equation}
  \sinh r_0 t_* \simeq r_0 t_* \simeq 2 e^{- r_0 \theta_{\infty} } \sinh r_0 t_\infty
\end{equation}
The $y_*$ is determined by Eq.\eqref{eqcond2}, which is much greater than $1$, 
\begin{equation}
  \left[\int_{y_*}^{y_\infty} dy \sqrt{\frac{f(y_*)}{f(y) (f(y) - f(y_*))}}\right] = r_0 \theta_\infty - \log \cosh r_0 t_\infty.
\end{equation}
In order for left hand side to be large, $y_*$ must be $y_* \ll 1$. 
In this limit, the integral can be evaluated as 
\ba
 && \left[\int_{y_*}^{y_\infty} dy \sqrt{\frac{f(y_*)}{f(y) (f(y) - f(y_*))}}\right] \no
 && = - \frac{1}{\sqrt{\kappa_+^2 - \kappa_-^2}} \log \left[ \frac{\kappa_+ + \sqrt{\kappa_+^2 - \kappa_-^2}}{4} y_*  \right],
\ea
where $\displaystyle \kappa_{\pm} := \sqrt{\frac{1 \pm \sqrt{1 - 2 \gamma^2}}{2}}$. 
The integral in Eq.\eqref{areasol} is approximated as 
\begin{equation}
  \begin{aligned}
      & \int_{y_*}^{y_\infty} dy \sqrt{\frac{f(y)}{f(y) - f(y_*)}} \\
      & \simeq \frac{\kappa_+}{\sqrt{\kappa_+^2 - \kappa_-^2}} \log \left[ \kappa_+ + \sqrt{\frac{\kappa_+^2 - \kappa_-^2}{4}} y_* \right] \\
      & \ \ \ - \log \left[ \frac{\kappa_+^2 - \kappa_-^2}{\sqrt{\kappa_+^2 - \kappa_-^2}} \right] + y_{\infty}. 
  \end{aligned}
\end{equation}
Combining them, the renormalized area reduces to 
\ba
 && A^{\mathrm{(ren)}} \underset{(t_\infty \ll \theta_\infty)}{\simeq} 2 \kappa_+ r_0 \theta_{\infty} + 2 (1 - \kappa_{+}) \log \cosh r_0 t_{\infty} \no
 &&\ \  - 2 \log \left[ \frac{\kappa_+ + \sqrt{\kappa_+^2 - \kappa_-^2}}{2} r_0 \right],
\ea
which is exactly Eq.\eqref{earlylim}. 

In the late time limit $(t_\infty \gg \theta_\infty \gg r_0)$, 
Eq.\eqref{eqcond1} and \eqref{eqcond2} lead to
\ba
&&  e^{r_0 (\theta_{\infty} - t_{\infty})} \no
 && \!\simeq\! 2\! e^{- r_0 t_*}\! \simeq\! \frac{1}{2}\! \exp \left[\!\int_{y_*}^{y_\infty}\! dy\! \sqrt{\frac{f(y_*)}{f(y) \!(\!f(y) \!- \!f(y_*)\!)}} \right]  (\!\ll \!1).\no
\ea
This implies $y_* \gg 1$, where the integrals in Eq.\eqref{eqcond2} and \eqref{areasol} reduce to
\begin{align}
  \int_{y_*}^{y_\infty} dy \sqrt{\frac{f(y_*)}{f(y) (f(y) - f(y_*))}} &\simeq \frac{2}{\sqrt[4]{1 - 2 \gamma^2}} e^{- y_*}, \\
  \int_{y_*}^{y_\infty} dy \sqrt{\frac{f(y)}{f(y) - f(y_*)}} &\simeq y_{\infty} - y_* + \log 2,
\end{align}
where we also used $y_{\infty} - y_* \gg 1$. 
the renormalized area becomes
\begin{equation}
  A^{\mathrm{(ren)}} \underset{(t_\infty \gg \theta_\infty)}{\simeq} 2 (r_0 \theta_{\infty} - \log r_0). 
\end{equation}
which is Eq.\eqref{latelim}. 

\begin{figure}[h]
  \centering
  \includegraphics[width=0.8\columnwidth]{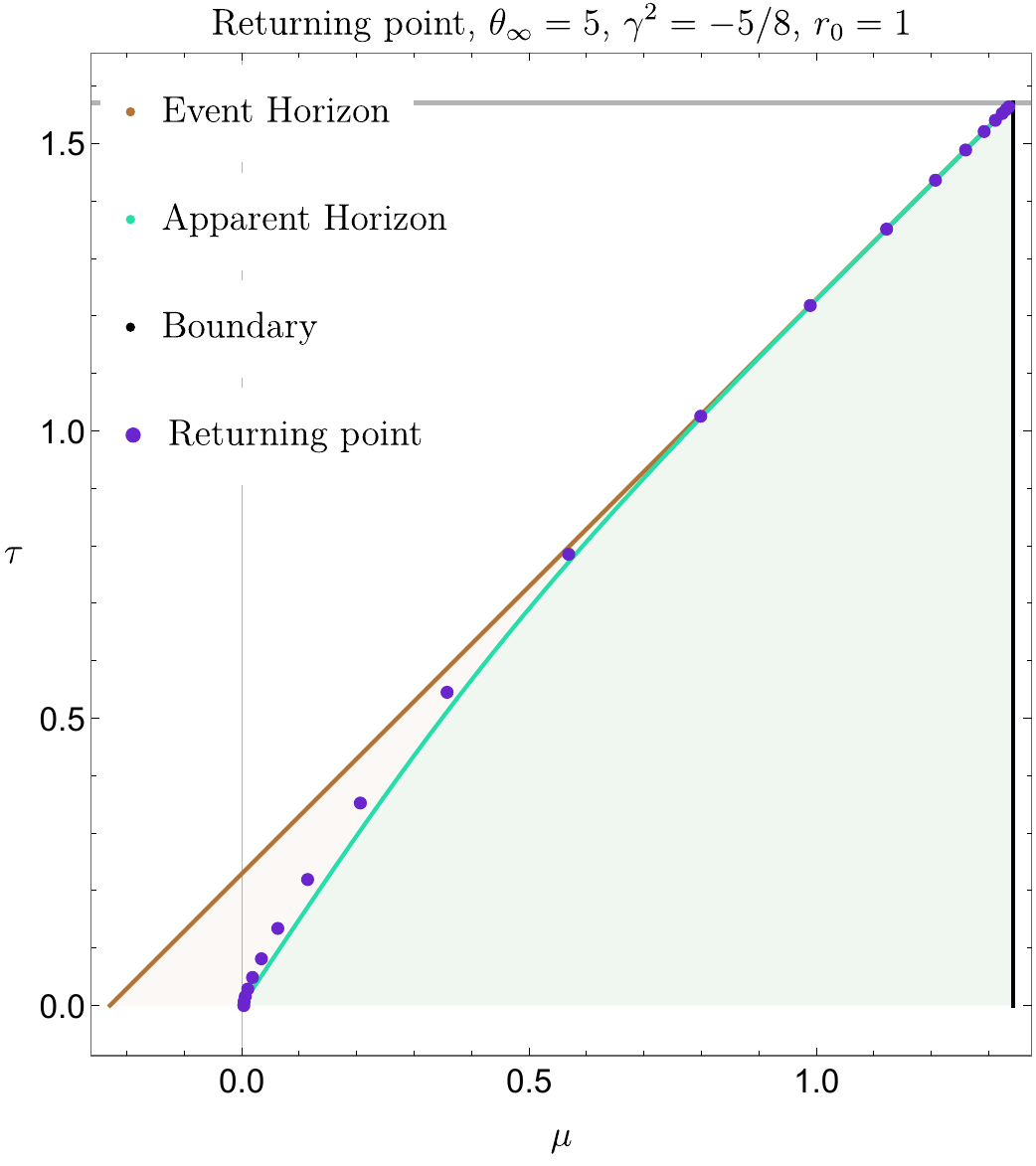}
  \caption{The location of returning point in imaginary Janus spacetime. At early time ($t_\infty \ll \theta_\infty$), the returning point is close to apparent horizon, which is located outside the event horizon. }
  \label{ijanre}
\end{figure}

The time evolution of renormalized area are shown in Fig.\ref{imjaen} in Sec.\ref{janholo}. 
We also show the time evolution of returning point in \ref{ijanre}.

\bibliography{NH.bib}


\end{document}